\documentclass[longauth]{aa}  

\newcommand \HII {\ion{H}{II}}
\newcommand \ha {\textup{H\ensuremath{\alpha}}}
\newcommand \SN {\ensuremath{\mathrm{S/N}}}

\newcommand \ngc {NGC\,6946}

\newcommand \sigsfr {\ensuremath{{\Sigma_\mathrm{SFR}}}}
\newcommand \sigsfrunit {\ensuremath{\mathrm{M_{\odot}\,yr^{-1}\,kpc^{-2}}}}
\newcommand \sfrunit {\ensuremath{\mathrm{M_{\odot}\,yr^{-1}}}}

\newcommand \sigmolmass {\ensuremath{\mathrm{\Sigma_{mol}}}}
\newcommand \sigmolmassunit {\ensuremath{\mathrm{M_{\odot}\,pc^{-2}}}}
\newcommand \tdepl {\ensuremath{\mathrm{\tau_{depl,\,150\,pc}^{mol}}}}
\newcommand \fdense {\ensuremath{f_\mathrm{{dense}}}}
\newcommand \sfedense {\ensuremath{\mathrm{SFE_{dense}}}}

\newcommand \hcn {\chem{HCN}{10}}
\newcommand \hco {\chem{HCO^{+}}{10}}
\newcommand \hnc {\chem{HNC}{10}}
\newcommand \cstwo {\chem{CS}{21}}
\newcommand \csthree {\chem{CS}{32}}
\newcommand \hcnten {\chem{HC_{3}N}{109}}
\newcommand \hcnsixt {\chem{HC_{3}N}{1615}}
\newcommand \ntwoh {\chem{N_{2}H^{+}}{10}}
\newcommand \cch {\chem{C_{2}H}{10}}
\newcommand \chohtwo {\chem{CH_{3}OH}{21}}
\newcommand \chohthree {\chem{CH_{3}OH}{32}}
\newcommand \htwoco {\chem{H_{2}CO}{21}}
\newcommand \co {\chem{CO}{10}}
\newcommand \cotwo {\chem{CO}{21}}

\newcommand{\uv}{{$u{-}v$\,}}
\newcommand{\hcnhnc}{{$\chem{HCN}/\chem{HNC}$}}
\newcommand{\hnchcn}{{$\chem{HNC}/\chem{HCN}$}}

\newcommand{\cmark}{\textcolor{green!80!black}{\ding{51}}}
\newcommand{\xmark}{\textcolor{red}{\ding{55}}}
\usepackage{graphicx}	
\usepackage{amsmath}	
\usepackage{multirow}
\usepackage{siunitx}
\usepackage{lipsum}
\usepackage{pifont}
\usepackage{xcolor}
\usepackage{mychem}
\usepackage{txfonts}
\usepackage{bm}	        
\usepackage{csvsimple}
\usepackage{siunitx} 
\usepackage[utf8]{inputenc}
\usepackage{ulem}
\usepackage{xparse}
\usepackage[pdfpagelabels=false]{hyperref}	
\hypersetup{colorlinks=true,linkcolor=blue,citecolor=blue,filecolor=blue,urlcolor=blue,}

\begin{document} 

\renewcommand{\figureautorefname}{Fig.} 
\renewcommand{\equationautorefname}{Eq.} 
\renewcommand{\sectionautorefname}{Section} 
\renewcommand{\subsectionautorefname}{Section}
\renewcommand{\subsubsectionautorefname}{Section}
\renewcommand{\appendixautorefname}{Appendix} 


   \title{A 2-3~mm high-resolution molecular line survey towards the centre of the nearby spiral galaxy NGC~6946\thanks{Based on observations carried out with the IRAM Plateau de Bure Interferometer (PdBI). IRAM is supported by INSU/CNRS (France), MPG (Germany) and IGN (Spain).}}

   \author{Cosima Eibensteiner \inst{1} \fnmsep\thanks{\email{eibensteiner@astro.uni-bonn.de}}
          \and
          Ashley T. Barnes \inst{1}
          \and
          Frank~Bigiel \inst{1}
          \and 
          Eva~Schinnerer \inst{2}
          \and
          Daizhong~Liu \inst{3}
          \and 
          David~S.~Meier \inst{4}
          \and \\
          Antonio~Usero \inst{5} 
          \and
          Adam~K.~Leroy \inst{6}
          \and
          Erik~Rosolowsky \inst{7}
          \and
          Johannes~Puschnig \inst{1}
          \and 
          Ilin~Lazar \inst{8}
          \and
          Jérôme~Pety \inst{9},\inst{10}
          \and
          Laura~A.~Lopez \inst{6}
          \and 
          Eric~Emsellem \inst{11},\inst{12}
          \and
          Ivana~Be{\v{s}}li{\'c} \inst{1}
          \and
          Miguel~Querejeta \inst{5}
          \and
          Eric~J.~Murphy \inst{13} 
          \and
          Jakob~den~Brok \inst{1}
          \and\\
          Andreas~Schruba \inst{3}
          \and 
          Mélanie~Chevance \inst{14}
          \and 
          Simon~C.~O.~Glover \inst{15}
          \and
          Yu~Gao \inst{14}
          \and
          Kathryn~Grasha \inst{16}
          \and
          Hamid Hassani \inst{7}
          \and
          Jonathan D. Henshaw \inst{2}
          \and
          Maria~J.~Jimenez-Donaire \inst{5}
          \and
          Ralf~S.~Klessen \inst{15,17}
          \and
          J.~M.~Diederik Kruijssen \inst{14}
          \and
          Hsi-An Pan \inst{19}
          \and\\
          Toshiki~Saito \inst{2} \and
          Mattia~C.~Sormani \inst{15}
          \and 
          Yu-Hsuan~Teng \inst{20}
          \and 
          Thomas~G.~Williams \inst{2}
          }

   \institute{Argelander-Institut f\"ur Astronomie, Universit\"at Bonn, Auf dem H\"ugel 71, 53121 Bonn, Germany
              \and
              Max Planck Institut f{\"u}r Astronomie, K{\"o}nigstuhl 17, D-69117 Heidelberg, Germany
              \and
              Max-Planck-Institut f\"{u}r extraterrestrische Physik, Giessenbachstra{\ss}e 1, D-85748 Garching, Germany
              \and
              New Mexico Institute of Mining and Technology, 801 Leroy Place, Socorro, NM 87801, USA; National Radio Astronomy Observatory, PO Box O, 1003 Lopezville Road, Socorro, New Mexico 87801, USA
              \and
              Observatorio Astron{\'o}mico Nacional (IGN), C/ Alfonso XII 3, E-28014 Madrid, Spain
              \and
              Department of Astronomy, The Ohio State University, 4055 McPherson Laboratory, 140 West 18th Avenue, Columbus, OH 43210, USA
              \and
              4-183 CCIS, University of Alberta, Edmonton, Alberta, T6G 2E1, Canada
              \and
              Centre for Astrophysics Research, School of Physics, Astronomy and Mathematics, University of Hertfordshire, College Lane, Hatfield AL10 9AB, UK
              \and
              Institut de Radioastronomie Millim\'{e}trique (IRAM), 300 Rue de la Piscine, F-38406 Saint Martin d'H\`{e}res, France; 
              \and
              LERMA, Observatoire de Paris, PSL Research University, CNRS, Sorbonne Universit\'es, 75014 Paris
              \and
              European Southern Observatory, Karl-Schwarzschild Straße 2, D-85748 Garching bei M{\"u}nchen, Germany
              \and
              Univ Lyon, Univ Lyon1, ENS de Lyon, CNRS, Centre de Recherche Astrophysique de Lyon UMR5574, F-69230 Saint-Genis-Laval France
              \and
              520 Edgemont Road, Charlottesville, VA 22903
              \and
              Astronomisches Rechen-Institut, Zentrum f\"{u}r Astronomie der Universit\"{a}t Heidelberg, M\"{o}nchhofstra\ss e 12-14, 69120 Heidelberg, Germany
              \and
              Universit\"{a}t Heidelberg, Zentrum f\"{u}r Astronomie, Instit\"ut  f\"{u}r Theoretische Astrophysik, Albert-Ueberle-Strasse 2, 69120 Heidelberg, Germany
              \and
              Department of Astronomy, Xiamen University, Xiamen, Fujian 361005, China; Purple Mountain Observatory, Chinese Academy of Sciences (CAS), Nanjing 210023, China
              \and
              Research School of Astronomy and Astrophysics, Australian National University, Canberra, ACT 2611, Australia
              \and
              Universit\"{a}t Heidelberg, Interdisziplin\"{a}res Zentrum f\"{u}r Wissenschaftliches Rechnen, INF 205, 69120 Heidelberg, Germany
              \and
              Department of Physics, Tamkang University, No.151, Yingzhuan Rd., Tamsui Dist., New Taipei City 251301, Taiwan
              \and
              Center for Astrophysics and Space Sciences, University of California San Diego, 9500 Gilman Drive, La Jolla, CA 92093, USA
              }

   \date{Received 09 November 2021 / Accepted 20 December 2021 }

 
  \abstract{The complex physical, kinematic, and chemical properties of galaxy centres make them interesting environments to examine with molecular line emission. We present new $2{-}4\arcsec$ (${\sim}75{-}150$~pc at $7.7$~Mpc) observations at 2 and 3~mm covering the central $50\arcsec$ (${\sim}1.9$~kpc) of the nearby double-barred spiral galaxy NGC~6946 obtained with the IRAM Plateau de~Bure Interferometer. We detect spectral lines from ten molecules: \chem{CO}, \chem{HCN}, \chem{HCO^+}, \chem{HNC}, \chem{CS}, \chem{HC_3N}, \chem{N_2H^+}, \chem{C_2H}, \chem{CH_3OH}, and \chem{H_2CO}. We complemented these with published 1mm CO observations and 33~GHz continuum observations to explore the star formation rate surface density \sigsfr\ on 150~pc scales.  In this paper, we analyse regions associated with the inner bar of NGC~6946 -- the nuclear region (NUC), the northern (NBE), and southern inner bar end (SBE) and we focus on short-spacing corrected bulk (CO) and dense gas tracers (\chem{HCN}, \chem{HCO^+}, and \chem{HNC}). We find that \chem{HCO^+} correlates best with \sigsfr, but the dense gas fraction (\fdense) and star formation efficiency of the dense gas (\sfedense) fits show different behaviours than expected from large-scale disc observations. The SBE has a higher \sigsfr, \fdense, and shocked gas fraction than the NBE. We examine line ratio diagnostics and find a higher $\chem{CO}{21}/\chem{CO}{10}$ ratio towards NBE than for the NUC. Moreover, comparison with existing extragalactic datasets suggests that using the $\chem{HCN}/\chem{HNC}$ ratio to probe kinetic temperatures is not suitable on kiloparsec and sub-kiloparsec scales in extragalactic regions. Lastly, our study shows that the $\chem{HCO^+}/\chem{HCN}$ ratio might not be a  unique indicator to diagnose AGN activity in galaxies.}


   \maketitle
%

\section{Introduction}
The study of star-forming regions within galaxy centres is of particular interest, as the extreme conditions within these regions are thought to greatly influence their host giant molecular cloud (GMC) populations, and, thus, the stellar populations that they may form. For example, the inner few kiloparsecs of galaxies typically have higher average gas and stellar densities, gas temperatures, levels of turbulence and magnetic field strengths relative to their discs. All of these characteristics have direct implications for the physical, dynamical and chemical properties of the star-forming `dense' \footnote{The majority of the discussed studies throughout this work, observed the base transitions of molecules such as HCN and HCO$^+$, which become excited at effective critical densities of $n > 3\times10^3$~cm$^{-3}$ \citep[e.g.][]{Shirley2015}. This is then typically referred to in the literature as `dense gas', relative to the density of gas traced by CO ($n\approx10^2$~cm$^{-3}$).} molecular gas within systems (see e.g. \citealt{McKee2007,lada2010star,lada2012star,Longmore2013,Klessen2016}). 

Molecular gas is denser in centres than is typically inferred within disc star-forming regions; the cause of this are strong compressive tidal forces, frequent molecular cloud interactions, short dynamical timescales, constant feeding of material via, for example, bars (e.g. \citealt{schinnerer2006molecular,Schinnerer2007}), and feedback from both young and old stars (e.g. \citealt{kruijssen19a,chevance20,chevance20b,Barnes2020}) along with energetic phenomena within the galaxy centres (e.g. AGN; \citealp{combes2013,Querejeta2016}) and this should favour star formation. However, it is seen both within the Milky Way (e.g. \citealp{Kruijssen2014a,Barnes2017}) and inferred at centres of nearby galaxies (e.g. \citealt{gallagher2018dense,Jimenez-Donaire2019EMPIRE}) that the gas over-density required to form stars is also higher. Averaged over the galaxy population (and thereby presumably over episodic variations, e.g. \citealt{Krumholz2017}), centres show an apparent deficit of star formation despite their large quantities of dense molecular gas (e.g. \citealt{Bigiel2015DenseGasFraction,usero2015variations,Jimenez-Donaire2019EMPIRE} among others). 
For a complete understanding of star formation, the study of these complex environments present in galaxy centres is crucial.

Observations and simulations of a Milky Way-like galaxy suggest that large-scale bars (a few kiloparsecs) could be responsible for a molecular gas mass build-up in the central few hundred parsecs (e.g. \citealt{Diaz-Garcia2020} and \citealt{Renaud2015, Emsellem2015, Sormani2019, Sormani2020}). The overlap between the end of those large-scale bars and the spiral arms are -- in addition to the centre -- active star-forming regions coupled with enhanced dense gas, for example the bar ends of NGC\,3627 \citep{Beuther2017, Beslic2021}. Secondary small-scale bars (a few hundred parsecs) of double barred galaxies (e.g. \citealt{Erwin2011}) were intensively studied through photometric analysis (e.g. \citealt{Mendez-Abreu2019,deLorenzo-Caceres2020}). However, secondary small-scale bars (we refer to them as `inner bar' in this work) have been hardly observed with emission from multiple rotational molecular line transitions. This leads us to question if we would find also enhanced molecular line emission (indicating increased gas volume densities) and/or star formation activity at the inner bar ends  compared to the surrounding environment (i.e. as with larger scale galactic bars). And how does this relate to the increased dense gas fraction (\fdense), and yet decreased dense gas star formation efficiency (\sfedense), typically observed in galaxy centres. 

The emission from rotational molecular line transitions observed within the radio domain is typically used to study the aforementioned properties of the molecular interstellar medium (ISM). Studies conducted over the past several decades found that emission of high-critical-density tracers, such as Hydrogen Cyanide (HCN), Formylium ion (HCO$^+$), Carbon Monosulfide (CS), and Diazenylium (N$_2$H$^+$) are prevalent across galaxy centres, suggesting the presence of large amounts of `dense' molecular gas \citep{Mauersberger1989CS, Mauersberger1989N2H, Mauersberger1991DenseGasN2H, Nguyen1992HCNandHCO, solomon1992DenseMolecularGas, Helfer1993DenseGasinBulges}. More recently, higher angular resolution observations have allowed the study of dense gas content of individual centres of galaxies on sub-kiloparsec scales (e.g. \citealt{Schinnerer2007,pan2015MolecularGasandSFR} and \citealt{Tan2018} for NGC~6946; \citealt{Martin2015} for NGC~253; \citealt{Murphy2015} for NGC~3627; \citealt{Salak2018} for NGC~1808; \citealt{Querejeta2019} for M\,51; \citealt{Bemis2019} for NGC~4038 and NGC~4039; \citealt{Callanan2021} for M~83). 

Galaxy centres also have a rich and diverse chemistry that could be used to probe the environmental conditions of gas beyond just the density (e.g. \citealt{Morris1996,Belloche2013,Rathborne2015,Barnes2019} and \citealt{Petkova2021} for the Milky Way;  \citealt{Meier2005CenterofIC342} for IC~342; \citealt{2012MeierChemistryinMaffei2} for Maffei\,2; \citealt{Aladro2013} for NGC\,1068; \citealt{2015MeierALMAImagingofStarburstNGC253, Leroy2018, Krieger2020} and \citealt{Holdship2021} for NGC\,253; \citealt{Martin2015} for NGC~1097; \citealt{Henkel2018} for NGC\,4945). These studies used molecular line ratio diagnostics to investigate the physics and chemistry of the ISM in these extreme environments. But many open questions still remain, such as: how the location of active formation of young massive stars correlate with that of the various molecular line tracers, or, what the ratios between molecular lines indicate about temperature, excitation and density structure in the gas.

\ngc\ is an ideal candidate to study the effect of large and small scale bars on the dense gas and chemistry through molecular line emission. \ngc\ has beside its large-scale stellar bar with a length of $100{-}120\arcsec$ ($\approx3.7{-}4.5$~kpc) \citep{Regan1995, Menendez-Delmestre2007, Font2014}, a small-scale bar (see below), and is one of the nearest large double-barred spiral galaxies ($D \approx 7.72$~Mpc; see Table~\ref{Tab: Properties}). It does not seem to host an AGN (see Section~\ref{Disc:HCO/HCN}) and leads the list of observed supernovae in a nearby galaxy disc (10 supernovae\footnote{\url{http://cbat.eps.harvard.edu/lists/Supernovae.html}}), which has led to its name \textit{Fireworks Galaxy}. 

\begin{figure*}
    \centering
    \includegraphics[width=0.89\textwidth]{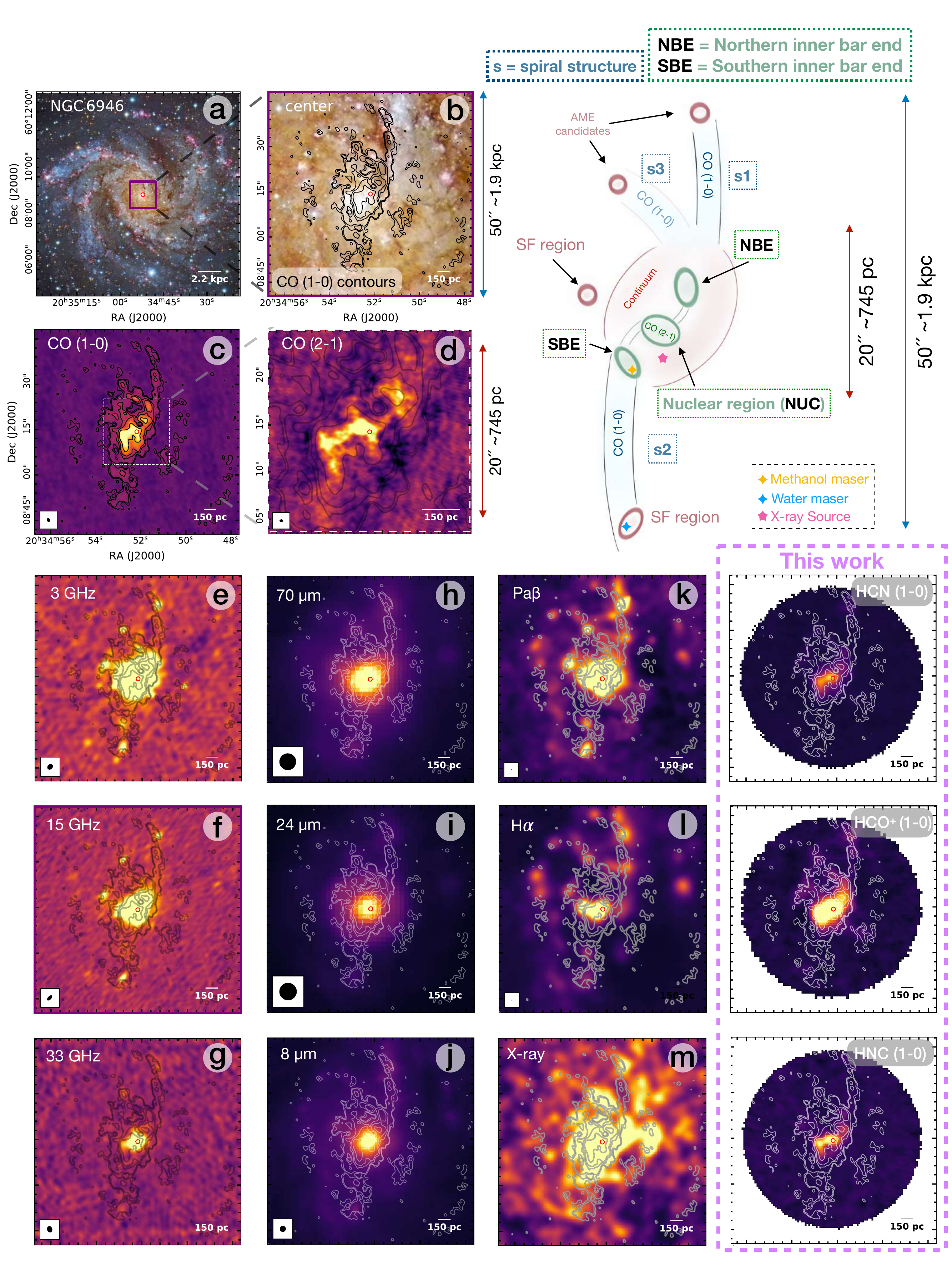}
    \caption{\textbf{Gallery of the multi-wavelength observation towards the centre of NGC\,6946. }
    {\it Left panels:} (a) three colour optical image, with the region of interest drawn as a purple box, (b) CO\,(1-0) emission shown in contours levels of 60, 90, 200, 300, 600 K~km$^{-1}$~s$^{-1}$, and are repeated on all panels for comparison, (c) integrated CO\,(1-0) emission, (d) CO\,(2-1) emission towards the inner $20\arcsec$ shown as colour scale; {\it Sketch (right):} denoting all observed features (see Section~\ref{sec:story-about-central-region} for an overview), (e)--(g) continuum emission, (h)--(j) infrared emission and (k)--(m) hydrogen recombination lines and $X$-ray emission (see Table~\ref{tab:fig1-ref} for references of all observations shown). The rightmost column shows integrated intensity maps of three of our fourteen detected emission lines for comparison. Optical image credits: NASA, ESA, STScI, R.~Gendler, and the Subaru Telescope (NAOJ).}
    \label{fig:color}
\end{figure*}

In this work, we study the inner $50\arcsec\approx1.9$~kpc of \ngc, which has previously been the target for many multi-wavelength observations which we can use to compare to, and build upon, for our analysis. We use ancillary $33$\,GHz continuum data \citep{Murphy2018SFRS, Linden2020SFRS} to obtain the SFR from the free-free emission part to compare dense gas, total molecular gas, and extinction-free estimates of recent star formation across the inner star-forming centre of \ngc. In the sketch in Figure~\ref{fig:color}, we see in green colours three distinct regions in $^{12}$CO (2-1) (hereafter CO(2-1)). The appearance of these three features as an S-shaped structure have been explained with modelling efforts in \cite{schinnerer2006molecular}. To do so, they obtained a three-dimensional representation of the luminosity and mass distribution of the galaxy using the multi-Gaussian expansion (MGE) method (see e.g. \citealt{Emsellem2003}). They then inferred the corresponding axisymmetric gravitational potential by using the Poisson equation and assuming a constant mass-to-light ratio, which is then perturbed by adding a $\sim$\,kpc bar-like structure to the MGE potential (as in \citealp{Emsellem2003}). They showed that their model reproduces the observed (straight) $\sim$\,kpc gas lane morphologies, when applying an \textit{inner bar} with a radius of $\sim$\,250\,pc (or $6.5{-}8\arcsec$ projected into the galaxy plane) with a pattern speed $\Omega_\mathrm{p}$ of $510{-}680$ km\,s$^{-1}$\,kpc$^{-1}$. They concluded that molecular gas flows inwards from the outer disc, shaping and driving the evolution of the centre of \ngc. Therefore, we refer to their `clumps' as the bar ends of the \textit{inner bar} (see green ellipses in the sketch of Figure~\ref{fig:color}). This allows us to study different regions in the centre: a) the nuclear region -- NUC, b) the northern inner bar end -- NBE, and c) the southern inner bar end -- SBE, 
%
%
which we refer to as abbreviated above throughout this work. 

In this work we present a suite of lines observed across the $1{-}3$~mm IRAM Plateau de~Bure Interferometer (PdBI) window [PI: E.~Schinnerer] to assess the physical and chemical structure of the inner ${\sim}50\arcsec$ of \ngc\ (see Figures~\ref{fig:intensities}). This gives us one of the most comprehensive, high resolution ($2{-}4\arcsec \approx 75{-}150$~pc) molecular line data set for a nearby galaxy centre in the northern hemisphere. In this first of a series of papers, we present the observations and especially focus on the short-spacing corrected bulk (CO) and dense gas tracers (HCN, HCO$^+$, and HNC).  

This paper is structured as follows: In \autoref{sec:Observation}, we present how the PdBI observations were taken, calibrated and imaged, along with information of the ancillary observations used within this work and how we convert observational measurements to physical quantities. \autoref{sec:Results} shows the results and \autoref{sec:3-SFRindicators-compare} the discussion of our extended data set. In \autoref{sec:ResultsB}, we show the results from our short-spacing corrected data set focusing on the nuclear region and inner bar ends. In \autoref{sec:Discussion}, we discuss line ratio diagnostics of the short-spacing corrected data set and possible implications for spectroscopic studies of other galaxy centres.


\section{Observations and ancillary data}
\label{sec:Observation} 
\begin{table}
    \begin{center}
    \caption{{Properties of \ngc}}
    \label{Tab: Properties}
    \begin{tabular}{l c c}
    \hline \hline
    Parameter            &  Value &  Notes \\ \hline
    Morphology           & SAB(rs)cd         &   (1) \\
    Nuclear type         & star-forming, \HII  & (2) \\
    Age of starburst     & $7{-}20$ Myr&   (3) \\
    Distance.            & $7.72$ Mpc      &   (4) \\
    Linear scale         & $37$ pc$/$arcsec &  \\
    Inclination          & $33\degr$        &  (5) \\
    P.A. major axis      & $242\degr$        &   (6) \\
    P.A. minor axis      & $152\degr$        &   (6) \\
    V$_{\rm LSR}$        & $50$ km\,s$^{-1}$&  (7)  \\
    \textit{Dynamical centre} & &  (7) \\
    \enspace R.A. (J2000)         & $20{:}34{:}52.35$  &  \\
    \enspace Decl. (J2000)        & $+60{:}09{:}14.58$ &  \\
    \textit{Inner bar} & & \\
    \enspace length & $6.5{-}8\arcsec$ & (7) \\
    \enspace NBE, R.A. &  $20{:}34{:}51.71$ & this work\\
    \enspace NBE, Decl.& $+60{:}09{:}17.21$ & this work\\
    \enspace SBE, R.A. &  $20{:}34{:}52.67$ & this work\\
    \enspace SBE, Decl.& $+60{:}09{:}11.53$ & this work\\
    \hline \hline
    \end{tabular}
    \end{center}
    \begin{minipage}{0.95\columnwidth}
        \vspace{1mm}
        {\bf Notes:} (1):  \cite{deVaucouleurs1991}; (2): \cite{Goulding2009}; (3): \cite{Engelbracht1996ObservationandModelingofNuclearStarburstinNGC6946}; (4): \cite{Anand2018Distance}; they applied the tip of the red giant branch method to measure the distance to NGC\,6946. The brightest red giant stars in the outer regions of NGC\,6946 were adopted as standard candles. Research on NGC\,6946 carried out before 2018 used smaller values for its distance, ranging from 5~Mpc (e.g. \citealt{schinnerer2006molecular}) to about 7~Mpc (e.g. \citealt{Murphy2011}). Where necessary, we convert measurements from the literature to our adopted distance. (5): \cite{deBlok2008Inclination}; (6): \cite{Crosthwaite2002}; (7): \cite{schinnerer2006molecular};
    \end{minipage}
\end{table}
\begin{table*}
\centering
\caption{{Summary of the molecular lines in our compiled data set towards the centre of NGC\,6946.}}
\label{tab:Obs-properties}
\resizebox{\textwidth}{!}{%
\begin{tabular}{llcccccccc}
\hline \hline
 &  & \multicolumn{1}{l}{} & $\nu_\mathrm{rest}$ & $\nu_\mathrm{obs}$ & $E_\mathrm{u}$ & $n_{\mathrm{H_2}}\,(\epsilon_\mathrm{max})$ & $n_\mathrm{eff}$ & project code & setup \\
& Molecules  & \multicolumn{1}{l}{} & [GHz] & [GHz] & [K] & [cm$^{-3}$] & [cm$^{-3}$]  &   &   \\
& (1) & (2) & (3) & (4) & (5) & (6) & (7) & (8) & (9) \\ \hline
\multirow{9}{*}{3\,mm} & C$_2$H & $N=(1{-}0)$, $J=3/2{-}1/2$ & 87.316 & 87.302 & 4.19 &             &              & Q059+R02C & A \\
& HCN        & $(1{-}0)$                & 88.632  & 88.617 & 4.25 & $2\times10^5$  & $4.5\times10^3$ & Q059+R02C+P069 & A \\
& HCO$^+$    & $(1{-}0)$                & 89.188  & 89.173 & 4.28 & $4\times10^4$  & $5.3\times10^2$ & Q059+R02C      & A \\
& HNC        & $(1{-}0)$                & 90.664  & 90.648 & 4.35 & $1\times10^5$  & $2.3\times10^3$ & Q059+R02C      & C \\
& HC$_3$N    & $(10{-}9)$               & 90.979  & 90.964 & 16.69&             & $4.3\times10^4$ & R09F           & C \\
& N$_2$H$^+$ & $(1{-}0)$                & 93.174  & 93.158 & 4.47 &             & $5.5\times10^3$ & R09F           & C \\
& CH$_3$OH   & $J_K=2_0{-}1_0$  & 96.741  & 96.725 & 6.96 &             &              & Q059+R02C      & B \\
& CS         & $(2{-}1)$                & 97.980  & 97.965 & 7.10 &             & $1.2\times10^4$ & R09F+S08E      & B \\
& $^{12}$CO  & $(1{-}0)$                & 115.271 & 115.252& 5.53 & $1\times10^2$  &              & M032           & D  \\ \hline
\multirow{4}{*}{2\,mm} & CH$_3$OH  & $J_K=3_0{-}2_0$ & 145.100 & 145.073 & 18.80 & &     & S02A           & B \\
& HC$_3$N    & $(16{-}15)$              & 145.561 & 145.534& 41.27 & 	 &                     & S02A           & B \\
& p-H$_2$CO    & $N=(2{-}1)$, $J=0/2{-}0/1$ & 145.603 & 145.576& 10.48 & & $6.3\times10^4$  & S02A           & B \\
& CS         & $(3{-}2)$                & 146.969 & 146.942& 14.11 &	  & $3.3\times10^4$       & S02A           & B \\ \hline
1.3\,mm  & $^{12}$CO  & $(2{-}1)$         & 230.538 & 230.499&	16.59 & $1\times10^3$  &          & M032+P069      & one mm   \\ \hline \hline
\end{tabular}
}
\begin{minipage}{2.0\columnwidth}
    \vspace{1mm}
    {\bf Notes:} (1--2): Molecular lines with their transition ordered by their rest frequencies; \textit{Throughout the paper, for simplicity, we refer to the transitions of C$_2$H as (1-0), of CH$_3$OH as (2-1) and (3-2), and of H$_2$CO as (2-1)}; (3--4): Rest and observational frequencies; (5): Upper energy levels (6): Minimum density at which the emissivity of the line reaches $95\%$ of its peak value for $T=25$~K (see table~2 within \citealt{Leroy2017} and references therein); (7): Taken from \citealt{Shirley2015} (table~1, for $20$~K), and have been defined by radiative transfer modelling as the density that results in a molecular line with an integrated intensity of 1~K\,km\,s$^{-1}$; (8--9): PdBI project code and setup. A = configuration C and~D; B = configuration A, B, C, D; C = configuration B, C, D. 
\end{minipage}
\end{table*}

\begin{table*}
\centering
\caption{{Properties of the \textit{PdBI only} data set.}}
\label{tab:PdBIonly-Characteristics}
\resizebox{\textwidth}{!}{%
\begin{tabular}{l|cc|cccc|cccc}\hline\hline
 &   \multicolumn{2}{c}{Native resolution} & \multicolumn{8}{c}{Characteristics at $4\arcsec$ resolution}\\    
 &  Beam size      & P.A.        & $I_\mathrm{line}$ & $T_\mathrm{peak}$ & Noise & \SN & \multicolumn{4}{c}{Ratios with}  \\
Molecules  &              [$\arcsec$] & [$\degr$]  & [K\,km\,s$^{-1}$] & [K] & [K\,km\,s$^{-1}$] & & CO\,(2-1) & CO\,(1-0) & HCN & \sigsfr \\
(1) & (2) & (3) & (4) & (5) & (6) & (7) & (8) & (9) & (10) & (11) \\\hline
$^{12}$CO  (1-0)         & 1.36 x 1.11  & 26    & 748.04 &  8.80 &	2.85 & 262.72 	& 2.52 & 1.00 &	13.04 & 8.31E+01   \\
$^{12}$CO  (2-1)         & 0.40 x 0.29  & 91    & 296.40 &  4.75 &	2.20 & 134.56 	& 1.00 & 0.40 &	5.17 &	3.29E+01   \\
HCN        (1-0)         & 2.61 x 2.12  & 97    & 57.35  & 	0.62 &	0.71 & 80.65	& 0.19 & 0.08 &	1.00 &	6.37E+00   \\
HCO$^+$    (1-0)         & 3.45 x 2.83  & -101  & 47.94  & 	0.53 &	0.52 & 92.43	& 0.16 & 0.06 &	0.84 &	5.31E+00   \\
HNC        (1-0)         & 2.45 x 2.00  & 62    & 22.06  & 	0.32 &	0.62 & 35.66	& 0.07 & 0.03 &	0.38 &	2.45E+00   \\
CS         (2-1)         & 1.89 x 1.60  & 97    & 29.39  & 	0.40 &	0.72 & 40.57	& 0.01 & 0.04 &	0.51 &	3.26E+00   \\
CS         (3-2)         & 1.98 x 1.76  & -180  & 21.07  & 	0.30 &	0.28 & 76.00	& 0.07 & 0.03 &	0.37 &	2.34E+00   \\
N$_2$H$^+$ (1-0)         & 2.39 x 1.96  & 65    & 6.43 	 &  0.21 &	0.49 & 13.02	& 0.02 & 0.01 &	0.11 &	7.14E-01   \\
HC$_3$N    (10-9)        & 2.41 x 1.98  & 64    & 4.06 	 &  0.09 &	0.68 & 5.93	    & 0.01 & 0.01 &	0.07 &	4.51E-01   \\
HC$_3$N    (16-15)       & 2.02 x 1.77  & 339   & 2.25 	 &  0.08 &	0.39 & 5.72	    & 0.01 & 0.01 &	0.04 &	2.50E-01   \\
CH$_3$OH   (2k-1k)       & 1.89 x 1.59  & 98    & 7.15 	 &  0.27 &	0.66 & 10.77	& 0.02 & 0.01 &	0.12 &	7.95E-01   \\
CH$_3$OH   (3k-2k)       & 1.98 x 1.76  & -17   & 6.12 	 &  0.28 &	0.55 & 11.09	& 0.02 & 0.01 &	0.11 &	6.80E-01   \\
C$_2$H     (1-0),3/2-1/2 & 3.54 x 2.84  & 77    & 12.78  &  0.14 &	0.72 & 17.76	& 0.04 & 0.02 &	0.22 &	1.42E+00   \\
H$_2$CO    (2-1)         & 1.99 x 1.75  & 343   & 2.29 	 &  0.08 &	0.35 & 6.61	    & 0.01 & 0.01 &	0.04 &	2.54E-01   \\ \hline\hline
\end{tabular}
}
\begin{minipage}{2.0\columnwidth}
    \vspace{1mm}
    {\bf Notes}: We arranged the \textit{PdBI only} data set as in \autoref{fig:central-sightline-spectra} and show them at their native and best-common $4\arcsec \approx 150$~pc spatial resolution. Column (1): Molecules and their transitions. (2--3): Characteristics of the native resolution observation. (4--11): We quote the mean for the central sight line, i.e. aperture of $4\arcsec \approx 150$~pc: (4): Integrated intensity. (5): Peak temperature of the spectrum. (6): Root mean square (rms) noise. (7): Signal to noise. (8--11): Ratios of molecular species with CO, HCN and \sigsfr\ . The I$_\mathrm{line}$/\sigsfr\ ratio results in units of K~km~s$^{-1}$/(\sigsfrunit).
\end{minipage}
\end{table*}

\begin{table*}
\centering
\caption{{Properties of the \textit{SSC + \uv trim} data set.}}
\label{tab:SSC-Characteristics}
\resizebox{\textwidth}{!}{%
\begin{tabular}{l|ccc|cccc|cccc}
\hline \hline
 &   \multicolumn{3}{c|}{Characteristic after SSC} & \multicolumn{4}{c|}{Characteristic at $4\arcsec$ resolution} &        &        &        &       \\
 &   Beam size      & P.A.      & Recov.~Flux     & $I_\mathrm{line}$   & $T_\mathrm{peak}$   & Noise   & $\mathrm{S/N}$  & \multicolumn{4}{c}{Ratios with} \\
Molecules   &[$\arcsec$]     & [$\degr$]      &   [\%]   & [K\,km\,s$^{-1}$]      & [K]   & [K\,km\,s$^{-1}$]   &         & CO(2-1) & CO(1-0) & HCN   & \sigsfr  \\
(1) & (2) & (3) & (4) & (5) & (6) & (7) & (8) & (9) & (10) & (11) & (12) \\\hline
$^{12}$CO  (1-0) & 1.37 x 1.12 & 25   & 52.8 & 892.75  & 9.33 & 3.60 & 247.75  & 1.52    & 1.00    & 13.13 & 97.46 \\
$^{12}$CO  (2-1) & 0.96 x 0.90 & -160 & 25.2 & 587.38  & 6.36 & 2.24 & 262.39  & 1.00    & 0.65    & 8.64  & 64.13\\
HCN        (1-0) & 2.90 x 2.57  & 93  & 60.2 & 67.97   & 0.67 & 0.58 & 116.90  & 0.12    & 0.07    & 1.00  & 7.42 \\
HCO$^+$    (1-0) & 3.62 x 3.07 & 82   & 53.2 & 54.74   & 0.57 & 0.51 & 106.57  & 0.09    & 0.06    & 0.81  & 5.98 \\
HNC        (1-0) & 2.90 x 2.57  & 55  & 68.1 & 25.16   & 0.34 & 0.62 & 40.52   & 0.04    & 0.03    & 0.37  & 2.75 \\ \hline \hline 
\end{tabular}
}
\begin{minipage}{2.0\columnwidth}
    \vspace{1mm}
    {\bf Notes}: We show the the short-spacing corrected and \uv trimmed data set at their native and best-common $4\arcsec \approx 150$~pc spatial resolution. Column (1): Molecules and their transition. (2--4): Characteristics after the short-spacing correction at their native resolution and recovered flux compared to \textit{PdBI only} data set (see Table~\ref{tab:PdBIonly-Characteristics}). (5--12): We quote the mean for the central sight line, i.e. aperture of $4\arcsec \approx 150$~pc: (5): Integrated intensity. (6): Peak temperature of the spectrum. (7) Root mean square (rms) noise. (8) Signal to noise. (9--12): Ratios of molecular species with CO, HCN and \sigsfr\ . The I$_\mathrm{line}$/\sigsfr\ ratio results in units of K~km~s$^{-1}$/(\sigsfrunit).
\end{minipage}
\end{table*}

%
\subsection{PdBI observations and data reduction}

The Plateau de~Bure Interferometer (PdBI) was used to image molecular lines at $1.3$ to $3$~mm  using a single pointing focusing on the central $50\arcsec$ of \ngc\ for a total of 132\,h from 2002 to 2009. Eight high-resolution spectral windows were used in each observation, each offering a bandwidth of $160~$MHz and channel width of $2.5~$MHz (i.e. $9$~km~s$^{-1}$ at $87$~GHz). The data, acquired from multiple observations (project codes: Q059, R02C, R09F, S02A, R09F, S08E, P069, M032 and S02A; PI: Schinnerer), were split between five spectral setups at reference frequencies of ${\sim}87$~GHz (setup~A), ${\sim}97$~GHz (setup~B), ${\sim}92$~GHz (setup~C), ${\sim}115$~GHz (setup~D) and ${\sim}230$~GHz (setup~one mm). We identified in the spectral setup~A: \cch, \hco\ and \hcn, in setup~B: \chohtwo, \cstwo, \chohthree, \hcnsixt, \htwoco\ and \csthree, in setup~C: \hnc, \hcnten\ and \ntwoh, in setup~D: \co, in setup~one mm: \cotwo\ (see  \ref{tab:Obs-properties}). The primary beam FWHM ranges from $\sim57\arcsec$ at $87$~GHz to $\sim43\arcsec$ at $115$~GHz and $\sim22\arcsec$ at $230$~GHz. The spectral window of each molecular line was centred on the redshifted frequency of the line assuming a systemic velocity of $50$~km\,s$^{-1}$ for \ngc\ \citep{schinnerer2006molecular}. The imaged width in each spectral window is the velocity width at zero level of the \chem{CO}{10} line ($300$~km\,s$^{-1}$, matching the Full Width at Zero Intensity of the HI line seen by THINGS in \citealt{Walter2008}) plus $50$~km\,s$^{-1}$ on each side of the line centre. These data were all calibrated and imaged using  \texttt{GILDAS}\footnote{\url{http://www.iram.fr/IRAMFR/GILDAS}}. 
For the bandpass calibration, observations of the bright quasars 3C454.3, 2145+067, 0923+392 and 3C273 have been used. The phase and amplitude calibrators were either both 2037+511 and 1928+738 or one of them. These calibrators were observed every 20 minutes. Most of the observations compared MWC\,349 observations with its IRAM flux model to calibrate the absolute flux scale. We expect an accuracy of the flux calibration of about 5$\%$ at 3mm 
and a typical antenna aperture efficiencies of about $25$~Jy\,K$^{-1}$ at $3$~mm, and $28$~Jy\,K$^{-1}$ at $2$~mm data. 

The continuum emission has been combined with the \texttt{UV MERGE} task after excluding the line channels. The spectral lines were resampled to $10$~km\,s$^{-1}$ for all the lines during the creation of UV tables in the \texttt{GILDAS CLIC} environment. The continuum has been subtracted in the \uv plane with the above continuum \uv table extracted from all line-free channels with the \texttt{UV SUBSTRACT} task before the imaging. Imaging of the visibilities used natural weighting, where each visibility is weighted by the inverse of the noise variance, as this maximises the sensitivity. The field of view of each image is twice the primary beam with pixel sizes $1/4$ of the synthesised beam major axis FWHM size (ranging from $0.4\arcsec~\times~0.29\arcsec$ to $2.0\arcsec~\times~1.7\arcsec$ and $3.6\arcsec~\times~2.9\arcsec$ , see \autoref{tab:PdBIonly-Characteristics}).

\texttt{Clean}ing was performed with a preconstructed clean mask which has been produced similar to the signal identification methodology used in \texttt{CPROPS} \citep{Rosolowsky2006,Rosolowsky2021,Leroy2021} using the high resolution (${\sim}1\arcsec$)  \chem{CO}{10} PdBI data without short-spacing correction \citep{schinnerer2006molecular}. Following Leroy et al. (\citealt{Leroy2021}; for the PHANGS-ALMA survey), the clean mask creation includes the following steps: (a) convolving the \chem{CO}{10} beam from $\sim1\arcsec$ to a coarser resolution of $33\arcsec$ and smoothing along the spectral axis with a boxcar function with a width of $20$~km~s$^{-1}$; (b) selecting peak pixels for each channel which have a S/N of at least 4; (c) expanding each peak down to a S/N of 2; (d) binary-dilating 4 channels along the spectral axis to account for edge effect. This produces a clean mask that covers all possible CO signal regions and ensures that the cleaning of other molecular gas tracers is less affected by noisy, no-signal regions outside the mask. The typical rms per channel observed is $\approx2$~mJy~ beam$^{-1}$.

Before being used in the deconvolution, the clean mask was regridded to match the astrometry of the dirty cube of each line. After imaging and deconvolution we corrected for the primary beam attenuation. We provide the observational characteristics of the \textit{PdBI only data} in \autoref{tab:PdBIonly-Characteristics} (and show the channel maps in \autoref{fig:app-hcn_cm}--\ref{app:channelmap_last}).  

Given that these data have different \uv coverage, we built a homogenised product by imaging only the visibilities that are within the \uv coverage of the \chem{CO}{10} PdBI data, that is $11.5$ to $153.8$~k$\lambda$. We do this to match the spatial scales of the CO data. This has been done using a \texttt{GILDAS} script where we loop through all the visibilities and flag the ones outside the \chem{CO}{10} \uv range by assigning negative weights to them. Then the \texttt{UV TRIM} function was used to discard the flagged data. 

Interferometric observations are insensitive to emission with low spatial frequencies due to the lack of the shortest baselines. For the short-spacing correction, we utilise IRAM 30\,m observations. We use data of the EMIR Multiline Probe of the ISM Regulating Galaxy Evolution \citep[EMPIRE;][]{Jimenez-Donaire2019EMPIRE} survey which observed in the $3{-}4$~mm regime, in particular, HCN across the whole disc of nine nearby spiral galaxies, among them \ngc. For the purpose of short-spacing correction (SSC) we use their $33\arcsec$ resolution \hcn, \hco\ and \hnc\ maps. For the \chem{CO}{21} observations, we use data from the HERA CO-Line Extragalactic Survey (HERACLES) \citep{Leroy2009Heracles}. For the remaining PdBI detected molecules we do not find public single-dish data with a significant \SN\footnote{We note that the 3mm emission lines of C$_2$H, N$_2$H$^{+}$, and HC$_3$N were covered by EMPIRE, but not included in the public release.}, therefore, no SSC was feasible for them.

SSC has been done in \texttt{GILDAS} in its \texttt{MAPPING} environment. Before performing SSC on the data available, we ensured that the 30\,m data used the same projection center and spectral grid as the interferometric data, before applying the \texttt{UVSHORT} task. This produces pseudo-visibilities that fill the short-spacings before imaging and deconvolution (see \citealt{Rodriguez2008,Pety2013} for details).
We summarise the observational characteristics of the \textit{SSC + \uv trim} data in \autoref{tab:SSC-Characteristics}.

In summary, within this work we make use of the following data sets: \textit{PdBI only}, which includes 14 molecular emission lines, and \textit{SSC + \uv trim}, which includes five molecular emission lines. 

\subsection{Integrated intensity maps}
\label{sec:Sampling}

As a next step, we converted our \textit{PdBI only} and \textit{SSC + \uv trim data} cubes from units of Jy\,beam$^{-1}$ to brightness temperature, $T_\mathrm{b}$, measured in Kelvin. Our observations include emission lines from a wavelength range of 1 to 3~mm which results in various spatial resolutions. We convolved our data to a common beam size of $4\arcsec$ (corresponding to $150$~pc at $7.72$~Mpc distance) and sampled the integrated intensities onto a hexagonal grid. The grid points are spaced by half a beam size to approximately Nyquist sample the maps. This approach has the advantage that the emission lines can be directly compared.

To improve the signal-to-noise ratio (\SN) we applied a masking routine for the determination of the integrated intensity maps. 
We used the bright \chem{CO}{10} emission line as a prior for masking and produced two \SN\ cuts: a low \SN\ mask ($\SN > 2$) and a high \SN\ mask ($\SN > 4$). Subsequently, the high \SN\ mask is expanded into connected voxels in the low \SN\ mask, and the integrated intensity is determined by integrating along the velocity axis for each of the individual sight lines multiplied by the channel width, $\Delta v_\mathrm{chan}$, of $10$~km\,$\rm s^{-1}$:
\begin{equation}
I = \sum_{n_\mathrm{chan}} T_\mathrm{b} \times \Delta v_\mathrm{chan}~.
\end{equation}
The uncertainty is calculated taking the square root of the number of masked channels (n$_\mathrm{chan}$) along a line of sight multiplied by the $1\sigma$ root mean square ($\sigma_\mathrm{rms}$) value of the noise and the channel width:
\begin{equation}
\sigma_{I} = \sqrt{n_\mathrm{chan}} \times \sigma_\mathrm{rms} \times \Delta v_\mathrm{chan}~.
\label{eq:unc}
\end{equation}
We calculate  $\sigma_\mathrm{rms}$ over the signal-free part of the spectrum using the \texttt{astropy} (\citealt{astropy:2013,astropy:2018}) function \texttt{mad\_std} that calculates the median absolute deviation and scales it by a factor of $1.4826$. This factor results from the assumption that the noise follows a Gaussian distribution. We show the integrated intensity maps of the \textit{PdBI only data} in \autoref{fig:intensities}.

\subsection{Ancillary data}
\label{sec:Ancillary data}

We include ancillary data for investigating the central regions of \ngc\ and analysing the relationship of our detected molecules to star formation rate (SFR) tracers. We also compare \ngc\ with other galaxy centers. Therefore, we include (1) $33$~GHz continuum emission, (2) EMPIRE dense gas observations of eight additional galaxy centres (angular resolution of 33$\arcsec\approx1$~kpc), (3) high resolution resolution dense gas observations of M~51 ($\sim$4$\arcsec\approx166$~pc) and NGC~3627 ($\sim$2$\arcsec\approx100$~pc), and (4) $0.5{-}7.0$~keV X-ray observations of \ngc. 


\subsubsection{SFR tracers}
\label{sec:2-SFRtracers}

Tracers of the number of ionizing photons, specifically free-free radio continuum emission and hydrogen recombination lines, are often regarded as 
good indicators of massive star formation. The radio continuum emission at low frequencies consists of two components: (1)~thermal free-free emission directly related to the production rate of ionizing photons in \HII\ regions and (2)~non-thermal emission arising from cosmic ray electrons or positrons which propagate through the magnetised ISM after being accelerated by supernova remnants or from AGN. Concentrating on the first case, only stars more massive than ${\sim}8$~M$_\odot$ are able to produce a measurable ionising photon flux (see e.g.\ \citealt{Murphy2012, KennicuttEvans2012}).

For \ngc\ we find recently published $33$~GHz continuum emission, which best match with the angular resolution of our molecular data set ($\sim$2.1$\arcsec$ angular resolution, see \autoref{fig:color}g and \autoref{fig:SFR-and-mask}). These data are from Very Large Array (VLA) observations within the Star Formation in Radio Survey (SFRS;~\citealt{Murphy2018SFRS,Linden2020SFRS}). In this work, we use the (1)~thermal free-free emission of the $33$~GHz continuum emission as a SFR tracer (see below). We discuss the reasons why we preferred $33$~GHz and check for consistency in the \autoref{appendix:sfr}.

\subsubsection{Calibration of the SFR} 
\label{sec:2-SFRCalibration}

The calibration of a variety of SFR indicators has been described in detail by \cite{Murphy2011}. They used \texttt{Starburst99} (\citealt{LEitherer1999Starburst99}) together with a \citet{Kroupa2001IMF} initial mass function (IMF). This type of IMF has a slope of $-1.3$ for stellar masses between $0.1$ and $0.5$~M$_{\odot}$, and a slope of $-2.3$ for stellar masses ranging between $0.5$ and $100$~M$_{\odot}$. Together with their assumptions of a continuous, constant SFR over ${\sim}100$~Myr, their \texttt{Starburst99} stellar population models show a relation between the SFR and the production rate of ionizing photons, $Q(H^{0})$, as:
\begin{equation}
     \frac{\mathrm{SFR}}{[\mathrm{M}_{\odot}~\mathrm{yr}^{-1}]} = 7.29 \times 10^{-54} \, \frac{Q(H^{0})}{[\mathrm{s}^{-1}]}~.
\end{equation}
For the $33$~GHz continuum map, we need to
separate the (1) thermal free-free and (2) non-thermal (synchrotron) parts of the emission. The thermal emission (denoted by~$^\mathrm{T}$) scales as $S_{\nu}^\mathrm{T} \propto \nu^{-\alpha^\mathrm{T}}~$, where $\nu$ refers to the frequency in GHz. To convert the thermal flux into a SFR, we follow Eq.\,(11) from \citealt{Murphy2011}:
\begin{equation}\label{eq:SFR}
\begin{split}
    \frac{\mathrm{SFR}_{\nu}^{\mathrm T}}{[\mathrm{M_{\odot}~yr^{-1}}]} = 4.6~\times~10^{-28} \, \left( \frac{T_\mathrm{e}}{[10^4~\mathrm{K}]} \right) ^{-0.45}\\ 
    \enspace \enspace \times~\left( \frac{\nu}{[\mathrm{GHz}]}\right) ^{{-\alpha^\mathrm{T}}} \, \frac{L^\mathrm{T}_{\nu}}{[\mathrm{erg\,s^{-1}\,Hz^{-1}}]}~,
\end{split}
\end{equation}
where $T_\mathrm{e}$ is the electron temperature in units of $10^4$~K, $\nu$ refers to the frequency in GHz, and $L^\mathrm{T}_{\nu}$ is the luminosity of the free-free emission at frequency~$\nu$ in units of $\mathrm{erg\,s^{-1}\,Hz^{-1}}$.

We adopt $\alpha^\mathrm{T} = 0.1$ and a thermal fraction of $f^{\mathrm{T}}_{33\,\mathrm{GHz}} = 0.62$ (as in \citealt{Murphy2011} for the nucleus of \ngc) and calculate the luminosity of the thermal free-free emission at frequency~$\nu$:

\begin{equation}
    \label{eq:lumthermal}
    \frac{L^\mathrm{T}_{{\nu}}}{[\mathrm{erg\,s^{-1}\,Hz^{-1}}]} = \frac{L}{[\mathrm{erg\,s^{-1}\,Hz^{-1}}]} \times f_{33\,\mathrm{GHz}}^{\mathrm{T}}~.
\end{equation}

Together with $T_\mathrm{e} = 0.42$ in units of $10^4$~K (again as in \citealt{Murphy2011} for the nucleus of \ngc) and Eq.~\eqref{eq:lumthermal} we can solve Eq.~\eqref{eq:SFR} and get a SFR$_\mathrm{33\,GHz}^\mathrm{T}$ map in units of ${\mathrm{M_{\odot}\,yr^{-1}}}$. We discuss the mean values within 150~pc sized apertures (i.e. the $4\arcsec$ working resolution) containing the NUC, SBE, or NBE regions in \autoref{sec:3-SFRindicators-compare}. In this work, we also use SFR surface densities ($\Sigma_\mathrm{SFR}$) in units of M$_{\odot}\,$yr$^{-1}\,$kpc$^{-2}$ for scaling relations in \autoref{sec:comp-moleculesandSFR}. We define $\Sigma_\mathrm{SFR}$ as:

\begin{equation}
    \frac{\Sigma_{\rm SFR}}{[\mathrm{M}_{\odot}~\mathrm{yr}^{-1} \mathrm{kpc}^{-2}]}   = \frac{\mathrm{SFR_{33\,GHz}}}{\mathrm{[M_{\odot}~yr^{-1}]}}  
    \left(\frac{\Omega}{\rm [kpc^{-2}]} \right)^{-1},
\end{equation}
where $\Omega$ is:
\begin{equation}
\begin{split}
    \frac{\Omega}{[\mathrm{kpc^{-2}]}} =  \bigg(~\pi~\left[\frac{\theta_{\mathrm{bmaj}}}{\mathrm{[arcsec]}}~\left(\frac{d}{\mathrm{[kpc]}}~\psi^{-1}\right)\right]\\
    \times~\left[\frac{\theta_{\mathrm{bmin}}}{\mathrm{[arcsec]}}~\left(\frac{d}{\mathrm{[kpc]}}~\psi^{-1}\right)\right]~\bigg)\\
    \times~\left[\left(4~\ln(2)\right)\right]^{-1}~.
\end{split}
\end{equation}
Here, $\theta_{\mathrm{bmaj}}$ and $\theta_{\mathrm{bmin}}$ are the major and minor axis of the beam in arcsec, $d$ the distance to \ngc\ in kpc and $\psi$ is the factor to convert from rad to arcsec (i.e. ($3600\times180$)/$\pi$).

\subsubsection{Molecular gas mass and depletion time}
\label{sec:SigmaMolandDepltime}

The molecular gas mass surface density can be estimated from the \chem{CO}{10} line emission in our data set. Given that H$_2$ is the most abundant molecule, the conversion of CO emission to molecular gas mass surface density is related to the CO-to-H$_2$ conversion factor $\alpha_\mathrm{CO}$. We adopt a fixed conversion factor $\alpha_\mathrm{CO} = 0.39$ M$_{\odot}$\,pc$^{-2}$ (K\,km\,s$^{-1}$)$^{-1}$ corrected for helium (\citealt{Sandstrom2013}; from their Table~6 for the centre of \ngc). This low, central $\alpha_\mathrm{CO}$ value of \ngc\ (a factor $\sim$10 lower than the canonical Milky Way value of 4.36 M$_{\odot}$\,pc$^{-2}$ (K\,km\,s$^{-1}$)$^{-1}$; see \citealt{Bolatto2013}) agrees with other studies \citep[e.g.][]{Cormier2018, Bigiel2020} and will not affect the main results in this paper. Then we convert the \chem{CO}{10} integrated intensity, $I_\mathrm{CO}^{1-0}$, to the molecular gas mass surface density, $\Sigma_\mathrm{mol}$, via:
\begin{equation}
    \label{eq:sigmamol}
    \frac{\Sigma_\mathrm{mol}}{\mathrm{[M_{\odot}\,pc^{-2}]}} = \alpha_\mathrm{CO} \, \frac{I_\mathrm{CO,\,150\,pc}^{1-0}}{\mathrm{[K\, km\,s^{-1}]}}~\cos{(i)}~.
\end{equation}
Here, $I_\mathrm{CO,\,150\,pc}^{1-0}$ stands for $I_\mathrm{CO}^{1-0}$ convolved to $150$~pc ($4\arcsec$) FWHM and the $\cos(i)$ factor corrects for inclination. We note, that the conversion from observed to physical quantity is subject to uncertainties for example, the low-J transition of \chem{^{12}CO} is known to be optically thick and does not necessarily peak in the 1-0 transition in many environments. These, together, may result in a less accurate CO to H$_2$ conversion, particularly towards starburst or AGN regions, yet this is most likely still secondary compared to the uncertainty on the $\alpha_\mathrm{CO}$ (e.g. due to metallicity). 
The ratio of the two profiles $\Sigma_\mathrm{mol}/$\sigsfr\ is the molecular gas depletion time, the time it takes (present day) star formation to deplete the current supply of molecular gas:
\begin{equation}\label{eq:depl}
\centering
    \frac{\tau_\mathrm{depl,\,150\,pc}^\mathrm{mol}}{[\mathrm{yr}^{-1}]} = \frac{\Sigma_\mathrm{mol}}{\mathrm{[M_{\odot}\,pc^{-2}]}} \left( \frac{\Sigma_\mathrm{SFR}}{{[\mathrm{M_{\odot}~yr^{-1}\,kpc^{-2}}]}} \times \frac{\gamma}{\mathrm{[kpc^{2}/pc^{2}]}} \right)^{-1}~.
\end{equation}
Here, $\gamma$ stands for the conversion factor to get from kpc$^{-2}$ to pc$^{-2}$.  \autoref{eq:depl} implies that all the molecular gas traced by CO will turn into star formation fuel which is an overestimate. We use CO to define \tdepl\ because we later compare it to literature values that used the same method to calculate $\tau_\mathrm{depl}^\mathrm{mol}$; this does not affect the main results of this work. 

\subsubsection{EMPIRE dense gas data, high-resolution M~51 and NGC~3627 data}
In \autoref{sec:Discussion} we discuss line ratio diagnostic plots of the dense gas tracers HCN, HCO$^{+}$ and HNC. To compare \ngc\ with other galaxy centres, we use available observations of additional eight galaxy centres (see legend in \autoref{fig:XDR-PDR}d) covered in the EMPIRE survey (\citealt{Jimenez-Donaire2019EMPIRE}). Those data products have a common angular resolution of 33$\arcsec$ ($\approx1$~kpc). For the analysis in \autoref{Disc:HCO/HCN} we also include high resolution observations of the dense gas tracers for: (i) M~51 and (ii) NGC~3627. 
The M~51 short-spacing corrected observations have an angular resolution of $\sim$4$\arcsec$ ($\approx166$~pc) and their reduction are described in \citealt{Querejeta2016}. We used the same technique as for the \ngc\ observations described in \autoref{sec:Sampling} to sample the data on a hexagonal grid and produce integrated intensity maps with a common angular resolution of 4$\arcsec$. We have also done this for the NGC~3627 observations, which have an angular resolution of $\sim$2$\arcsec$ ($\approx100$~pc); their short-spacing correction and reduction are described in \citealt{Beslic2021}.

\subsubsection{X-ray}
We also compare our observations to \textit{Chandra} \mbox{ACIS-S} observations (ObsIDs 1054 and 13435) from 2001 (PI: S.~Holt; \citealt{Holt2003}) and 2012 (PI: C.~Kochanek) totalling $79$~ks. We reduced the data using the \textit{Chandra} Interactive Analysis of Observations (\texttt{ciao}) Version 4.9 and produced exposure-corrected images using the \texttt{ciao} command \textit{merge\_obs}. Point sources were identified using \textit{wavdetect} on the merged, broad-band ($0.5{-}7.0$~keV) image and were removed with \textit{dmfilth} to create the $X$-ray image shown in \autoref{fig:XDR-PDR}. We also show a version of the $X$-ray diffuse, hot gas that has been smoothed with a Gaussian kernel (FWHM of 3) in \autoref{fig:color}\,(m).
%
%
%
%
%
%
%
%

\subsection{Multi-wavelength gallery of the central region of NGC~6946}
\label{sec:story-about-central-region}

\autoref{fig:color} shows a compilation of the various observations towards the inner $50\arcsec
~\approx1.86$~kpc of \ngc\ (see \autoref{tab:fig1-ref} for details on the observations). The centre appears to have several branches that connect to the outside environment. The two more pronounced spiral-like structures running north and south (extending $50\arcsec$) are most apparent in \chem{^{12}CO}{10} (published in  \citealt{schinnerer2006molecular}) (c) and to some extent also in the infrared emission (h)--(j). Another spiral-like structure could be presumed from these observations. Preceding studies even assumed four spiral features \citep{Regan1995}, which are not apparent in this \chem{CO} map. 
The sketch on the top right shows the two prominent spiral-like structures and the third presumable one as blue coloured arcs. 

The red coloured ellipse in the sketch (extending $20\arcsec \approx750$~pc) denotes the radio continuum emission (e)--(g). In the $70~\mu$m, $24~\mu$m, $8~\mu$m, Pa$\beta$, H$\alpha$ and $X$-ray emission (h)--(m), we likewise find emission in this region. In maps (e)--(m), however, no substructures are visible within the red ellipse due to high saturation or limited angular resolution. In contrast to that, \chem{CO}{21} reveals the nuclear region and two additional features to the north and south, illustrated as green ellipses in the sketch (central ${\sim}10\arcsec
~\approx372$~pc). \cite{schinnerer2006molecular} showed with dynamical models that the structures of \chem{CO}{21} can be explained with an inner bar (first observed and proposed via FIR by \citealt{Elmegreen1998}). The two inner bar ends (NBE and SBE) are connected to the nuclear region by straight gas lanes running along a position angle of $\mathrm{P.A.} {\sim}40\degr$. Those regions are also bright in emission in \chem{HCN}, \chem{HCO^+} and \chem{HNC} (see \autoref{sec:Results}). 

The four additional red ellipses in the sketch, most visible in the continuum emission (e) -- (f), were identified by \cite{Linden2020SFRS}. The southern two are associated with star formation (SF) regions and the northern two are suggested to be anomalous microwave emission (AME) candidates \citep{Linden2020SFRS}. The exact mechanism causing AMEs is not entirely understood, but the most promising explanation is that they occur due to small rotating dust grains in the ISM (see the review on AMEs by \citealt{Dickinson2018AMEReview}). One of the techniques to identify AMEs is to investigate the $33~\mathrm{GHz}/\ha$ flux ratio \citep{Murphy2018SFRS}. Larger ratios of $33$~GHz flux to \ha\ line flux would arise by an excess of non-thermal radio emission (see fig.~4 in \citealt{Murphy2018SFRS}; ratios of ${\sim}10^9$). Using $7\arcsec$ apertures in diameter for the two AME candidates we find $33~\mathrm{GHz}/\ha$ ratios of ${\sim}10^9$ expected for AME emission. This would confirm them being AME candidates. 

Within the southern SF region, at the end of the \chem{CO}{10} southern spiral structure, \citet{Gorski2018} found a water maser (\chem{H_2O}\,($6_{16}{-}5_{23}$) at $22.235$~GHz) and within the southern inner bar end two methanol masers (\chem{CH_3OH}\,($4_{-1}{-}3_0$) at $36.169$~GHz), marked as blue and orange stars in the sketch. The water maser is associated with one of the identified SF regions of \citet{Linden2020SFRS}.
Water masers can be in general variable on timescales of a few weeks and could indicate YSOs or AGB stars \citep{Palagi1993}. \citet{Billington2020} showed within the Milky Way the relationship between dense gas clumps and water masers. The location of the water maser matches well with the SF region seen in the \ha\ and $33$~GHz maps.


\section{Results -- Molecules in different environmental regions in the centre of NGC 6946}
\label{sec:Results}
\begin{figure*}[ht!]
    \centering
    \includegraphics[width=1.0\textwidth]{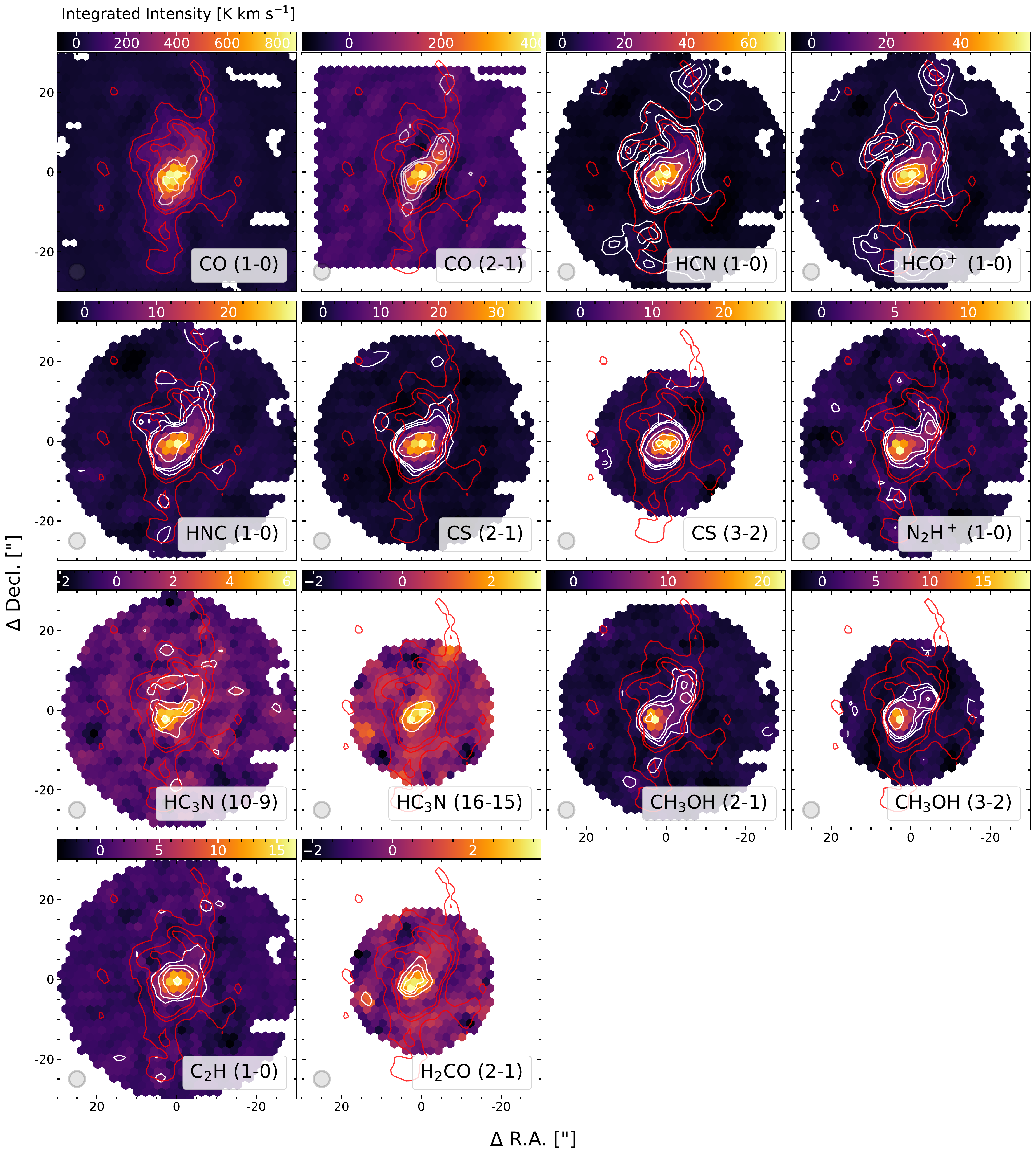}
    \caption{\textbf{Integrated intensity maps (moment~0) for 14 molecular emission lines at a common resolution of $\bm{4\arcsec \approx 150}$~pc.} The maps were created using a hexagonal sampling with half a beam spacing, where \chem{CO}{10} was used as a mask. The grey shaded circle in the lower left corner marks the beam size and the red contours show \chem{CO}{10} \SN\ levels of $30, 60 \text{ and } 90$. The white contours in the first panel show \SN\ levels of $200$ and $300$, and in the second panel \SN\ levels of $30, 60 \text{ and } 90$. The following panels (from \chem{HCN} to \chem{H_2CO}{21}) show \SN\ levels of $3, 6, 9, 30, 60 \text{ and } 90$. The colourbar indicates the integrated intensity of each line in units of K\,km\,s$^{-1}$. See Table \ref{Tab: Properties} for the properties of this data set.}
    \label{fig:intensities}
\end{figure*}

\begin{table}
\centering
\caption{{Spatial distribution comparison of the molecular species.}}
\begin{tabular}{l|cccccc} \hline \hline  
    \multicolumn{1}{c|}{}    & s1 & s2 & s3 & NBE & SBE & NUC  \\
    \hline
    $^{12}$CO  $(1{-}0)$         & \cmark & \cmark & \cmark & \cmark & \cmark & \cmark$^p$ \\
    $^{12}$CO  $(2{-}1)$         & \cmark & \cmark & \cmark & \cmark & \cmark & \cmark$^p$ \\
    HCN        $(1{-}0)$         & \cmark & \cmark & \cmark & \cmark & \cmark & \cmark$^p$ \\
    HCO$^+$    $(1{-}0)$          & \cmark & \cmark & \cmark & \cmark & \cmark & \cmark$^p$ \\
    HNC        $(1{-}0)$         & \cmark & \cmark & \cmark & \cmark & \cmark & \cmark$^p$ \\
    CS         $(2{-}1)$         & \xmark & \xmark & \cmark & \cmark & \cmark & \cmark$^p$ \\
    CS         $(3{-}2)$          & \xmark & \xmark & \xmark & \cmark & \cmark & \cmark$^p$ \\
    N$_2$H$^+$ $(1{-}0)$         & \xmark & \cmark & \cmark & \cmark & \cmark$^p$ & \cmark \\
    HC$_3$N    $(10{-}9)$        & \xmark & \xmark & \cmark & \cmark & \cmark$^p$ & \cmark \\
    HC$_3$N    $(16{-}15)$       & \xmark & \xmark & \xmark & \xmark & \cmark$^p$ & \cmark \\
    CH$_3$OH   $\mathrm{(2k{-}1k)}$       & \xmark & \xmark & \cmark & \cmark & \cmark$^p$ & \cmark \\
    CH$_3$OH   $\mathrm{(3k{-}2k)}$       & \xmark & \xmark & \xmark & \cmark & \cmark$^p$ & \cmark \\
    C$_2$H     $(1{-}0)$,\, $3/2{-}1/2$ & \xmark & \xmark & \xmark & \cmark & \cmark & \cmark$^p$ \\
    H$_2$CO    $(2{-}1)$         & \xmark & \xmark & \xmark & \xmark & \cmark$^p$ & \cmark \\
    \hline \hline
\end{tabular}
\begin{minipage}{0.9\columnwidth}
    \vspace{1mm}
    {\bf Notes}: The comparison is based on their $\mathrm{S/N} > 5$ peaks in the integrated intensity maps (see Figure~\ref{fig:sn5} in the Appendix). We consider the three spiral structures (s1--s3), the inner bar ends (NBE and SBE) and the nuclear region (NUC). \cmark\ denotes that we find within this region a $\mathrm{S/N}>5$ detection, \xmark\ not;  \cmark$^p$ refers to the location of the peak/maximum in integrated intensity in the inner $50\arcsec\approx1.9$~kpc.
\end{minipage}
    \label{tab:ticks}
\end{table}
\begin{figure}[ht!]
    \centering
    \includegraphics[width=1.0\linewidth]{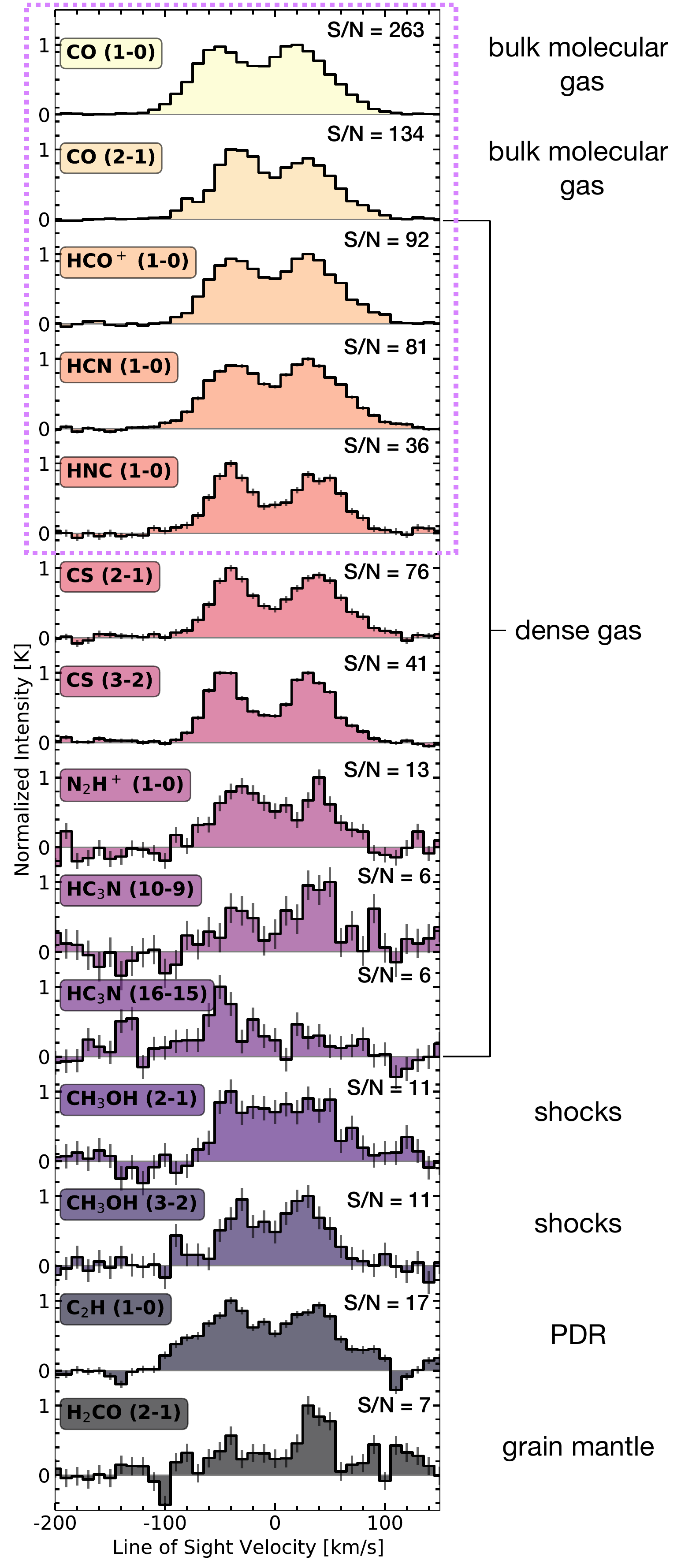}
    \caption{\textbf{Spectra of fourteen emission lines.} We show the spectra of all the emission lines detected within this survey for the central sight line, i.e. aperture of $4\arcsec \approx 150$~pc. The pink dotted rectangle shows the ones we use for discussing line ratio diagnostic plots later in this work (\autoref{sec:Discussion}). We normalise all of them to the maximum value in their spectra. The right-hand side indicates what each molecule \textit{could} be tracing according to literature (we refer the reader to the cautionary statements \autoref{sec:Mol_indicating}). Within each group we ordered the spectra by their \SN\ (see~\autoref{tab:PdBIonly-Characteristics}).} 
    \label{fig:central-sightline-spectra}
\end{figure}

We investigate how the molecular species in our data set spatially vary in the inner kiloparsec of \ngc. Our compiled data set contains 14 molecular emission lines obtained with the IRAM PdBI covering the inner $50\arcsec \approx 1.9$~kpc of \ngc, which are summarised in \autoref{tab:PdBIonly-Characteristics}. Their intensity maps can be seen in \autoref{fig:intensities} showing the integrated intensities for the detected molecular lines.

\subsection{The substructures in the inner kiloparsec of NGC 6946}\label{sec:Mol_struc}

In every map of \autoref{fig:intensities} we detect significant emission in the inner $10\arcsec \approx 375~\mathrm{pc}$. We note that we have for the first time detected ethynyl \cch, cyanoacetylene \hcnten\ and \hcnsixt\ in the inner kiloparsec of \ngc. All other molecules have been detected previously, mostly at lower resolution (e.g. ${\sim}30\arcsec$ \hnc\ and \ntwoh: \citealt{Jimenez-Donaire2019EMPIRE}; $25\arcsec$ and $17\arcsec$ \chohtwo, \chohthree, \htwoco: \citealt{Kelly2015MappingCSinNGC6946}; $8\arcsec$ \hco: \citealt{Levine2008}; $1\arcsec$ \hcn: \citealt{Schinnerer2007}).

\textbf{The spiral structures:} 
We see several branches that seem to be connected to the outside environment. The two more prominent spiral-like structures are best visible in the integrated intensity map of \co\ tracing the bulk molecular gas. For comparison with the other molecular species, we plot red contours of CO\,(1-0) with \SN\ levels of $30, 60, 90$ showing the spirals to the north and south in all of the 14~maps. The white \SN\ contours of \hcn, \hco\ and \hnc\ overlap with the northern spiral structure (labeled as `s1' in the sketch in \autoref{fig:color}). The southern spiral (s2) is visible in \hcn, \hco, \hnc\ and \ntwoh. We denoted a third spiral (s3), which is present in \hcn, \hco, \hnc, \ntwoh, \cstwo\ and \chohtwo.

\textbf{The inner bar and nuclear region:}
Already inferred in the sketch in \autoref{fig:color} from the native resolution \cotwo\ data, we know the location of the inner bar ends (NBE and SBE) and the nuclear region (NUC), and are able to investigate them with our other molecular emission lines. 
Considering only the integrated intensities with $\mathrm{S/N} > 5$ in each of the maps in \autoref{fig:intensities}, we see that the small-scale distributions of molecular species vary in these environments and that they do not necessarily peak at the same locations (see \autoref{tab:ticks}). 

While in the strongest lines, \co, \cotwo, \hcn, \hco\ and \hnc, all these regions are evident and their integrated intensities are peaking in the NUC, the situation becomes different for molecular species with higher effective critical densities.\footnote{The effective critical densities, $n_\mathrm{eff}$, are defined in \autoref{tab:Obs-properties}.} In particular, \chem{CS}{32} and \cch\ show emission concentrated to the inner $5\arcsec$ in radius. Contrary to that, the highest integrated intensities are not peaking at the very centre for \hcnten, \ntwoh, \chohtwo\ and \chohthree. Their peaks in emission are in the southern inner bar end. Interestingly, in the CS maps we do not see a peak in the SBE, although their $n_\mathrm{eff}$ is higher than \chem{N_2H^+} which peaks in the SBE. Of course, the different distributions of the molecular emissions may not be a direct result of the density difference, but also temperature (e.g. driven by the embedded star formation) that could drive both excitation and abundance variations. \autoref{tab:ticks} compares the spatial distribution of various molecular species.

\subsection{Molecular line profiles towards the nuclear region and what they are indicating}\label{sec:Mol_indicating}

\autoref{fig:central-sightline-spectra} shows the spectra of all the detected molecules for the central sight line ($4\arcsec \approx 150$~pc aperture). For visualisation purposes we normalised them to the maximum value in their spectra. We investigate what the molecular species in our data set reveal about the physical state of the molecular gas at the centre of \ngc.

CO is tracing the bulk molecular gas content and its line brightness is up to a factor of $14$ higher than other molecular lines within this work (see \autoref{tab:PdBIonly-Characteristics}). Relative to CO, molecules such as HCN are tracing denser gas, as they get excited at effective densities of $n \gtrsim 10^{3}$~cm$^{-3}$. As we go up in $n_\mathrm{eff}$, HNC, N$_2$H$^+$, CS and HC$_3$N are potentially tracing even denser molecular gas. This means, HC$_3$N is tracing the densest molecular gas in our data set. The spectrum in \autoref{fig:central-sightline-spectra}, however, shows that HC$_3$N has in the central $150$~pc the lowest \SN\ among all our molecular emission lines (\SN\ of~$6$). However, as we could see in \autoref{tab:ticks}, their maximum integrated intensities are located in the SBE, and there we find a \SN\ of~$11$ and~$36$. 

Among our molecular species, there are some that reveal more about the chemistry of the gas. The formation path of ethynyl (C$_2$H) is favoured in PDRs by the reaction of C$^+$ with small hydrocarbons and additionally through photodissociation of C$_2$H$_2$ \citep[][and references therein]{Meier2005CenterofIC342}. This means enhanced C$_2$H in PDRs associated with massive hot cores indicates hidden star formation \citep{Martin2015, Harada2018}. Recently, \citet{Holdship2021} showed in the nucleus of NGC\,253 that high abundances of C$_2$H can be caused by a high cosmic ray ionization rate. We detect \cch\ towards the nuclear region of \ngc\ and the line shows similar to CO and dense gas tracers a broad line profile (FWHM${\sim}200$~km\,s$^{-1}$) including two peaks.

Strong methanol (CH$_3$OH) emission is considered a tracer of shocks \citep[e.g.][]{Saito2017}. The reason for this is the formation process of CH$_3$OH in the gas phase is not efficient enough to produce higher amounts of CH$_3$OH \citep{Lee1996}. 
Intense CH$_3$OH emission is believed to arise from a series of hydrogenations of CO on dust grain surfaces under a low-temperature condition \citep{Watanabe2003}. After production on dust, it needs energetic heating mechanisms -- shock chemistry -- \citep[e.g.][]{Viti2011, James2020} to heat the dust and then sublimate CH$_3$OH into the gas phase. However, methanol emission can also be enhanced in non-shocked environments, such as towards massive stars or sources of cosmic ray or X-rays that heat the dust to ${\sim}100$~K and allow methanol to evaporate into the gas phase. In our data set, both methanol transitions are stronger in their line brightness than the faintest dense gas tracers -- N$_2$H$^+$ and HC$_3$N. Also, both methanol tracers have their maximum integrated intensity peaks in the SBE, where their ratios with CO are higher ($I_{\chem{CH_3OH}} / I_{\chem{CO}{21}}$). Methanol is also known to be a good kinetic temperature probe \citep[e.g.][]{Beuther2005}. Furthermore, para-H$_2$CO transitions can also be used as a temperature indicator which is sensitive to warmer ($T > 20$~K) and denser ($n {\sim}10^{4-5}$~cm$^{-3}$) gas \citep[e.g.][]{Ginsburg2016, Mangum2019}. H$_2$CO can be formed in the gas phase as well as on the surface of dust grains (e.g. \citealt{TerwisschavanScheltinga2021}). In our data set, \htwoco\ is one of the faintest lines with $\SN~ {\sim}7$ towards the nuclear region. 

We note, however, that each molecule is not a unique tracer of a particular process or sets physical conditions in a galaxy. To highlight this point, \cch, for example, in the Milky Way is a tracer of PDRs, yet in NGC~1068 or NGC~253 it is tracing a completely different type of gas (\citealt{Garcia-Burillo2017,Holdship2021}). 
In NGC~1068 it appears to trace a turbulent extended interface between outflows and ISM, yet in NGC~253 it appears to trace the dense gas that is subject to enhanced cosmic-ray ionization rates. 
Modelling can shed some light on the interpretation, and will be the subject of future work.

\cite{schinnerer2006molecular} found that the profiles of \co\ and \cotwo\ are double peaked in \ngc. In \autoref{fig:central-sightline-spectra} we also notice double-peaked profiles in the spectra of \hcn, \hco, \hnc, \cstwo, \csthree, \hcnten, \ntwoh, \cch, \chohtwo\ and \chohthree. These could be due to galactic orbital motions or in/outflows in the central $4\arcsec \approx 150$~pc. A more detailed analysis of the kinematics and dynamics of these spectral features are not discussed in this paper but it is planned for a future publication. 

In \autoref{fig:central-sightline-spectra} we denote the peak \SN\ and what condition each of the molecules can potentially indicate on the right-hand side.

\section{The environmental variability of the star formation rate in NGC 6946}
%
%
\label{sec:3-SFRindicators-compare}
\begin{figure*}[ht!]
\centering
    \includegraphics[width=1.0 \linewidth]{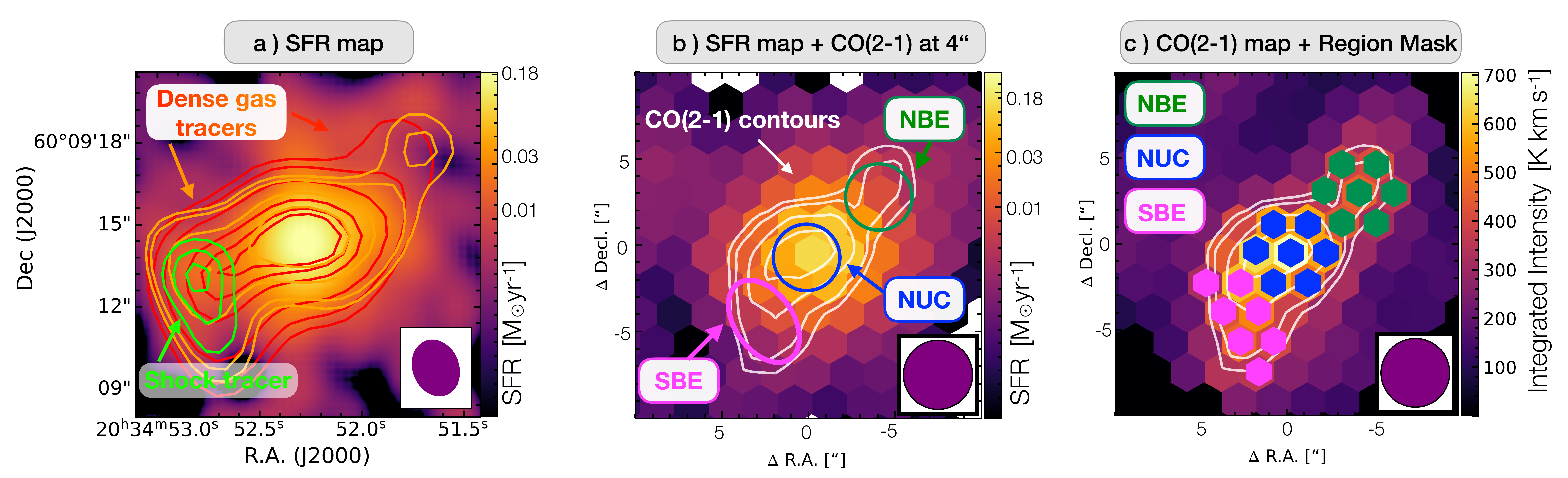}
    \caption{\textbf{Star formation rate map and region mask.} {\it Panel~(a):} SFR map with red, orange and green contours of \hcn, \hnc\ and \chohthree\ integrated intensities (we used the thermal part of the 33GHz continuum emission at $2\arcsec.12\times1\arcsec.70$~resolution; see Section~\ref{sec:2-SFRCalibration} for SFR calibration). {\it Panel~(b):} SFR map at $4\arcsec\approx150$~pc resolution on hexagonal grid with white contours of \chem{CO}{21} integrated intensities.  Overploted are the identified regions -- nuclear region, northern and southern inner bar end. {\it Panel~(c):} The region mask showing the chosen hexagonal points in colours of green for the northern inner bar end, blue for the nuclear region and pink for the southern inner bar end. To each of the regions we associate -- equivalent to our beam size of $4\arcsec$ -- seven individual lines of sight (hexagonal points). In the background we show the integrated intensity map of \chem{CO}{21} at $4\arcsec$ resolution. We detect in the southern inner bar end higher SFR in combination with shock tracers (e.g. CH$_3$OH\,(3k--2k)) and higher concentration of denser gas (e.g. \chem{HCN}{10}) than in the northern inner bar end.} 
    \label{fig:SFR-and-mask}
\end{figure*}

We seek answers to how the molecular lines in our data set relate to the current star formation rate (SFR) and if the different environments vary in their SFR on 150~pc scales, or if they have similar characteristics in this respect. We show a map of SFR in panel~(a) of \autoref{fig:SFR-and-mask} with contours of tracers of denser gas \hcn\ and \hnc, and shock tracer \chohthree\ at their native resolution (see \autoref{tab:Obs-properties}). SFR is peaking at the very centre and drops towards the NBE and SBE. 

To further investigate these differences and the correlations between SFR and the molecular lines in our data set, we convolve the SFR map in units of \sfrunit\ to our working resolution of $4\arcsec \approx 150$~pc and sampled it onto a hexagonal grid to match with our integrated intensity maps.\footnote{We do this also with our \sigsfr\ map in units of \sigsfrunit\ to calculate star formation efficiencies of the dense molecular gas (see Eq. \eqref{eq:sigmamol}). } 
The white contours in panel~(b) of ~\autoref{fig:SFR-and-mask} show the \cotwo\ emission indicating the three distinct features -- the nuclear region (NUC) and the northern and southern inner bar ends (NBE and SBE). Since integrated intensities of our dense gas tracers are higher in the SBE than in NBE, we expect to find higher SFR in SBE.

In the following we quote SFR values for the NUC, SBE and NBE which have been averaged over these 150~pc sized regions. 

As expected, we find the highest SFR within the NUC, $0.089$~\sfrunit. But we see differences in the other two regimes: higher SFR in the SBE than in NBE, $0.013$ and $0.006$~\sfrunit, respectively\footnote{We calculate the uncertainties for these SFR values using the uncertainty $33$GHz map and apply the equations as in  \autoref{sec:2-SFRCalibration}. We then get over a $4\arcsec$ region a SFR uncertainty of  $\pm1.35\times10^{-3}$~\sfrunit.}. This agrees with our extended molecular data set, where we find that molecules representing even denser gas, such as \ntwoh\ or \hcnten, reach a maximum in their integrated emission in SBE. Additionally we find the shock tracer CH$_3$OH peaking in their integrated intensities in SBE. This suggests that maybe weak shocks lead to a higher SFR in the SBE compared to the NBE. 

The inner bar ends also differ in their molecular masses, as already noted by \cite{schinnerer2006molecular}. We find in our $150$~pc sized apertures using \autoref{eq:sigmamol} molecular gas mass surface densities (\sigmolmass) of $\approx189{\pm}0.26$~\sigmolmassunit\ in the NBE and $\sigmolmass\approx236{\pm}5.09$~\sigmolmassunit\ in the SBE. We find that the bar ends have a molecular gas depletion time (\tdepl) that is a factor of 10 higher than NUC; the highest in the SBE $\tdepl\approx3.05\times10^{8}$yr (which is rather short, see discussion below).

Towards the NUC we find that stars are formed at a rate of $0.089$~\sfrunit\ over the past ${\sim}10$~Myr\footnote{Recall that the approximate age sensitivity of the thermal fraction of the 33~GHz continuum emission as a SFR tracer is ${\sim}10$~Myr} which translates into $0.89\times10^{6}$~M$_{\odot}$ of recently formed stars.
We find $\sigmolmass\approx298{\pm}1.98$~\sigmolmassunit\ and at the current SFR it will take $\tdepl\approx3.43\times10^{7}$ years to deplete the present supply of molecular gas. This is shorter (factor of ${\sim}10^{2}$) than found by previous studies on $\sim$kpc scales, for example a median $\sigmolmass~{\sim}2.2$~Gyr for the HERACLES galaxies in \citet{leroy2013}, or $1.1$~Gyr in \citet{usero2015variations}. These studies used a Milky Way-like $\alpha_\mathrm{CO}$ value (a factor of $10$ smaller), which is however, not applicable for \ngc\ (see e.g. \citealt{Sandstrom2013,Bigiel2020} and \autoref{sec:SigmaMolandDepltime}). The bar ends have a depletion time of a factor of 10 higher than NUC, and if we calculate \tdepl\ with a MW value, we would also get into the Gyr range for the bar ends (i.e. $\approx1.6$~Gyr for NBE and $\approx3.3$~Gyr for SBE). This demonstrates the importance of adopting a proper $\alpha_\mathrm{CO}$ factor. The nuclear region of \ngc\ was studied in terms of its SFR by e.g. \citet{Meier2004Nucleusof6946CO}, \citet{Schinnerer2007} and \citet{Tsai2013} using different SFR indicators over different spatial scales. 

For example, \citet{Schinnerer2007} found within a $3\arcsec \times 3\arcsec$ square box region a SFR of ${\sim}0.18$~\sfrunit\ using Pa$\alpha$, 6\,cm and 3\,mm continuum as SFR tracers.\footnote{They adopted a distance of $5.5$~Mpc to \ngc\ for their SFR calculations. We have re-scaled their value to the distance of $7.7$~Mpc using a factor of 1.8, see \autoref{app:SFRdistance}.} We find with our 33~GHz based star formation in the same region a SFR of ${\sim}0.11$~\sfrunit, that is slightly lower than that of \cite{Schinnerer2007}.

We remind the reader that the SFR we calculate is based on free-free emission, which is not subject to the extinction problems that complicate the estimation of SFR at optical and ultraviolet wavelengths. The free-free emission is a tracer of high-mass star formation, as it is only sensitive to stars that are able to ionise the surrounding gas and produce an \HII\ region. However,  we have to mention some uncertainties with this SFR. A different choice of initial mass function,\footnote{As well as any potential stochasticity due to IMF sampling.} or stellar population models during the calibration of 33\,GHz (see \autoref{sec:2-SFRCalibration}), could introduce a systematic offset in the derived SFR but this should not vary from region to region. The parameter T$_{e}$ in Eq. \eqref{eq:SFR} has an impact of at most ${\sim}7\%$ on the SFR and regional variations in the thermal fraction, $f^\mathrm{T}$, are likely to be too small to explain the observed variations in SFR among the bar ends (see e.g. Section 4.1 in \citealt{Querejeta2019}). Another factor that could potentially vary from region to region is the escape fraction of ionising photons (i.e. which part is absorbed in a region outside the 150~pc aperture and therefore does not contribute to the observed free-free emission). However, these uncertainties probably do not explain the factor of 2 higher SFR in the SBE than in the NBE.

\section{Results - Line ratios and relationships among molecular species}
\label{sec:ResultsB}
Molecular line ratios open up the possibility to investigate the physical and chemical state of the ISM. Integrated line ratios among certain molecules were proposed to provide insights into the environment (e.g. $\chem{HCO^+}/\chem{HCN}$; \citealt{Loenen2008}), the temperatures ($\chem{HCN}/\chem{HNC}$; \citealt{Hacar2020}) and density variations (e.g. line ratio pattern to CO; \citealt{Leroy2017}). The question can be asked whether we see differences between the inner bar ends and the nuclear region, and if these line ratio diagnostics hold for an extragalactic extreme environment like the centre of \ngc. For a meaningful analysis of the line ratios and their relationships, we need to correct the interferometric data for missing short-spacing information. Therefore, we use the five emission lines (\chem{CO}{10}, \chem{CO}{21}, \chem{HCN}, \chem{HCO^+} and \chem{HNC}) for which single-dish observations were available (see \autoref{tab:SSC-Characteristics}). The line ratios are calculated from integrated line intensities in units of $\mathrm{K}\,\mathrm{km}\,\mathrm{s}^{-1}$ throughout the entire paper and we refer to them simply as `line ratio'. 

\subsection{Ratios of integrated intensities}
\label{Sec:Res-LineRatios}

\begin{figure}[ht!]
\centering
    \includegraphics[width=1.0 \linewidth]{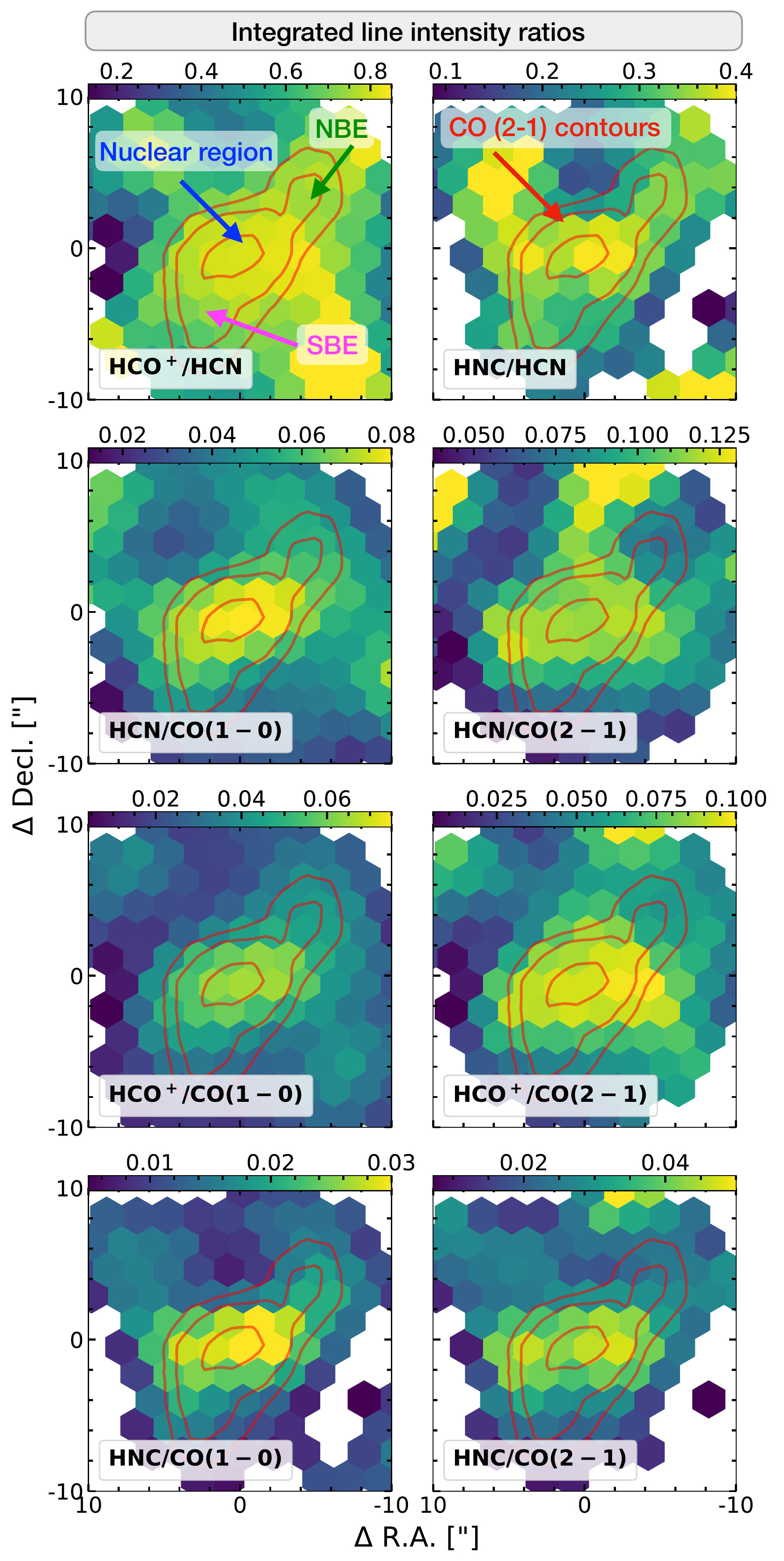}
    \caption{\textbf{Line ratio maps of the central $\bm{20\arcsec}$.} We show the line ratio maps for the central $20\arcsec \approx 750$~pc. We overplot on each map contours of the \chem{CO}{21} integrated intensity map in red and denote the features associated with the inner bar (see Figure~\ref{fig:SFR-and-mask} for the region mask).}
    \label{fig:RatioMaps-uvtrimmed-SSC}
\end{figure}

In \autoref{fig:RatioMaps-uvtrimmed-SSC} we show line ratio maps for our \textit{SSC + \uv trim data} set (see \autoref{tab:SSC-Characteristics}). We specify the line ratios such that the generally brighter line is in the denominator, while the overall weaker line is in the numerator. The line ratio maps were calculated by only taking integrated intensities with $\SN > 5$ for the fainter dense gas tracers (DGTs: \chem{HCN}, \chem{HCO^+} and \chem{HNC}) and $\SN > 15$ for the bulk molecular gas tracers (CO); non-detections were discarded from the line ratio maps:

\begin{equation}
    \mathrm{Ratio\,map} = \frac{I_\mathrm{line_1} \left[ I_\mathrm{line_1}/\sigma_\mathrm{line_1} > \epsilon \right]}{I_\mathrm{line_2} \left[ I_\mathrm{line_2}/\sigma_\mathrm{line_2} > \epsilon \right]} \
    \begin{cases}
      \epsilon = 5 & \text{for DGTs}\,,\\
      \epsilon = 15 & \text{for CO}\,.
      \label{eq:Line ratios}
    \end{cases}   
\end{equation}
In the next section, we include non-detections in our analysis. For the case of no significant detection we replaced the values with the upper limits ($2\sigma$ in \autoref{eq:unc}) and include the propagated errors ($\sigma_\mathrm{prop}$). We derive them as: 
\begin{equation}
    \sigma_\mathrm{prop} = \frac{|I_\mathrm{line_1}|}{|I_\mathrm{line_2}|} \sqrt{{\left(\frac{\sigma_\mathrm{line_1}}{I_\mathrm{line_1}} \right)}^2 + {\left(\frac{\sigma_\mathrm{line_2}}{I_\mathrm{line_2}}\right)}^2}~.
\end{equation}
The errors are expressed on a logarithmic scale of base~10 as:
\begin{equation}
    \sigma_\mathrm{log} = \frac{1}{\ln(10)} \times \left( \sigma_\mathrm{prop} / I_\mathrm{ratio} \right) \approx 0.434 \times \left( \sigma_\mathrm{prop} / I_\mathrm{ratio} \right)~.
\end{equation}

In \autoref{fig:RatioMaps-uvtrimmed-SSC} the line ratio maps for the central $20\arcsec \approx 750$~pc show differences in line ratios between the environments NUC, SBE and NBE. In the following we report straight mean values over these regions by applying the mask in \autoref{fig:SFR-and-mask}. The ratios $\chem{HCO^+}/\chem{HCN}$ and $\chem{HNC}/\chem{HCN}$ both show values below unity in all three regions; and over the entire field of view. In both cases, the ratio to HCN is greater in the NUC than in the SBE or NBE. 

Looking at the NUC reveals values of $0.81\pm0.01$ and $0.37\pm0.01$ for $\chem{HCO^+}/\chem{HCN}$ and $\chem{HNC}/\chem{HCN}$, respectively. Ratios to CO indicate higher values in the SBE than in NBE; best visible in the case of $\chem{HNC}/\chem{CO}{21}$. We discuss the implications of these line ratios in \autoref{Disc:HNC/HCN}, \ref{Disc:HCO/HCN} and \ref{Disc:lineratiopattern}.

\subsection{The relationship between molecular lines and SFR surface density}
\label{sec:comp-moleculesandSFR}

\begin{figure*}[ht!]
    \includegraphics[width=0.95\linewidth]{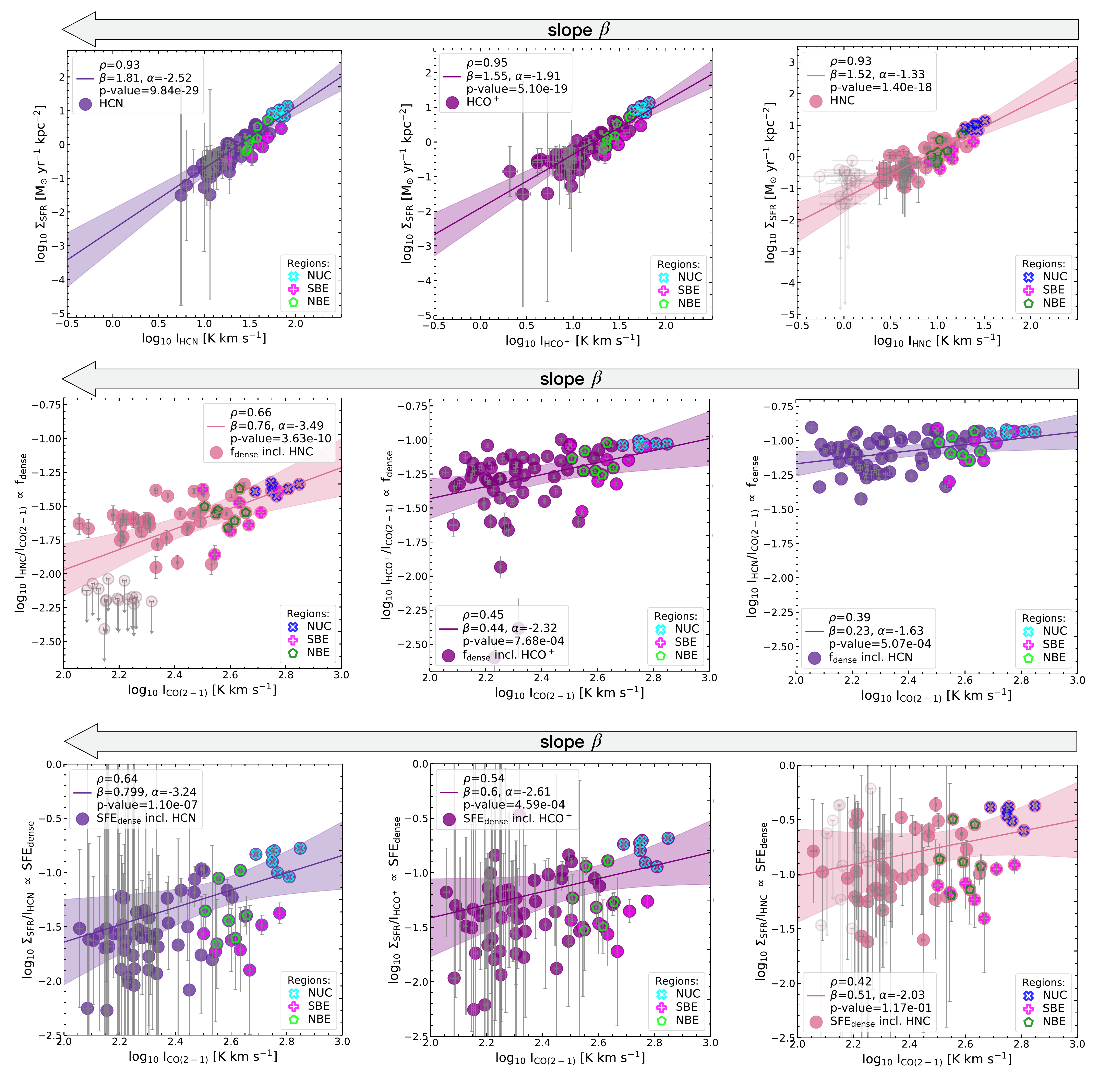}
    \caption{\textbf{Correlation plots ordered by their slope, $\bm \beta$.} Dark purple colours show HCN, purple HCO$^+$ and pink HNC. {\it Top row}: Integrated intensities of the dense gas tracers (ordered by their slope, $\beta$) versus \sigsfr. {\it Middle row}: Integrated intensity of \chem{CO}{21} -- an indicator of the mean volume density -- versus the line ratio with \chem{CO}{21} -- a~spectroscopic tracers of \fdense\ ;  {\it Bottom row:} Integrated intensity of \chem{CO}{21} -- an indicator of the mean volume density -- versus the ratio of \sigsfr\ with the dense gas tracers -- a~spectroscopic tracers of \sfedense. The \texttt{linmix} fits (accounting for upper limits) are shown as solid lines surrounded by $3\sigma$ confidence intervals (shaded regions of corresponding colours). Absolute uncertainties are plotted on each data point, which are generally small towards the regions NUC, NBE and SBE. We find the highest uncertainties in the outskirts in our \sigsfr\ map. Correlations including HNC result in 58 significant sight lines and 15 upper limits (denoted as light pink markers). We display in each panel Pearson's correlation coefficient and the power-law slope, intercept and $p$-value (see Table~\ref{tab:Corr}).}
    \label{fig:Corr}
\end{figure*}
\begin{table*}
\centering
\caption{ Scaling relations and correlation coefficients for the central ${ 20\arcsec \approx 745}$~pc of \ngc\ at ${150}$~pc scale.}
\label{tab:Corr}
\begin{tabular}{cccccccc}\hline\hline
$\log(y)$            & $\log(x)$       & $\beta$           & $\alpha$          & $\rho$  & $p$-value  & scatter & Cov($\log(x)$,$\log(y)$) \\ 
& & \textit{slope} & \textit{intercept} & \textit{Pearson} & & &\\
& & & & (1) & & (2) & \\\hline
\sigsfr      & HCN     & 1.81$\pm$0.13  & -2.52$\pm$0.20 & 0.93$\pm$6.6E-04 & 9.84E-29 & 0.04 & 9.75   \\
\sigsfr      & HCO$^+$ & 1.55$\pm$0.11  & -1.91$\pm$0.15 & 0.95$\pm$1.7E-02 & 5.10E-19 & 0.04 & 8.24  \\
\sigsfr      & HNC     & 1.52$\pm$0.14  & -1.33$\pm$0.14 & 0.93$\pm$1.1E-02 & 1.40E-18 & 0.07 & 3.30  \\
\sigsfr      & CO(1-0) & 2.46$\pm$0.22  & -6.56$\pm$0.60 & 0.89$\pm$2.2E-03 & 7.90E-23 & 0.06 & 112.00  \\
\sigsfr      & CO(2-1) & 1.93$\pm$0.22  & -4.68$\pm$0.54 & 0.80$\pm$2.4E-03 & 5.90E-16 & 0.09 & 68.70  \\\hline
HCN / CO(2-1)     $\propto$ \fdense & CO(2-1) & 0.23$\pm$0.06 & -1.63$\pm$0.16 & 0.39$\pm$3.4E-04 & 5.07E-04 & 0.01 & 0.01  \\
HCO$^+$ / CO(2-1) $\propto$ \fdense & CO(2-1) & 0.44$\pm$0.11 & -2.32$\pm$0.26 & 0.45$\pm$6.5E-04 & 7.68E-04 & 0.03 & 0.02 \\
HNC / CO(2-1)     $\propto$ \fdense & CO(2-1) & 0.76$\pm$0.13 & -3.49$\pm$0.31 & 0.66$\pm$8.5e-08 & 3.63E-10 & 0.03 & 0.04\\\hline
\sigsfr / HCN     $\propto$ \sfedense & CO(2-1) & 0.80$\pm$0.22  & -3.24$\pm$0.56 & 0.64$\pm$2.9E-01 & 1.10E-07 & 0.04 & 0.04   \\
\sigsfr / HCO$^+$ $\propto$ \sfedense & CO(2-1) & 0.60$\pm$0.20  & -2.61$\pm$0.52 & 0.54$\pm$3.2E-01 & 4.59E-04 & 0.04 & 0.03 \\
\sigsfr / HNC     $\propto$ \sfedense & CO(2-1)  & 0.51$\pm$0.24  & -2.03$\pm$0.62 & 0.42$\pm$2.8E-01 & 1.17E-01 & 0.05 & 0.02\\ \hline\hline  
\end{tabular}
\begin{minipage}{2.0\columnwidth}
    \vspace{1mm}
    {\bf Notes}: (1): We perform a Monte Carlo analysis perturbing the x and y data points to get the uncertainty of $\rho$; (2): The regression intrinsic scatter; 
\end{minipage}
\end{table*}




In this section we investigate how our molecular species correlate with \sigsfr\ and how the dense gas fraction traced by $\chem{HCN}/\chem{CO}$ (we investigate also $\chem{HCO^+}/\chem{CO}$ and $\chem{HNC}/\chem{CO}$) responds to the integrated intensity of CO (an indicator of the mean volume density, see below).  We study how these quantities relate to conditions in the centre of \ngc. \autoref{fig:Corr} shows the relationships we investigate. 

We characterise scaling relations by including upper limits and measurement errors, using the hierarchical Bayesian method described in \cite{Kelly2007}. This approach is available as a python package: \texttt{linmix}\footnote{\url{https://linmix.readthedocs.io/en/latest/index.html}}. It performs a linear regression of $y$ on $x$ while having measurement errors in both variables and being able to account for non-detections (upper limits) in~$y$. The regression assumes a linear relationship in the form of:
\begin{equation}
    \log(y) = \beta \times \log(x) + \alpha~,
\end{equation}
where $\beta$ is the slope, and $\alpha$ is the $y$-intercept\footnote{We specify the covariance between the measurement errors in x and y (xycov parameter) and set K$=2$.}. We find Pearson's correlation coefficients $\rho$ of the data sets for each fitted relationship, and the $3\sigma$ confidence intervals are estimated via Markov chain Monte Carlo (MCMC). For a detailed description we refer to \citet{Kelly2007}. We provide all the correlations in \autoref{tab:Corr}.\\ \\
%
%
%
Dense molecular gas traced by, e.g., HCN emission, has been observed to correlate with SFR (e.g. \citealt{gao2004hcn,lada2010star,lada2012star}). \citet{Kauffmann2017} showed that HCN traces more extended gas and therefore can have an impact on the observed SF trends in galaxies (also see e.g. \citealp{Pety2017,Barnes2020}). \citet{Krumholz2007} showed that star formation correlates with any line with a critical density comparable to the median molecular cloud density. Therefore, we expect to see positive correlations between \sigsfr\ and the surface density of \chem{HCN}, \chem{HCO^+} and \chem{HNC}. In our observations this is confirmed with the additional characteristic that the molecular line with the lowest effective critical density ($n_\mathrm{eff}$) shows the strongest correlation ($n_\mathrm{eff}$ ordered as $\chem{HCO^+} < \chem{HNC} < \chem{HCN}$; see \autoref{tab:Obs-properties}). HCO$^+$ shows the strongest ($\rho~{\sim}0.95$, see \autoref{tab:Corr} for uncertainties of $\rho$) correlation with a slope of $\beta = 1.55$ and a small intrinsic scatter:
\begin{equation}
    \log(\sigsfr) = 1.55\pm0.11 \times \log(I_{\chem{HCO^+}}) - 1.91\pm0.15~.
\end{equation}
We note that our slopes are all higher ($\beta>1$) than those found at global scales ($\beta<1$; e.g. \citealt{gao2004hcn,Krumholz2007}). We speculate that this is due to two contributing factors. Firstly, in this work we are focussing on the centre of \ngc\ (central $20\arcsec\approx745$~pc), and not the whole galaxy disc. Hence, this could be due to the limited dynamic range in environmental conditions we are including within the analysis - that is focussing on the densest and most actively star-forming gas within the galaxy. Secondly, this could be a result of the resolution obtained with our PdBI observations. At around 150
~pc, we are close to resolving individual discrete star-forming and/or quiescent regions, which could result in the different slope compared to lower resolution studies that include an average on small scale conditions within each sample point; somewhat akin to following a branch of the tuning fork within the recent `uncertainty principle for star formation' framework \citep{kruijssen18,kruijssen19a,chevance20,Kim2021}.
%
%
 %
The dense gas fraction, \fdense\footnote{In this work, we take the \chem{CO}{21} data for our \fdense\ estimates because they have a higher S/N and better quality than \chem{CO}{10}.}, usually traced by the integrated intensity of \chem{HCN}{10} over \chem{CO}, has been observed to increase towards the centres of galaxies (e.g. \citealt{usero2015variations,bigiel2016empire,gallagher2018dense,Jimenez-Donaire2019EMPIRE,Jiang2020,Beslic2021}). In turbulent cloud models, increasing the mean volume density of a molecular cloud results in a shift of the gas density distribution to higher densities (e.g. \citealt{Federrath2013}). 
In addition, the velocity dispersion (or Mach number) widens the density PDF, which causes a larger fraction of the mass to be at higher gas densities. 
Combined, these increase the fraction of gas above a fixed density (e.g. the effective critical density of HCN), and consequently, these models predict a positive correlation between the volume density and \fdense\ (see \citealt{Padoan2014} for a review). Such trends are, in particular, interesting to study within galaxies centres environments thanks to their higher average densities and broader (cloud-scale) line widths compared to typical disc star-forming regions (e.g. \citealp{Henshaw2016,Krieger2020}). 
In the following we use the integrated intensity of \chem{CO}{21} as an indicator of the mean volume density (\citealt{Leroy2016,Sun2018,gallagher2018spectroscopic}, see also \autoref{Disc:lineratiopattern}) and explore the dense gas fraction using \chem{HCN}, \chem{HCO^+} and \chem{HNC}. All three \fdense\ versus \chem{CO}{21} fits in \autoref{fig:Corr} present sub-linear power-law indices (i.e. $\beta < 1.0$). We find that the correlations are weakest for the dense gas fraction using HCN ($\rho~{\sim}0.39$) and strongest for $\chem{HNC}/\chem{CO}{21}$ ($\rho~{\sim}0.66$):
\begin{equation}
    \log \left( \frac{I_{\chem{HNC}}}{I_{\chem{CO}{21}}} \right) = 0.76\pm0.13 \times \log \left( I_{\chem{CO}{21}} \right) - 1.63\pm0.16~.
\end{equation}
The particular order in the slopes and correlation coefficients do not follow the order of $n_\mathrm{eff}$ as given in \citet{Shirley2015}. Instead, they show the order of $\beta_\mathrm{HNC}$ > $\beta_\mathrm{HCO^+}$ > $\beta_\mathrm{HCN}$ (see second row in \autoref{fig:Corr}). Possible explanations for this behaviour could be anomalous excitation for one of the species, for example IR pumping of HCN and HNC levels and/or peculiar filling factors.
%
%
%
The star formation efficiency of dense gas (\sfedense) -- the ratio of SFR to the integrated intensity of \chem{HCN}{10} -- has been observed to decrease towards galaxy centres (same studies as above). This is because the critical overdensity (relative to the mean density) is higher due to the higher Mach number, but also that the absolute critical density (i.e. that obtained from the above line ratios) is higher due to the higher (1) mean density and (2) Mach number (in the context of the CMZ of the Milky Way, see e.g. \citealt{Kruijssen2014a}). 
Hence, as the mean density of a molecular cloud approaches $n_\mathrm{eff}$ of the molecular line, the line's intensity will increasingly trace the bulk mass of the cloud, and not exclusively the overdense, star-forming gas; in effect reducing the apparent SFE.
From that we would expect to see \sfedense\ to drop with $I_{\chem{CO}{21}}$ ($\propto$~volume density) as it has been observed over the disc regions of galaxies (e.g. \citealt{gallagher2018dense,Jimenez-Donaire2019EMPIRE}). We use the ratio of \sigsfr\ to the integrated intensities of the dense gas tracers as \sfedense, and find a different picture of \sfedense\ with galactocentric radius -- increasing towards the NUC; as it shows higher efficiencies in the NUC than in the bar ends (see third row in \autoref{fig:Corr}). We find a moderate\footnote{Here we refer to a moderate correlation if $\rho$ lies in the range of $0.5{-}0.7$.}, approximately linear correlation between $\sigsfr / I_{\chem{HCN}}$ and $I_{\chem{CO}{21}}$. The other two \sfedense\ measurements show weak sub-linear relationships with $\rho~{\sim}0.46$ and $\rho~{\sim}0.19$ for ratios using \chem{HCO^+} and \chem{HNC}, respectively. We find that \sfedense\ increases with increasing $I_{\chem{CO}{21}}$, contrary to the trends found by \citet{gallagher2018dense, Jimenez-Donaire2019EMPIRE}. The reason for that could be the special environment of \ngc\ (inner bar) and/or the different used SFR tracers (we use the free-free emission of 33~GHz continuum compared to H$\alpha$+24\,$\mu$m and total infrared). Also the above studies could not resolve the inner bar of \ngc\ (\citealt{Jimenez-Donaire2019EMPIRE}, angular resolution of $33\arcsec\approx~$1~kpc) or \ngc\ was not in the nearby disc galaxy sample \citep{gallagher2018dense}. However, we notice, as expected, that stronger correlations in \fdense\ result in weaker correlations in \sfedense\ and vice versa (e.g. $I_{\chem{HNC}} / I_{\chem{CO}{21}}$ with $\rho = 0.76\pm0.13$ and $\sigsfr / I_{\chem{HNC}}$ with $\rho = 0.51\pm0.24$; see \autoref{tab:Corr}).   

\section{Implications for spectroscopic studies of other galaxies -- from bulk molecular gas to dense gas}
\label{sec:Discussion}
Here we discuss some of the common integrated line ratios in more detail and what they can be used for. We compare our high-resolution ($150$~pc) integrated line ratios of dense gas tracers towards the central region of \ngc\ with available dense gas tracers from the EMPIRE survey, which include eight additional galaxy centers. It is worth noting that the central regions in the EMPIRE sample are ${\sim}1$~kpc sized areas which result in one data point per galaxy (see \citealt{Jimenez-Donaire2019EMPIRE} for more details). Nevertheless, it provides us with an understanding of how \ngc\ compares to other centers. 


\subsection{\texorpdfstring{$\bm{R_{21}}$}{R21} variations in the nuclear region and inner bar ends}
\label{sec:R21}

The \chem{CO}{21}-to-\chem{CO}{10} line ratio, $R_{21}$, is widely used to convert \chem{CO}{21} emission to \chem{CO}{10}, which then can be further converted via the $\alpha_\mathrm{CO}$ conversion factor to the molecular gas mass. It has been shown that higher $R_{21}$ values are expected within the central kpc in individual galaxies \citep{Leroy2009Heracles, leroy2013, koda2020}. The same trend has been found by \citet{denBrok21} and \citet{yajima2021} studying a galaxy sample. \ngc\ is in both of the aforementioned studies. They find within the central kpc region $R_{21}~{\sim}0.7$ (at an angular resolution of $33\arcsec~{\sim}1$~kpc; \citealt{denBrok21} using the EMPIRE sample) and $R_{21}~{\sim}1.1$ (angular resolution of $17\arcsec~{\sim}0.6$~kpc; \citealt{yajima2021}). However, these studies are not able to resolve variations within the centre between different sub-features. From a physical point of view, $R_{21}$ should depend on the temperature and density of the gas, as well as on the optical depths of the lines (e.g. \citealt{Penaloza2018}). Therefore, understanding how $R_{21}$ varies in response to the local environment also has the prospect of providing information about the physical conditions of the molecular gas.
 
\begin{figure}[ht!]
\centering
    \includegraphics[width=1.0 \linewidth]{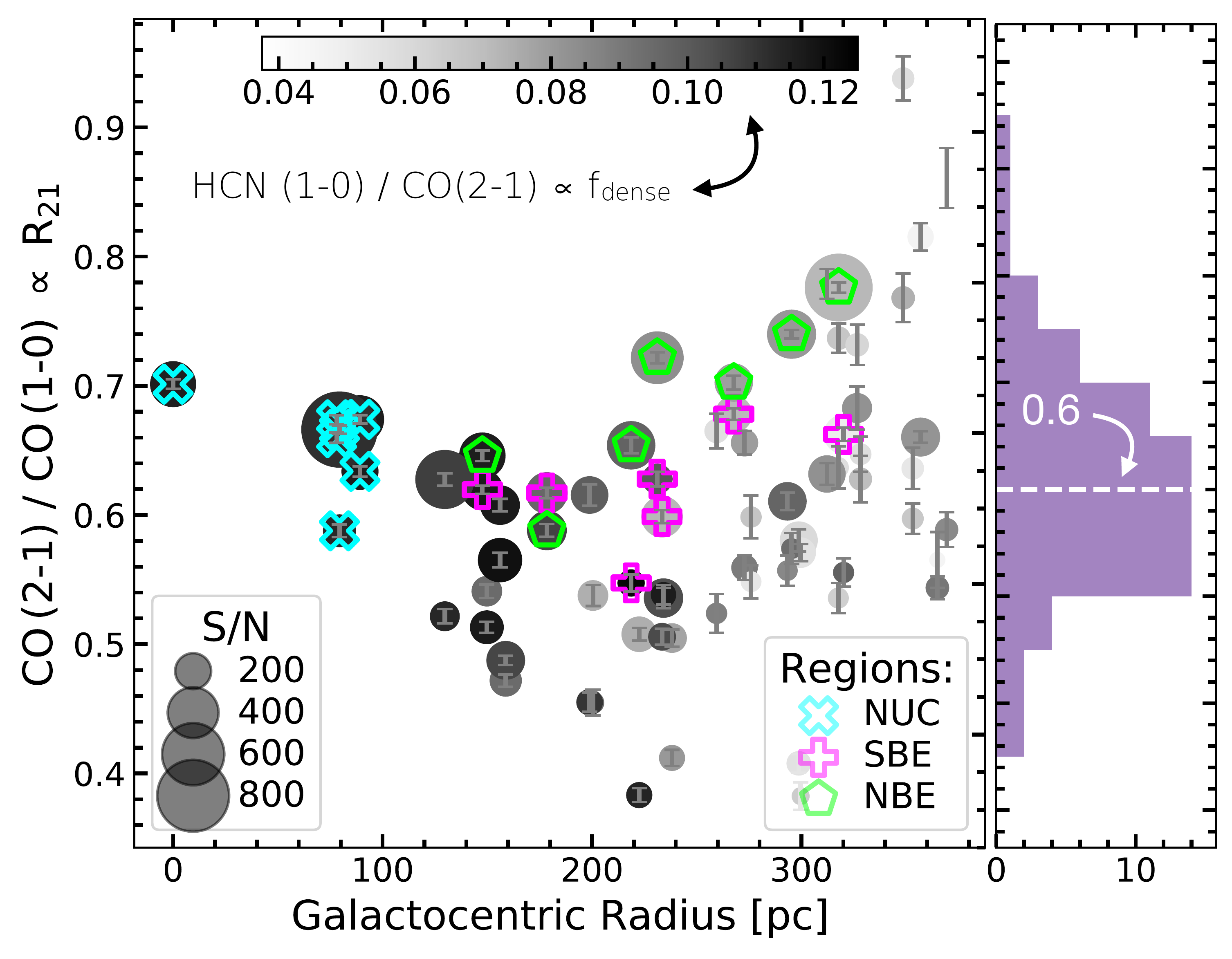}
    \caption{ \textbf{$\bm{R_{21}}$ against galactocentric radius.} The circular shaped markers represent all of the $R_{21}$ values for the central $20\arcsec\approx745$~pc coloured by their \fdense. The different sizes show their varying signal-to-noise ratio in \cotwo. Additionally, we overplot the contours of the three different regions: The cyan colour refers to the nuclear region (NUC), the green to the northern bar end (NBE) and the magenta to the southern bar end (SBE). Along the galactocentric radius, we find in the centre higher $R_{21}$ values, an initial steady decrease followed by a gradual increase. $R_{21}$ towards NBE is higher than expected (more in Section~\ref{sec:R21}).}
    \label{fig:R21}
\end{figure}

With our higher resolution ($4\arcsec \approx 150$~pc) observations we are able to resolve smaller-scale structures and thus have the opportunity to investigate variations of $R_{21}$ within the centre. In \autoref{fig:R21} we show $R_{21}$ against galactocentric radius, highlighting NUC, NBE and SBE in different colours. 

We see along the galactocentric radius, in the centre higher $R_{21}$ values, an initial steady decrease followed by a gradual increase. We find average $R_{21}$ of $0.65\pm2$E$-3$ within NUC and even lower values in the southern end of the bar $R_{21}=0.62\pm3$E$-2$ (see \autoref{tab:regions-charac}). Interestingly, however, we observe higher $R_{21}$ values towards the northern end of the bar ($R_{21}=0.70\pm3$E$-2$) compared to the NUC (factor of~${\sim}1.08$), despite the higher \sigsfr\ in the SBE compared to the NBE (see \autoref{sec:3-SFRindicators-compare}). The reason for a higher $R_{21}$ value could be related to denser gas and/or warmer gas with higher temperatures.  We colour-code points in \autoref{fig:R21} by their $\chem{HCN}{10}/\chem{CO}{21}$ line ratio ($\propto \fdense$) and find them showing higher \fdense\ towards the NUC. Furthermore, we would expect to observe higher HCN-to-CO ratios towards the NBE. However, we do not see an increase in the denser gas in the NBE, suggesting a different physical driver for the increased $R_{21}$ in the NBE. We analyse the three regions regarding their molecular gas density in more detail in \autoref{Disc:lineratiopattern}.

In summary, we find higher $R_{21}$ values towards one of the inner bar ends of \ngc\ compared to the nuclear region. If substructures such as small-scale bar ends are to be observed and analysed, $R_{21}$ may possibly deviate in a minimal way from the kpc-sized $R_{21}$ values from the literature.    

\subsection{\texorpdfstring{${\mathrm{HNC}/\mathrm{HCN}}$}{HNC/HCN}: sensitive to kinetic temperatures in extragalactic environments?}
\label{Disc:HNC/HCN}

In interstellar space, isomers do not necessarily share similar chemical or physical properties. HNC (hydrogen iso-cyanide) and HCN (hydrogen cyanide) isomers are both abundant in cold clouds, but at temperatures exceeding ${\sim}30$~K, HNC begins to be converted to HCN by reactions with atomic~H. These isomers exhibit an abundance ratio of unity at low temperatures \citep{Schilke1992,Graninger2014}. A major study to understand this ratio, which was focused on the galactic SF region Orion Molecular Cloud 1 (OMC-1), was carried out by \citet{Schilke1992}. They found that the \hnchcn\ ratio is ${\sim}1/80$ in the direction of Orion Kleinmann-Low (Orion-KL) but increases to $1/5$ in regions with lower temperatures near Orion-KL. In the coldest OMC-1 regions, the ratio rises further to~1. The temperature dependence suggests that the ratio must be kinetically controlled \citep{Herbst2000}, so that the integrated intensity line ratio \hnchcn\ should decrease at higher temperatures \citep{Pety2017}. 

Whether this ratio is sensitive to temperatures in extragalactic sources is uncertain (e.g.  \citealt{Aalto2002, Meier2005CenterofIC342, 2012MeierChemistryinMaffei2}). For example, \cite{Meier2005CenterofIC342} found a kinetic temperature for the centre of IC~342 using the \hcnhnc\ ratio and the empirical relation of \cite{Hirota1998} of a factor of~2 less than the dust temperature and the kinetic temperature of the gas using \chem{CO}{21} and ammonia (\chem{NH_3}). They suggested that there might be an abundant dense component in IC~342 that is significantly cooler and more uniform than the more diffuse CO, but this was not consistent with the similar distribution of CO, HNC and HCN, unless such a dense component directly follows the diffuse gas. Alternatively, they assumed that this line relationship might not capture temperature, for the nuclear region of IC~342. Also \cite{Aalto2002} find overluminous HNC in many of the most extreme (and presumably warm) (U)LIRGs and suggest that its bright emission cannot be explained by the cool temperatures demanded.

Recently, however, this line ratio has come back into focus. \cite{Hacar2020} demonstrated the strong sensitivity of the \hnchcn\ ratio to the gas kinetic temperature, $T_\mathrm{k}$, again towards the Orion star-forming region. 
They compared the line ratio with \chem{NH_3} observations (\citealt{Friesen2017}) and derived $T_\mathrm{k}$ from their lower inversion transition ratio $\chem{NH_3~(1,1)} / \chem{NH_3~(2,2)}$. In particular, they found that $T_\mathrm{k}$ can be described by a two-part linear function for two conditions (we show only Eq.~(3) in \citealt{Hacar2020}):
\begin{equation} \label{eq:hcn-hnc}
    T_\mathrm{k} ~ [\mathrm{K}] = 10 \times \frac{I_{\chem{HCN}}}{I_{\chem{HNC}}} \quad \text{for} \quad \frac{I_{\chem{HCN}}}{I_{\chem{HNC}}} \leq 4~.
\end{equation}
However, since the $\chem{NH_3~(1,1)} / \chem{NH_3~(2,2)}$ transitions are only sensitive to $T_\mathrm{k} \lesssim 50$~K (see e.g. Fig.~1 of \citealt{Mangum2013}), the calibration shown above only represents the low temperature regime. The challenge of apply such concepts in nearby galaxies is that the concentrations of dense gas studied by \citet{Hacar2020} in the local Milky Way environment (i.e. Orion) are very compact (${\sim}0.1{-}1$pc; \citealt{Lada2003}), and representative of solar neighbourhood environmental conditions (e.g. chemistry, and average densities, Mach numbers and kinetic temperatures). Achieving such a resolution is currently extremely difficult in an extragalactic context (e.g. 1\,pc = 0.025\arcsec\ at the distance of the NGC 6946), potentially limiting our capability to determine kinetic temperatures (e.g. when using \chem{H_2CO}; see \cite{Mangum2019} and below). That said, galaxy centres present an ideal regime in which to focus our efforts. As densities similar to the concentrations observed within local star-forming regions are not compact, but can span (nearly) the entire CMZ (so up to 100\,pc), leading to luminous and, importantly, extended HCN and HNC emission (\citealt{Longmore2013,Rathborne2015,Krieger2017,Petkova2021}). Hence, testing this temperature probe within galaxy centres overcomes the requirement for such extremely high-resolution observations, and should be possible with $\sim$\,100\,pc scale measurements presented in this work. 

We find in the nuclear region of \ngc\ a mean \hnchcn\ ratio of ${\sim}0.37$. For the bar ends we find lower ratios: ${\sim}0.31$ for NBE and ${\sim}0.32$ for the SBE (see \autoref{tab:regions-charac}). \cite{Jimenez-Donaire2019EMPIRE} reported over kpc-scales a ratio of ${\sim}0.31$. There are no other ratios of HNC and HCN in the literature for a comparison, since HNC has hardly been observed towards \ngc.  If we assume that the ratio of HCN and HNC traces kinetic temperature and adopt the \cite{Hacar2020} relation\footnote{The equation in \citealt{Hacar2020} uses HCN over HNC ($\chem{HCN}/\chem{HNC}$), therefore we have for e.g. NUC = $1/0.37 = 2.7$.}, then we would infer a $T_\mathrm{k} (\chem{HCN}/\chem{HNC})$ of ${\sim}27$~K for the NUC. For the bar ends we calculate slightly higher temperatures of ${\sim}31$~K and ${\sim}32$~K (on $4\arcsec \approx 150$~pc scales).
\cite{Meier2004Nucleusof6946CO} predict $T_\mathrm{k}~{\sim}20{-}40$~K (for $n_{\chem{H_2}} = 10^{3}$~cm$^{-3}$) based on CO and its isotopologues using a large velocity gradient (LVG) radiative transfer model (building on the models presented in \citealp{Meier2000}). Including HCN into their LVG models, favoured higher $T_\mathrm{k}$ (${\sim}90$~K) and $n_{\chem{H_2}}$ (${\sim}10^{4} - 10^{4.5}$~cm$^{-3}$), but these numbers are sensitive to whether \chem{^{13}CO} and HCN trace the same gas component. With the inverse transition of \chem{NH_3}, \cite{Mangum2013} found $T_\mathrm{k}~{\sim}47{\pm}8$~K (using the $\chem{NH_3~(1,1)} / \chem{NH_3~(2,2)}$ ratio) which is a factor of ${\sim}2$ higher than our inferred $T_\mathrm{k}$~(\hnchcn) . The higher excitation $\chem{NH_3~(2,2)} / \chem{NH_3~(4,4)}$ ratio, which monitors $T_\mathrm{k} \lesssim 150$~K, was not yet detected towards \ngc\ \citep{Mangum2013, Gorski2018}. Those studies already showed that an unambiguous determination of the kinetic temperature is challenging.

Comparing our obtained $T_\mathrm{k}$~(\hnchcn) to typical $T_\mathrm{k}$ measurements towards the central molecular zone (CMZ) in the Milky Way, reveals higher gas temperatures in the CMZ ($T_\mathrm{k} > 40$~K; \citealt{Ao2013,Ott2014,Ginsburg2016,Krieger2017}). Investigating individual CMZ clouds, \cite{Ginsburg2016} used \chem{para{-}H_2CO} transitions as a temperature tracer which is sensitive to warmer ($T_\mathrm{k} > 20$~K) and denser ($n~{\sim}10^{4-5}$~cm$^{-3}$) gas. They determined gas temperatures ranging from ${\sim}60$ to ${>}100$~K. We know from extragalactic studies that high kinetic temperatures ($50$ to ${>}250$~K) can be produced by both cosmic ray and mechanical (turbulent) heating processes \citep{Mangum2013,Gorski2018}. The CMZ of our own Galaxy seems to be different in this respect where the mismatch between dust and gas temperature at moderately high density ($n~{\sim}10^{4-5}$~cm$^{-3}$) is better explained by mechanical heating \citep{Ginsburg2016}. However, it is not clear what might be the reason for observing low $T_\mathrm{k}$ in an environment where we expect mechanical heating processes. 

Compared to the better studied extragalactic nuclear source NGC~253, \cite{Mangum2019} found kinetic temperatures on $5\arcsec\approx85$~pc scales of $T_\mathrm{k} > 50$~K using ten transitions of \chem{H_2CO}, while on scales ${<}1\arcsec$ (${\sim}17$~pc) they measure $T_\mathrm{k} > 300$~K. Using \chem{NH_3} as a thermometer indicates the presence of a warm and hot component with $T_\mathrm{k} = 75$~K and $T_\mathrm{k} > 150$~K, respectively \citep{Gorski2018, Perez-Beaupuits2018}. The reported $\chem{HCN}/\chem{HNC}$ ratio over the whole nucleus is ${\sim}1$, which if \cite{Hacar2020} were true would imply $T_\mathrm{k} (\chem{HNC}/\chem{HCN})~{\sim}10$~K. This is in contrast to the aforementioned warm component a factor of $7$ lower. It indicates that this ratio provides no reliable information about $T_\mathrm{k}$ in the extragalactic region of NGC~253.

We find for the eight kpc-sized galaxy centers in the EMPIRE sample -- using Eq.~\eqref{eq:hcn-hnc} -- $T_\mathrm{k}$~(\hnchcn) lower than $50$~K for almost all galaxies\footnote{For NGC~3627, NGC~4254 and NGC~5055 we had to take for the calibration the second part of the two-part linear function in \cite{Hacar2020}.}. NGC~3627 and NGC~5055 exhibit higher kinetic temperatures, $58$~K and $61$~K, respectively. \cite{Beslic2021} found towards NGC~3627 on 100~pc scales (using the same framework) lower $T_\mathrm{k}$~(\hnchcn) of ${\sim}34$~K. 

In summary, the \hnchcn\ ratio results in low inferred kinetic temperatures in galaxy centres ($T_\mathrm{k} < 50$~K) if \cite{Hacar2020} prescription can be applied. However, in the absence of other accurate kinetic temperature measurements (with \chem{NH_3} or \chem{H_2CO}) against \ngc\ and the EMPIRE galaxies, we speculate that the isomer ratio is not a suitable $T_\mathrm{k}$ probe for large sized extragalactic regions, and, in particular, towards galaxy centres that can also have high optical depths and complex chemistry (also AGN activity and prominent additional excitation mechanisms). A comparison with similarly high resolution observations towards a galaxy centre using kinetic temperatures derived from ammonia emission would be worthwhile to further investigate the \hnchcn\ temperature sensitivity framework of \cite{Hacar2020}.



\subsection{Examining ratios among HCN, \texorpdfstring{HCO$\bm{^{+}}$}{HCO+} and HNC as a diagnostic of AGN state}
\label{Disc:HCO/HCN}

Clouds of gas in the inner kpc of galaxies are exposed to intense radiation, which can emanate from an active galactic nucleus (AGN), seen as hard $X$-rays with $E > 1$~keV; from starburst regions, dominated by radiation of O~and B~stars; or from both. Excess of $X$-ray emission affects the thermal and chemical balance of the surrounding ISM, which in turn could influence molecular line emission (see below). The centre of \ngc\ exhibits no clear indication for the presence of an AGN. \citet{Holt2003} studied the distribution of the X-ray emission over the full disc of \ngc\ and found several low-luminosity point-like sources, one of which coincides with the dynamical centre determined by \cite{schinnerer2006molecular}.

Theoretical modelling  of ratios between \chem{HCO^+} and HCN suggested it as a diagnostic tool to distinguish between photon-domi\-nated regions (PDRs) and $X$-ray-domi\-nated regions (XDRs) for a given column density of $N~{\sim}10^{23}$~cm$^{-2}$ in the presence of ionizing radiation (\citealt{Meijerink2005,Meijerink2007}). In their models the $\chem{HCO^+}/\chem{HCN}$ ratio seems to systematically vary with gas density, the incident ultraviolet and infrared radiation field. 
Also mechanical heating and cosmic ray ionization could be possible sources of variations in $\chem{HCO^+}/\chem{HCN}$ (\citealt{Bayet2010,Meijerink2011}). Including HNC to the analyses, \cite{Loenen2008} claims that in XDRs HNC is always stronger than the HCN line, whereas the inverse trend is seen in PDRs (resulting in line ratios lower than unity). For their analyses they used observations obtained with the IRAM \mbox{30-m} telescope for the HCN, HNC, \chem{HCO^+} line emission of 37 infrared luminous galaxies \citep{Baan2008} and additional 80 sources from the literature (see \citealt{Loenen2008} and references therein); all unresolved measurements. Then they compared the observational data with the predictions of PDR and XDR models \citep{Meijerink2007} with varying volume densities, ranging from $10^{4.5}$ to $10^{6.0}$~cm$^{-3}$.

A low $\chem{HCO^+}/\chem{HCN}$ ratio was proposed as the signature of an AGN. Since studies of galaxies hosting an AGN have found evidence for enhanced emission from HCN, relative to \chem{HCO^+} (e.g. \citealt{Kohno2001DenseMolecularGas,Imanishi2007,Davies2012}). Recently, however, this statement has been under discussion, for example, in \cite{Privon2020}. They investigated \textit{NuSTAR} hard $X$-ray emission together with literature \chem{HCO^+} and HCN observations and found no correlation between the $\chem{HCO^+}/\chem{HCN}$ ratio and the $X$-ray to IR luminosity ratio or the AGN luminosity. Thus, observing enhanced HCN relative to \chem{HCO^+} against a galaxy centre is not convincingly linked to currently observed AGN activity.

We now apply the PDR versus XDR framework by \cite{Baan2008} and \cite{Loenen2008} to our observed dense gas tracers towards the centre of \ngc\ and investigate \textit{Chandra} $X$-ray observations.

\begin{figure*}[ht!]
\centering
    \includegraphics[width=1.0 \linewidth]{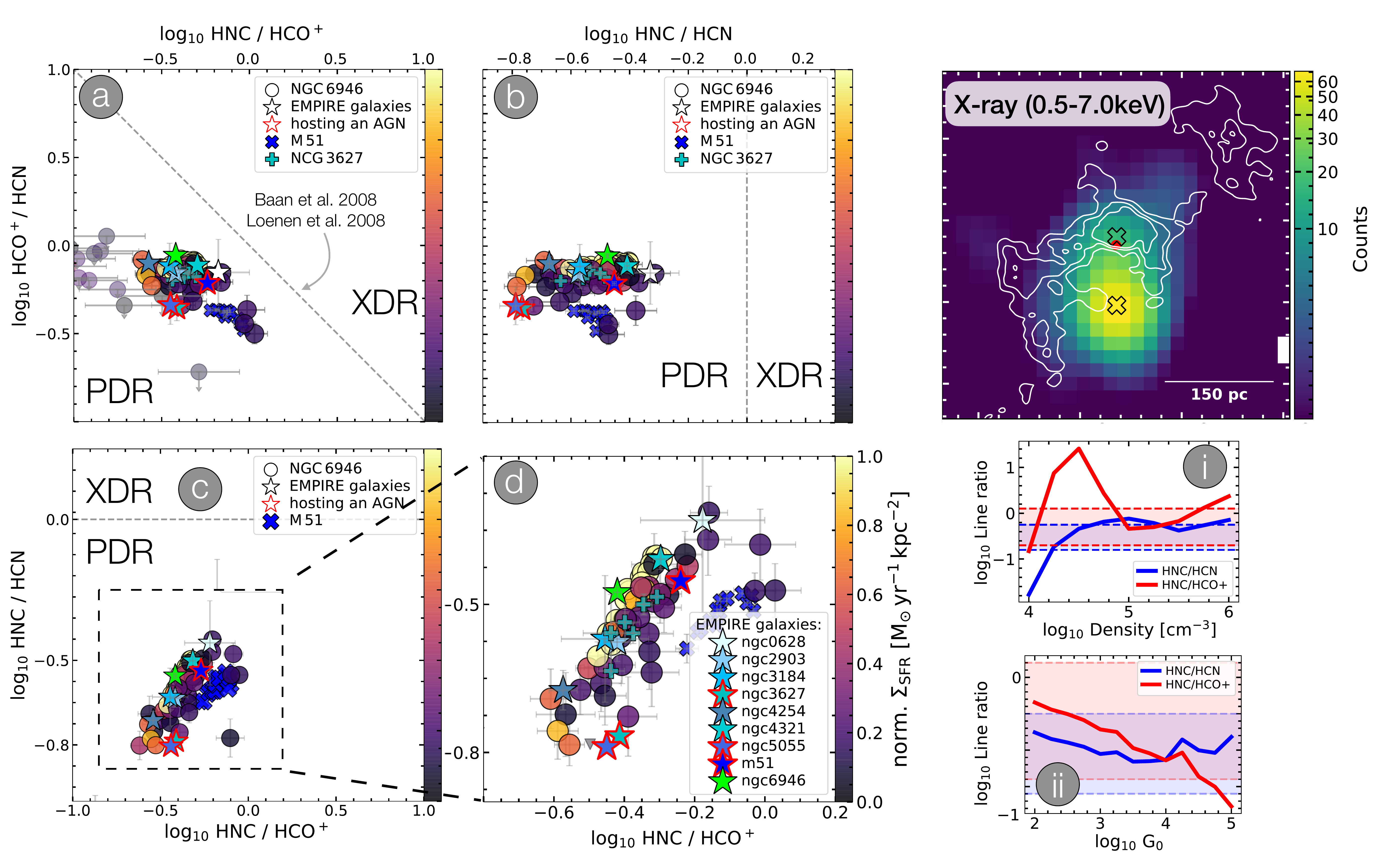}
   
    \caption{\textbf{Diagnostic plots using integrated line ratios of HCN, HNC and HCO$^+$ versus each other.} \textit{Left panels:} (a): Integrated $\chem{HCO^+}/\chem{HCN}$ versus $\chem{HNC}/\chem{HCO^+}$ ratios. (b): Integrated $\chem{HCO^+}/\chem{HCN}$ versus $\chem{HNC}/\chem{HCN}$ ratios. (c): Integrated $\chem{HNC}/\chem{HCN}$ versus $\chem{HNC}/\chem{HCO^+}$ ratios. The grey dashed lines marks the boarder between XDR and PDR \citep{Baan2008, Loenen2008}. Circles indicate reliable values ($\SN > 5$) and upper limits for the central ${20\arcsec \approx 745}$~pc towards NGC~6946. Stars show the EMPIRE galaxies and stars with red contours present galaxy centres hosting an AGN \citep{Goulding2009}. For two of them we find ancillary data: the cross shaped markers show the $\sim$4$\arcsec$ ($\approx166$~pc) M~51 observations (\citealt{Querejeta2016}) and the $\sim$2$\arcsec$ ($\approx100$~pc) NGC~3627 observations (\citealt{Beslic2021}); (d): The enclosed integrated $\chem{HNC}/\chem{HCN}$ versus $\chem{HNC}/\chem{HCO^+}$ with the kpc-sized centres of the EMPIRE survey (see \citealt{Jimenez-Donaire2019EMPIRE} for details). \ngc\ and the EMPIRE galaxies are in the PDR regime in all panels. Interestingly, EMPIRE galaxies harbouring an AGN do not show an enhancement of HCN as suggested in many studies (see text for a discussion). We run \citealt{Meijerink2005} PDR models for the observed ranges of line ratios (blue and red shaded areas) by (i) fixing the radiation field and (ii) fixing the density. The red and blue lines show the model output (more in text). The colours in (a--d) indicate \sigsfr. We have normalised them to the mean \sigsfr\ value and find no trend with the integrated line ratios. \textit{Upper right panel:} $X$-ray $0.5{-}7.0$~keV in counts for \ngc\ with integrated \chem{CO}{21} contours on top. The black crosses mark the positions of the detected $X$-ray sources by \citet{Holt2003} which are not strong enough to trigger a XDR.}
    \label{fig:XDR-PDR}
\end{figure*}

We see from the top panels of \autoref{fig:RatioMaps-uvtrimmed-SSC} that the $\chem{HCO^+}/\chem{HCN}$ and $\chem{HNC}/\chem{HCN}$ ratios exhibit values lower than unity in all three regions (see \autoref{tab:regions-charac}). In \autoref{fig:XDR-PDR} we investigate the diagnostic plots proposed by \cite{Baan2008} and \cite{Loenen2008} to visually discriminate between XDR and PDR by comparing the line ratios between \chem{HCO^+}, HCN and HNC in the central $20\arcsec\approx745$~pc towards \ngc. In all these diagnostic plots (panels a--d), \ngc\ is in the PDR regime. Panel~(d) could indicate a linear relationship between $\chem{HNC}/\chem{HCN}$ and $\chem{HNC}/\chem{HCO^+}$. We see $\log_{10}$($\chem{HNC}/\chem{HCN}$) in the range of $-0.80$ to $-0.25$ and $\log_{10}$($\chem{HNC}/\chem{HCO^+}$) between $-0.70$ and $+0.10$. We investigate which mechanism could cause the observed ranges of line ratios. For this purpose we run the \cite{Meijerink2005} models for the PDR case. In particular, we investigate two scenarios: (i) fixing the radiation field ($G_0 = 10^2$) with varying densities ranging of $n = 10^5 - 10^6$~cm$^{-3}$, and (ii) fixing the density ($n = 10^{5.5}$~cm$^{-3}$) with varying radiation field\footnote{$ G_0 = 10^2$ and $G_0 = 10^5$ are the default minimum and maximum $G_0$ in their model outputs.} of $ G_0 = 10^2 - 10^5$ (see panel (i) and (ii) in \autoref{fig:XDR-PDR}). For the first scenario, we find consistent ratios for n$~{\sim}10^{5.25} - 10^{5.75}$~cm$^{-3}$, but the predicted ratios span a narrow range (from $\log_{10}$($\chem{HNC}/\chem{HCO^+}$=$-0.25$ to $+0.10$) compared to the observations (red shaded areas). On the other hand,  $\log_{10}$($\chem{HNC}/\chem{HCN}$) remains rather constant. In the second case, $\log_{10}$($\chem{HNC}/\chem{HCO^+}$) decreases roughly linearly with $G_0$, from $-0.2$ to $-0.7$. This corresponds quite well to our observed ranges. From this we conclude that the scatter we observe in $\log_{10}$($\chem{HNC}/\chem{HCO^+}$) (panel~d) is mainly due to variations in the radiation field strength, with smaller ratios for stronger radiation fields. From all these diagnostic plots (panels a--d), it seems that the centre of \ngc\ is dominated by photons (PDR) rather than X-rays (XDR).

For \ngc\ -- as mentioned above -- there is no clear evidence for an AGN. In the following we investigate whether there are $X$-ray sources that are strong enough to trigger an XDR, although the diagnostic diagrams favour a PDR. The \textit{Chandra} $X$-ray map ($0.5{-}7.0$~keV) in counts towards the central $20\arcsec$ of \ngc\ shows us that most of the diffuse $X$-ray emission is coming from a region not associated with the NUC, SBE and NBE (see \autoref{fig:XDR-PDR}). The stronger of the two detected $X$-ray sources near NUC by \cite{Holt2003} (shown as black crosses) is ${\sim}2\arcsec$ away from the dynamical centre position (shown as red circle). They find a flux of $2.8\times10^{-13}$ erg~s$^{-1}$~cm$^{-2}$ and classified its hardness as `medium' (see their Table~2 for SourceID~45). Scaling this flux to our working resolution results in $4.53\times10^{-3}$ erg~s$^{-1}$~cm$^{-2}$. \footnote{Assuming the source is point-like, we scale the flux by (D/d)$^2$, where D is the distance to \ngc\ and d is the distance from the $X$-ray source, 2$\arcsec$.} 
This is similar to the lowest values in \cite{Meijerink2007}. This suggests that the brightest $X$-ray source might affect the gas within a beam scale, but hardly beyond that. 

Interestingly, all the kpc-sized central regions of the EMPIRE galaxies in \autoref{fig:XDR-PDR} lie in the PDR regime. Their $\log_{10}(\chem{HNC}/\chem{HCN})$ ratios are in the same range as those observed for \ngc. On the other hand, $\log_{10}(\chem{HNC}/\chem{HCO^+})$ varies only from $-0.60$ to $-0.20$. The galaxies NGC~3627, NGC~5055 and M51 are known to host an AGN classified as LINER \citep{Goulding2009} and are still located in the PDR regime in \autoref{fig:XDR-PDR}. There is no strong enhanced HCN emission relative to \chem{HCO^+} towards these three galaxy centres which would `move' them to the XDR regime of the diagnostic plots. The fact that they do not lie within the XDR region of \cite{Loenen2008} could be because: a) their models are not quantitatively accurate; b) the EMPIRE AGNs are faint and their effects are diluted when averaging over $1$~kpc regions; and c) a different model that does not require the significant variations in the line ratio to be driven by a PDR or XDR (e.g. \citealt{Viti2017}).
We test the dilution effects in a way that we include available high-resolution dense gas observations of M~51 and NGC~3627 (i.e. ${\sim}$4$\arcsec\approx166$~pc from \citealt{Querejeta2016} and ${\sim}$2$\arcsec\approx100$~pc from \citealt{Beslic2021}, respectively; see \autoref{sec:Ancillary data}). For both, we show line ratios for the central beam size; they are all in the PDR regime. From that we speculate that the diagnostic plots shown in \autoref{fig:XDR-PDR} and in particular the $\chem{HCO^+}/\chem{HCN}$ line ratio might not be a unique indicator to diagnose the presence of an AGN in galaxies on kpc/sub-kpc scales. This finding is consistent with the previous studies by \citet{Privon2020} and \citet{Li2021}.




\subsection{Density variations at the inner bar ends and nuclear region}
\label{Disc:lineratiopattern}

\begin{figure}[ht!]
    \centering
    \includegraphics[width=0.85 \linewidth]{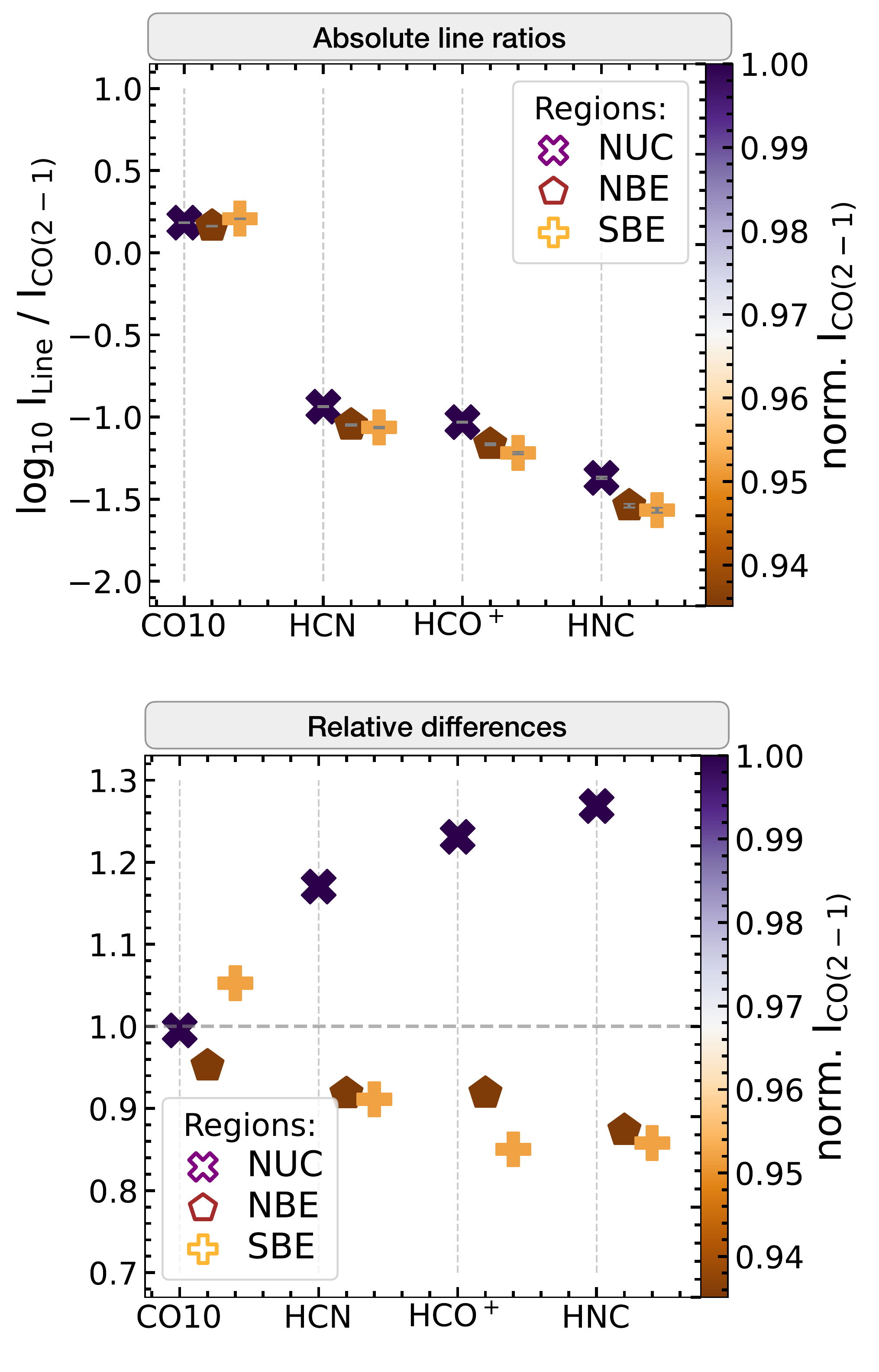}
    \caption{\textbf{Line ratios for the nuclear region and inner-bar ends.} \textit{Top}: Scatter plot showing the mean line ratios of the 7~hexagonal points for each molecule on the $y$-axis with respect to \chem{CO}{21}. For visualisation purpose we spaced the markers along the $x$-axis. The different shapes of the markers present the defined regions.  The colourbar shows the normalised \chem{CO}{21} integrated intensity to the mean \chem{CO}{21} value.  
    We plot the propagated errors in grey and find that they are very small. For the ratios we applied a signal-to-noise cut (see Section~\ref{Sec:Res-LineRatios}). \textit{Bottom}: The relative differences of the line ratios with respect to the mean of the line ratios among the different regions (linear-scale).}
    \label{fig:regions}
\end{figure}

Already in \autoref{fig:intensities} we saw that the \co\ and \cotwo\ emission is spatially more extended than \hcn, \hco\ or \hnc, which could be a sign that the CO lines trace a lower density regime. That said, line intensities depend not only on density, but also on optical depth, elemental abundance variations and IR pumping (see e.g. \citealt{Shirley2015, Barnes2020b}), and all of these effects can thus drive the relative line intensity ratios.

\cite{Leroy2017} analysed how changes in the sub-beam density distributions affect the beam-averaged line emissivity, by applying non-LTE radiative transfer models coupled with a parametrised density probability distribution. They found that the strength of tracer emission for dense gas is more sensitive to changes in gas density than for example CO. More precisely, the line can still be emitted at densities below the critical density, but with lower efficiency. As a result, a small increase in gas density can significantly increase the efficiency of the emission. This is not the case for lines with a lower critical density (e.g. bulk molecular gas tracer -- CO). The density of the gas exceeds the critical density, so varying the gas density does not significantly affect the efficiency of the emission.

We investigate the scenario from \cite{Leroy2017} that line ratios can reflect changes in density distributions. In Figure~\ref{fig:regions} we investigate the dependence of the observed molecular line ratios. The colourbar shows the integrated intensity of CO (an indicator of the volume density at cloud scales \citealt{Leroy2016,Sun2018}) towards the 150~pc sized NUC, NBE and SBE (see \autoref{fig:SFR-and-mask}; we take the mean over the 7 hexagonal points). The upper panel shows on the $y$-axis logarithmic molecular line ratios with \cotwo. At first glance, within NUC we find line ratios being enhanced by at least $20\%$~compared to the inner bar ends. 
Identifying ratio variations by eye in the two bar ends is challenging. The lower panel shows the relative differences. Here we divide for example $\chem{HCN}/\chem{CO}$ by the mean $\chem{HCN}/\chem{CO}$ of all the 3 regions. We ordered them by their flaring appearance. We see the trend is not monotonic with $n_\mathrm{eff}$, instead they show an order of $\chem{HCN} < \chem{HCO^+} < \chem{HNC}$; that is the order of decreasing intensity. We find that the highest ratios are associated with NUC (purple marker), except for the $\chem{CO}{10}/\chem{CO}{21}$ ratio. Comparing the two bar ends we find (i) higher ratio values towards the SBE (orange marker) for ratios including \co\ and \chem{HCN}, (ii) higher ratio values towards the NBE (brown markers) for ratios including \chem{HCO^+} and \chem{HNC} and, (iii) that the highest differences between the SBE and NBE is ${\sim}10\%$ which are ratios including \co\ and \chem{HCO^+}. The \chem{HCO^+} intensity in the SBE seems to be under-luminous compared to the other two environments. The \chem{HNC} intensity provides the largest dynamic range, with the highest value seen in the NUC and the lowest one in the SBE.

We compare the observed line ratios of CO, HCN, \chem{HCO^+} and HNC in the three regions using radiative transfer models to estimate the mass-weighted mean gas density. From the temperature analyses (see \autoref{Disc:HNC/HCN}) we would expect potential higher densities in the bar ends, whereas from the $R_{21}$ examinations (see \autoref{sec:R21}) we would anticipate higher molecular densities in the NBE than in the SBE. Furthermore, from our derived \sigsfr\ (see \autoref{sec:3-SFRindicators-compare} and \autoref{tab:regions-charac}) we would expect higher molecular gas densities in the SBE than in the NBE. For that purpose we use the radiative transfer code \texttt{Dense Gas Toolbox} \citep{Puschnig2020} which is based on the approach by \cite{Leroy2017} including that emission lines emerge from an isothermal ensemble of gas densities that follow a log-normal distribution (with or without power-law tail) in combination with \texttt{RADEX} calculations \citep{vanderTak2007}. It works as follows: (i) Using Bayesian inference, model parameters (i.e. temperature and density) are inferred from a number of integrated input line intensities. (ii) The assumed fixed line optical depths and abundances are calibrated through observations of the EMPIRE survey (\citealt{Jimenez-Donaire2019EMPIRE}, which includes \ngc). (iii) Then it solves for line emissivities in each density bin which are calculated using expanding-sphere escape probabilities (large velocity gradient approximation) as implemented in \texttt{RADEX} \citep{vanderTak2007}. (iv) Given the estimated kinetic temperature in \autoref{Disc:HNC/HCN}, we are assuming a fixed temperature of $30$~K. A more detailed description of the \texttt{Dense Gas Toolbox} will be provided in J.~Puschnig et al. (in prep.). The models suggest that the density in NUC is highest with a mass-weighted mean density of ${\sim}10^{4.0}$~cm$^{-3}$, while the lowest one is found in the NBE with a value of ${\sim}10^{3.7}$~cm$^{-3}$. 

In summary, these models broadly agree with our empirical findings of \autoref{fig:regions}. The higher mass-weighted mean densities in the SBE agree with the enhanced \sigsfr\ in the SBE (compared to the NBE). However, the model results do not explain the elevated $R_{21}$ in NBE (see \autoref{fig:R21}) and, therefore, an additional physical mechanism has to be responsible for it. A note of caution is appropriate, as the mass-weighted mean density differences found among the three regions are of small magnitude and might not be significant in conjunction with the model assumptions (see \autoref{tab:regions-charac}). For example, the kinetic temperature in the model is fixed and the same for the two bar ends and the nuclear region. An accurate measurement of kinetic temperatures on parsec scales could unveil the duality of the bar ends.   

\begin{table*}
\centering
\caption{Summary of the characteristics of the nuclear region and inner bar ends of NGC\,6946 based on the analyses in this work.}
\label{tab:regions-charac}
\begin{tabular}{lcccc|cc} \hline \hline
                  & Unit & NUC &  NBE & SBE & Section & Notes\\ \hline
I$_{\mathrm{CO(1-0)}}$ $\propto$ bulk mol. gas  & [K km s$^{-1}$]& 892.75$\pm$2.43& 565.91$\pm$2.23& 705.35$\pm$2.25& \S\ref{sec:Results} & \\
I$_{\mathrm{ CO(2-1)}}$ $\propto$ bulk mol. gas & [K km s$^{-1}$]& 587.38$\pm$1.51& 388.65$\pm$0.86& 438.52$\pm$1.68& \S\ref{sec:Results}& \\
I$_{\mathrm{ HCN}}$ $\propto$ dense gas         & [K km s$^{-1}$]& 67.96$\pm$0.39 &35.19$\pm$0.37 &40.12$\pm$0.36 &\S\ref{sec:Results} &\\
I$_{\mathrm{ CH_3OH}}$ $\propto$ shock          & [K km s$^{-1}$]& 7.15$\pm$0.45& 5.17$\pm$0.40& 11.53$\pm$0.35& \S\ref{sec:Results}& (1) \\
I$_{\mathrm{ CH_3OH}}$/I$_{\mathrm{CO}}$ $\propto$ f$_{\mathrm{shocked}}$ & [$\%$]& 2.43$\pm$0.24 & 2.98$\pm$0.34 & 8.85$\pm$0.50 & \S\ref{sec:Results}& (1) \\
SFR                 & [\sfrunit]        & 8.9E-2$\pm$1.33E-3   & 6.0E-3$\pm$1.33E-3   & 1.3E-3$\pm$1.33E-3 &\S\ref{sec:3-SFRindicators-compare}& \\
\sigmolmass\        & [\sigmolmassunit] & 298.07$\pm$1.98 & 188.95$\pm$0.26 & 235.50$\pm$5.09 &\S\ref{sec:3-SFRindicators-compare}&\\
\sigmolmass/\sigsfr\ $\propto$ \tdepl\  &  [yr$^{-1}$] & 3.43E+7 & 1.46E+8 & 3.05E+8 & \S\ref{sec:3-SFRindicators-compare}& \\
I$_{\mathrm{HCN}}$/I$_{\mathrm{CO}}$ $\propto$ f$_{\mathrm{dense}}$ & [$\%$]& 11.56$\pm$0.11 & 9.07$\pm$0.15 & 9.00$\pm$0.14 & \S\ref{sec:comp-moleculesandSFR} &\\
I$_{\mathrm{HNC}}$/I$_{\mathrm{CO}}$ $\propto$ f$_{\mathrm{dense}}$ & [$\%$]& 4.27$\pm$0.11  & 2.94$\pm$0.11 & 2.89$\pm$0.12 & \S\ref{sec:comp-moleculesandSFR}& \\
\sigsfr/I$_{\mathrm{HCN}}$ $\propto$ SFE$_{\mathrm{dense}}$ & & 0.13$\pm$3E-3 & 0.05$\pm$7E-3 & 0.02$\pm$7E-3 & \S\ref{sec:comp-moleculesandSFR}& (2) \\
I$_{\mathrm{CO(2-1)}}$/I$_{\mathrm{CO(1-0)}}$ $\propto$ R$_{21}$ &    & 0.65$\pm$2E-3 & 0.70$\pm$3E-3 & 0.62$\pm$3E-3 & \S\ref{sec:R21}&\\
I$_{\mathrm{HNC}}$/I$_{\mathrm{HCN}}$    &            &0.37$\pm$0.01   &  0.32$\pm$0.01  & 0.31$\pm$0.01  & \S\ref{Disc:HNC/HCN} &\\
T$_k$                          & [K]        & 27.19$\pm$0.06 & 31.19$\pm$0.09  & 31.99$\pm$0.11 & \S\ref{Disc:HNC/HCN}& (3) \\
I$_{\mathrm{HCO^+}}$/I$_{\mathrm{HCN}}$  &            &  0.81$\pm$0.01 & 0.76$\pm$0.01   & 0.70$\pm$0.01  & \S\ref{Disc:HCO/HCN}&\\
n$_{\mathrm{H2}}$                   &[cm$^{-3}$] & 10$^{3.7-4.1}$ & 10$^{3.3-4.1}$  & 10$^{3.4-4.2}$ & \S\ref{Disc:lineratiopattern}&\\

\hline \hline 
\end{tabular}
\begin{minipage}{1.8\columnwidth}
    \vspace{1mm}
    {\bf Notes:} We quote the mean values over each region that are equivalent to our beam size of $4\arcsec \approx 150$~pc. \\ NUC = Nuclear Region; NBE = northern inner bar end; SBE = southern inner bar end. \\ (1): Taken from our extended molecular data set -- \textit{PdBI only}. CH$_3$OH is the (2k-1k) transition and CO the (2-1) transition. \\ (2): In units of K km s$^{-1}$/(\sigsfrunit) \\ (3): If the \cite{Hacar2020} framework holds valid in \ngc; see detailed discussion in Section \ref{Disc:HNC/HCN}.
\end{minipage}
\end{table*}



\section{Conclusion}
\label{sec:Conclusions}

In this paper we present the high-resolution ($2{-}4\arcsec \approx 75{-}150$~pc at $7.7$~Mpc distance) IRAM PdBI multi-molecule observation towards the inner $50\arcsec
~\approx1.9$~kpc of the Fireworks Galaxy, \ngc\ 
(see \autoref{fig:color}). Our compiled data set includes a total of 14 detected molecular lines in the mm-wavelength range.
For the analyses in this paper, we convolve our molecular lines to a common beam size of $4\arcsec$, and analyse three distinct environmental regions associated with the inner bar of \ngc: the nuclear region (NUC), the northern inner bar end (NBE) and the southern inner bar end (SBE) (see \autoref{fig:SFR-and-mask} for mask and \autoref{tab:regions-charac}). We find the following results:
\begin{enumerate}
   \item[1)] We report the first detection of \cch, \hcnten\ and \hcnsixt\ towards \ngc.
   
   \item[2)] The different lines show distinct morphology. More extended emission is seen with \co, \cotwo\ (associated with bulk molecular gas), \hco, \hcn\ and \hnc. The latter three are typically used to trace denser gas. We find that even denser gas (e.g. as seen by \ntwoh) and shocks (e.g. as seen by \chohtwo) are more concentrated in the SBE than in the NUC or NBE (see \autoref{fig:intensities}).
   
   \item[3)] We discover a higher star formation rate (SFR) in the SBE. Shocks and denser gas in this bar end could possibly enhance star formation compared to the lower SFR in NBE.
\end{enumerate}
We use our \uv trimmed and short-spacing corrected data set (\textit{SSC + \uv trim data}) which includes \co, \cotwo, \hcn, \hco\ and \hnc\ to investigate how they relate to \sigsfr\ and how the dense gas fraction (\fdense) and the star formation efficiency (\sfedense) respond to the integrated intensity of \cotwo\ (an indicator of mean volume density).
\begin{enumerate}
    \item[4)] We find among our molecules that \hco\ correlates best with \sigsfr, with Pearson's correlation coefficient $\rho~{\sim}0.95$. Additionally, we find a relation of $\rho$ to the effective critical density ($n_\mathrm{eff}$) of the dense gas tracers.
    
\item[5)] The dense gas fraction fits do not follow the order of $n_\mathrm{eff}$, instead they show the order of $\beta_\mathrm{HNC}$ > $\beta_\mathrm{HCO^+}$ > $\beta_\mathrm{HCN}$. The strongest correlation among our mean volume density versus different \fdense\ is I$_{\mathrm{CO(2-1)}}$ versus $\chem{HNC}/\chem{CO}{21}$ with $\rho~{\sim}0.76$.
    
    \item[6)] We find that \sfedense\ increases with increasing gas surface density (i.e. $I_{\chem{CO}{21}}$ emission) within the inner $\sim$\,1\,kpc of NGC\,6946, contrary to the opposite trends found within low-resolution observations covering the whole disc \citep{gallagher2018dense, Jimenez-Donaire2019EMPIRE}. The reason for this is that we are resolving structures within this region, such as the nuclear region (NUC; inner $\sim$\,0.15\,kpc) that appears to exhibit \sfedense\ enhancements relative to the rest of the centre. 
\end{enumerate}
We test line ratio diagnostic plots and compare our high-resolution ($150$~pc) dense gas tracer towards \ngc\ with those of eight other galaxy centres in the EMPIRE survey (angular resolution of 33$\arcsec\sim$1~kpc) and discuss implications for other galaxy centre studies.
\begin{enumerate}    
    \item[7)] From previous studies higher \co\ to \cotwo\ ($R_{21}$) values are expected in the inner kiloparsec compared to their disc. We find $R_{21}$ variations in the centre; higher $R_{21}$ in NBE (${\sim}0.7$) than in the NUC or SBE (${\sim}0.65$ and ${\sim}0.62$). The reason for this could be, for example, denser gas that is in the NBE. However, we do not find an increase of higher \fdense\ in NBE (see \autoref{fig:R21} and \autoref{tab:regions-charac}).
    
    \item[8)] Whether the ratio between HCN and HNC is temperature dependent has been discussed by several authors in the past. We speculate that the isomer ratio is not an accurate probe for kinetic temperatures for kiloparsec and sub-kiloparsec sized extragalactic regions. This is because, if the prescription of \cite{Hacar2020} can be applied here, the $\chem{HCN}/\chem{HNC}$ ratio results in low $T_\mathrm{k}$ (${<}50$~K, for \ngc\ as well as for six EMPIRE galaxies, IC\,342 and NGC\,253) which is up to a factor of~$7$ lower compared to other existing $T_\mathrm{k}$ measurements (\chem{NH_3} or \chem{H_2CO}). 
    
    \item[9)] The observed $\chem{HCO^+}/\chem{HCN}$ ratios have been proposed to be a diagnostic tool to distinguish between XDR and PDR. The diagnostic plots for the centre of \ngc\ favour a PDR. We find that an $X$-ray source about $2\arcsec$ away from the dynamical centre does not affect the molecular gas properties in the nucleus. However, when we include the EMPIRE galaxy centres in our analysis, we find that those galaxies that have an active galactic nucleus (AGN) fall within the PDR range in these diagnostic diagrams. We find for two of these AGN host galaxies higher resolution observations (${\sim}100$pc for NGC~3627 and ${\sim}166$pc for M51) which also lay in the PDR regime. Thus, the $\chem{HCO^+}/\chem{HCN}$ ratio might not be a  unique indicator to diagnose AGN activity in galaxies at (sub-)kiloparsec scales. This ambiguity is consistent with recent studies (see \autoref{fig:XDR-PDR} and \autoref{Disc:HCO/HCN}).
    
    \item[10)] Molecular line ratios may reflect changes in the (unresolved) gas density distributions \citep{Leroy2017}. The ratios to CO, show enhanced line ratios within the NUC compared to the inner bar ends. \hco\ in NBE seems to be under-luminous compared to the NUC and SBE. On the other hand, HNC provides the largest dynamic range, with the highest values in the NUC and the lowest in the SBE (see \autoref{fig:regions}). We compared that with radiative transfer models and find mass-weight mean densities of ${\sim}10^{3.7-4.3}$~cm$^{-3}$ for the NUC, while the NBE shows the lowest mean density of ${\sim}10^{3.3-4.1}$~cm$^{-3}$. The higher mean densities in the SBE compared to the NBE agree with the higher integrated line intensities and the enhanced SFR in the SBE.
\end{enumerate}
This study reflects the importance of analysing molecular lines to better understand galactic centres. Also, it shows that bar ends in galaxies can vary in their dense gas fraction, star formation rate, integrated line ratios and molecular gas densities. In future work we follow up on the investigation of shocks, the densest gas tracers in our extended data set and examine the kinematic and dynamic of the inner bar in more detail.

\begin{acknowledgements}
     We would like to thank the anonymous referee for their insightful comments that helped improve the quality of the paper.
     CE gratefully acknowledges funding from the Deutsche Forschungsgemeinschaft (DFG) Sachbeihilfe, grant number BI1546/3-1.
      FB, AB, IB, JP and JdB acknowledge funding from the European Research Council (ERC) under the European Union’s Horizon 2020 research and innovation programme (grant agreement No.726384/Empire).
      ES, DL, HAP, TS and TGW acknowledge funding from the European Research Council (ERC) under the European Union’s Horizon 2020 research and innovation programme (grant agreement No. 694343).
      IL acknowledges funding from the Deutsche Forschungsgemeinschaft (DFG) Sachbeihilfe, grant number SCHI 536/11-1.
      MC and JMDK gratefully acknowledge funding from the Deutsche Forschungsgemeinschaft (DFG) in the form of an Emmy Noether Research Group (grant number KR4801/1-1) and the DFG Sachbeihilfe (grant number KR4801/2-1), and from the European Research Council (ERC) under the European Union’s Horizon 2020 research and innovation programme via the ERC Starting Grant MUSTANG (grant agreement number 714907).
      HH acknowledges the support of the Natural Sciences and Engineering Research Council of Canada (NSERC), funding reference number RGPIN-2017-03987 and the Canadian Space Agency funding reference 21EXPUVI3.
      SCOG and RSK acknowledge financial support from the German Research Foundation (DFG) via the collaborative research center (SFB 881, Project-ID 138713538) `The Milky Way System' (subprojects A1, B1, B2, and B8). They also acknowledge funding from the Heidelberg Cluster of Excellence `STRUCTURES' in the framework of Germany’s Excellence Strategy (grant EXC-2181/1, Project-ID 390900948) and from the European Research Council via the ERC Synergy Grant `ECOGAL' (grant 855130). 
      The work of AKL is partially supported by the National Science Foundation under Grants No. 1615105, 1615109, and 1653300.
      JP acknowledges support from the Programme National “Physique et Chimie du Milieu Interstellaire” (PCMI) of CNRS/INSU with INC/INP co-funded by CEA and CNES.
      MQ acknowledges support from the Spanish grant PID2019-106027GA-C44, funded by MCIN/AEI/10.13039/501100011033.
      ER acknowledges the support of the Natural Sciences and Engineering Research Council of Canada (NSERC), funding reference number RGPIN-2017-03987.
      TS acknowledges funding from the European Research Council (ERC) under the European Union’s Horizon 2020 research and innovation programme (grant agreement No. 694343).
      MCS acknowledges financial support from the German Research Foundation (DFG) via the collaborative research center (SFB 881, Project-ID 138713538) ”The Milky Way System” (subprojects A1, B1, B2, and B8). 
      AU acknowledges support from the Spanish grants PGC2018-094671-B-I00, funded by MCIN/AEI/10.13039/501100011033 and by "ERDF A way of making Europe", and PID2019-108765GB-I00, funded by MCIN/AEI/10.13039/501100011033. 
      Y-HT acknowledges funding support from NRAO Student Observing Support Grant SOSPADA-012 and from the National Science Foundation (NSF) under grant No. 2108081.

\end{acknowledgements}

\bibliographystyle{aa}
\bibliography{references.bib} 


\appendix

\section{Star formation rates}
\label{appendix:sfr}
In this work, we used the $33$~GHz continuum emission as a SFR tracer. We also found published $15$~GHz and H$\alpha$ observations in comparable angular resolution of our molecular data set ($2~\arcsec$; see Table~\ref{tab:fig1-ref}). We chose the $33$~GHz over the $15$~GHz as a SFR tracer because the $33$~GHz observations are: (1) higher in sensitivity  ($9.72$ compared to $21.44$~mK, \citealt{Linden2020SFRS}), and (2) probe less non-thermal emission (38$\%$ compared to 63$\%$, see Equations below). We did not use extinction corrected  H$\alpha$ data (from \citealt{Kessler2020}) for our main SFR in this paper because: (1) they used 2$\arcsec$ sized apertures across the whole galaxy covering the bright H$\alpha$ emitting regions (1$\sigma$ threshold in H$\alpha$) for the extinction correction (using Pa$\beta$) and those apertures do not fully cover the inner kpc of \ngc, and (2) causes them to miss a handful of heavily attenuated regions in the centre of \ngc\ (see Fig. 2 in \citealt{Kessler2020}).

In the following we check the consistency of our SFR derived from the $33$~GHz thermal part of the radio continuum emission with the thermal part of the $15$~GHz as a \sigsfr\ tracer (observations from \citealt{Murphy2018SFRS} and \citealt{Linden2020SFRS}) for the inner kpc-sized region of \ngc. 

\begin{enumerate}
    \item \textit{33~GHz continuum:} Using 33~GHz for the inner 1~kpc gives us a mean \sigsfr\ of 0.792 \sigsfrunit. 
    \item \textit{15~GHz continuum:} Before we can use Eq.~\eqref{eq:SFR}, we first have to compute the thermal fraction ($f^\mathrm{T}$) at $15$~GHz. We recall that we denote thermal as $^\mathrm{T}$, non-thermal as $^\mathrm{NT}$ and the frequency as $\nu$. Knowing that $S^\mathrm{T}_{\nu} \propto \nu^{-\alpha^\mathrm{T}}$ with $\alpha^\mathrm{T} \sim 0.1$ and $S^{\rm NT}_{\nu} \propto \nu^{-\alpha^{\rm NT}}$ with $\alpha^\mathrm{NT} \sim 0.74$ (from \citealt{Murphy2011} for the nucleus of \ngc), we find:
\begin{equation}
    f^\mathrm{T}_{\mathrm{15\,GHz}} = \left( \frac{f^\mathrm{T}_{\rm 33\,GHz}} {\frac{S_{\rm 33 GHz}^\mathrm{T}}{S_{\rm 33 GHz}^{\rm NT}}}\right) \times \left(\frac{S^\mathrm{T}_{\rm 15 GHz}}{S_{\rm 15 GHz}^{\rm NT}} \right)~.
    \label{app:eq}
\end{equation}
We take as in Section \ref{sec:2-SFRCalibration} $f^\mathrm{T}_{33~\mathrm{GHz}} = 0.62$ and find using Eq.~\eqref{app:eq} that $f^\mathrm{T}_{15~\mathrm{GHz}} = 0.37$. Now we are able to calculate $L_{\nu^\mathrm{T}}$ (Eq.~\eqref{eq:lumthermal}) and use Eq.~\eqref{eq:SFR} to convert the thermal fluxes into a SFR and further to \sigsfr\ . Within the central kpc-sized region we get a mean \sigsfr\ of $0.774$~\sigsfrunit. \\
\end{enumerate}
We conclude, that the \sigsfr\ for the inner kpc derived from $33$~GHz and $15$~GHz are within the margin of error (${\sim}1\sigma$).

\subsection{Correcting for distance}
\label{app:SFRdistance}

\cite{Schinnerer2007} found within a $3\arcsec \times 3\arcsec$ region a SFR of ${\sim}0.1$ \sfrunit\ adopting a distance of $5.5$~Mpc. Using $33$~GHz as a SFR tracer and taking the same distance and region, we get $0.06$ \sfrunit. Adopting the updated distance to \ngc\ of $7.72$~Mpc results in $0.11$ \sfrunit. This is only a factor of $1.8$ higher.

\begin{table*}
\caption{References of the data shown in Fig~\ref{fig:color}.}
\label{tab:fig1-ref}
\resizebox{\textwidth}{!}{%
\begin{tabular}{ccccl}
\hline\hline
Panel in Fig\ref{fig:color}&$\lambda$, L & Instrument & Res. & Survey, Reference \\ \hline 
(a,\,b) & optical      &  \textit{HST}, Subaru & & NASA, ESA, STScI, R. Gendler and the Subaru Telescope (NAOJ) \\
(c) & 115\,GHz     &  PdBI & 1.4$\arcsec$& \cite{schinnerer2006molecular} \\
(d) & 230\,GHz     &  PdBI & 0.4$\arcsec$ & \cite{schinnerer2006molecular} \\

(e) & 3\,GHz      & VLA & 2.0$\arcsec$& SFRS, \cite{Murphy2018SFRS,Linden2020SFRS} \\
(f) & 15\,GHz      & VLA & 2.1$\arcsec$ & SFRS, \cite{Murphy2018SFRS,Linden2020SFRS} \\
(g) & 33\,GHz      & VLA & 2.1$\arcsec$& SFRS, \cite{Murphy2018SFRS,Linden2020SFRS}  \\
(h) & 70\,$\mu$m      & \textit{Herschel}/PACS & 5.5$\arcsec$ & KINGFISH, \cite{Kennicutt2011} \\
(i) & 24\,$\mu$m      & \textit{Spitzer}/MIPS  & 5.7$\arcsec$ & SINGS, \cite{Kennicutt2003}  \\
(j) & 8\,$\mu$m       & \textit{Spitzer}/IRAC  & 2.0$\arcsec$ & SINGS, \cite{Kennicutt2003} \\
(k) & Pa$\beta$   & \textit{HST}           & 2.0$\arcsec$  &   \cite{Kessler2020}   \\ 
(l) & H$\alpha$   & 3.5-m WIYN telescope           & 2.0$\arcsec$  &   \cite{Long2019,Kessler2020}   \\ 
(m) & $X$-ray   & \textit{Chandra}            &   &    ObsIDs 1054 and 13435, \cite{Holt2003}  \\ 
 \hline \hline
\end{tabular}
}
\begin{minipage}{1.8\columnwidth}
    \vspace{1mm}
    {\bf Notes:} To overlay the contours of CO data on the optical images, we used \url{http://nova.astrometry.net/}
\end{minipage}
\end{table*}
\begin{figure*}
    \centering
    \includegraphics[width=1.0\textwidth]{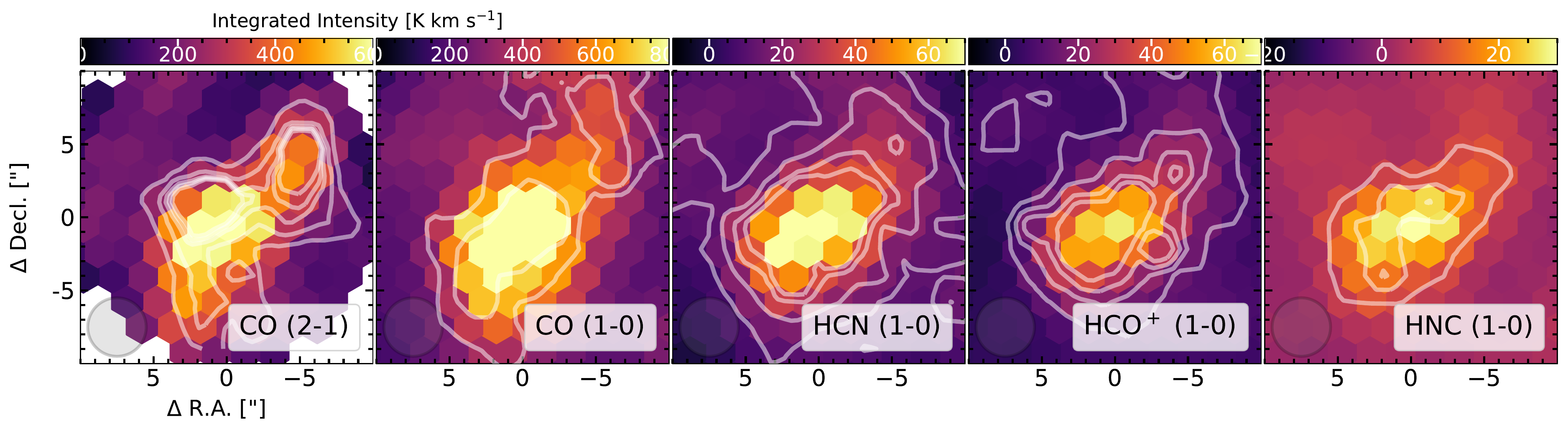}
    \caption{\textbf{SSC + uv\,trim integrated intensity maps:} The way of presenting the data is done as in Figure~\ref{fig:intensities}. Here, however, we performed a \uv cut and corrected for the short-spacing using available EMPIRE data for the three typical dense gas tracers. In the first two panels we show contours of $200, 300\sigma$ and in the remaining three panels contours of $3, 6, 9, 30, 60, 90\sigma$.}
    \label{fig:intensities-uvtrimmed-SSC}
\end{figure*}
\begin{figure*}
    \centering
    \includegraphics[width=\textwidth]{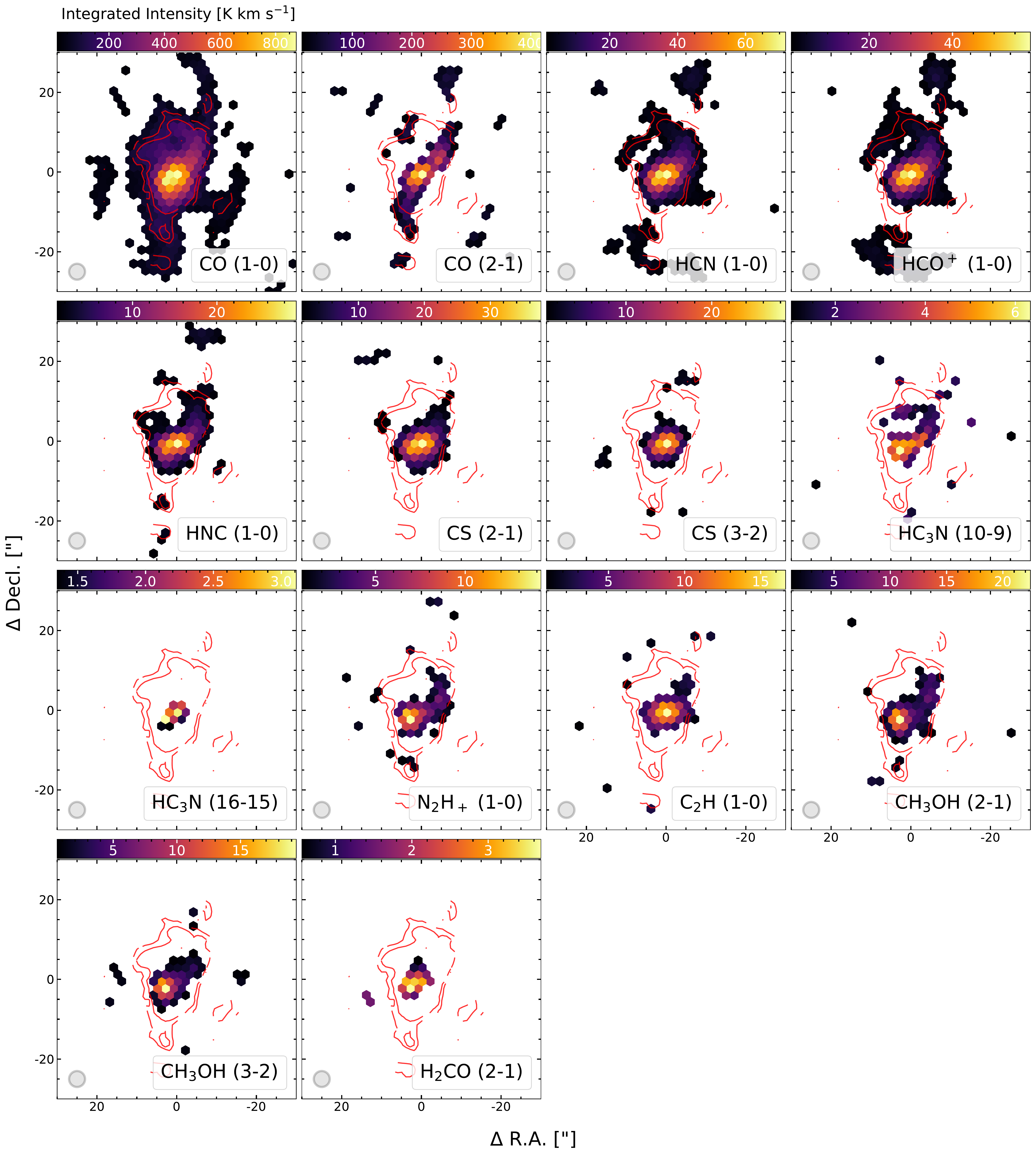}
    \caption{Same as Figure~\ref{fig:intensities} but only showing hexagonal points with $\SN > 25$ for \co\ and \cotwo, and with $\SN > 5$ for all other lines.}
    \label{fig:sn5}
\end{figure*}

\begin{figure*}
    \centering
    \includegraphics[width=1.0\textwidth]{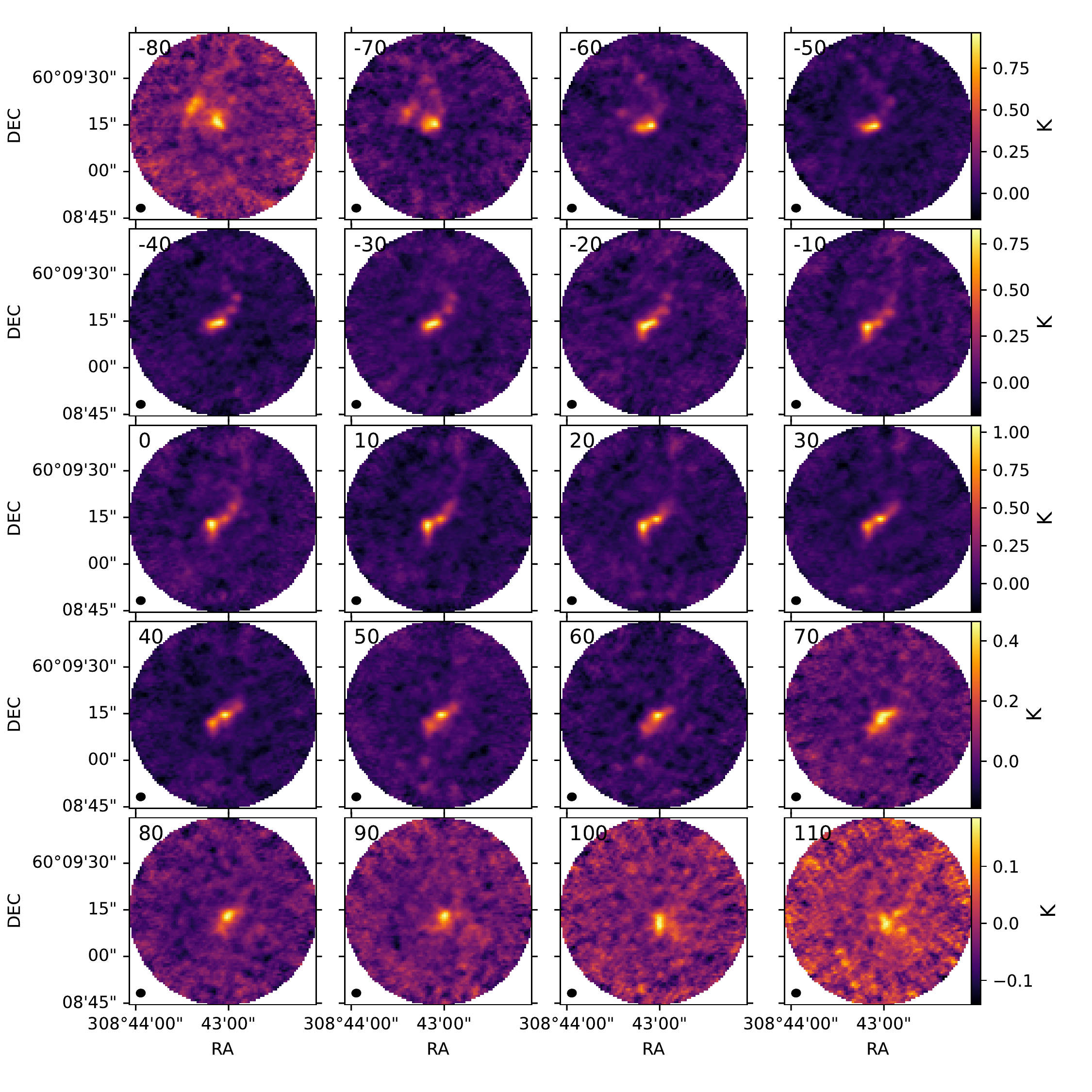}
    \caption{HCN channel maps, from $-80$ to $110$ km\,s$^{-1}$.}
    \label{fig:app-hcn_cm}
\end{figure*}
\begin{figure*}
    \centering
    \includegraphics[width=1.0\textwidth]{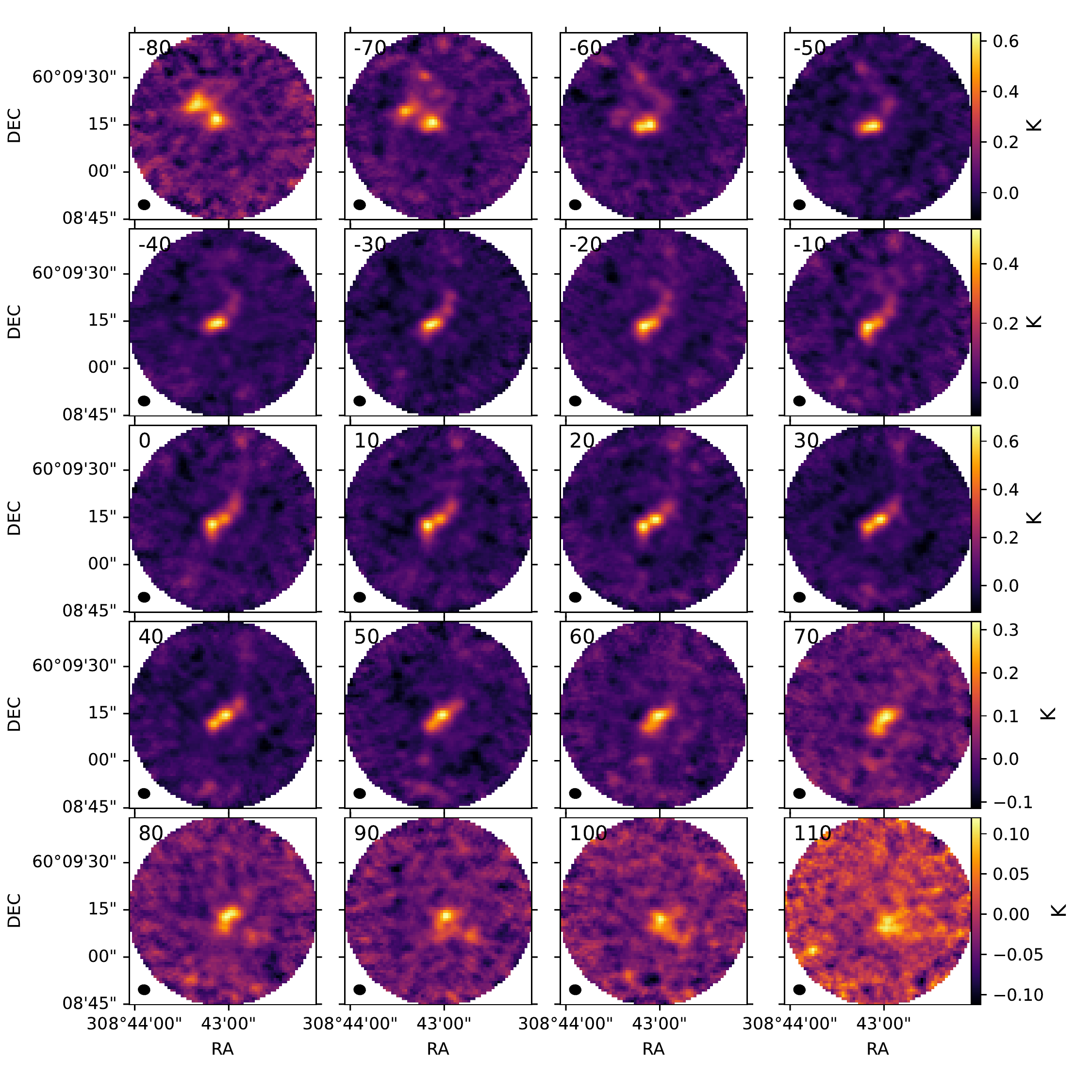}
    \caption{HCO$^+$ channel maps, from $-80$ to $110$ km\,s$^{-1}$.}
\end{figure*}
\begin{figure*}
    \centering
    \includegraphics[width=1.0\textwidth]{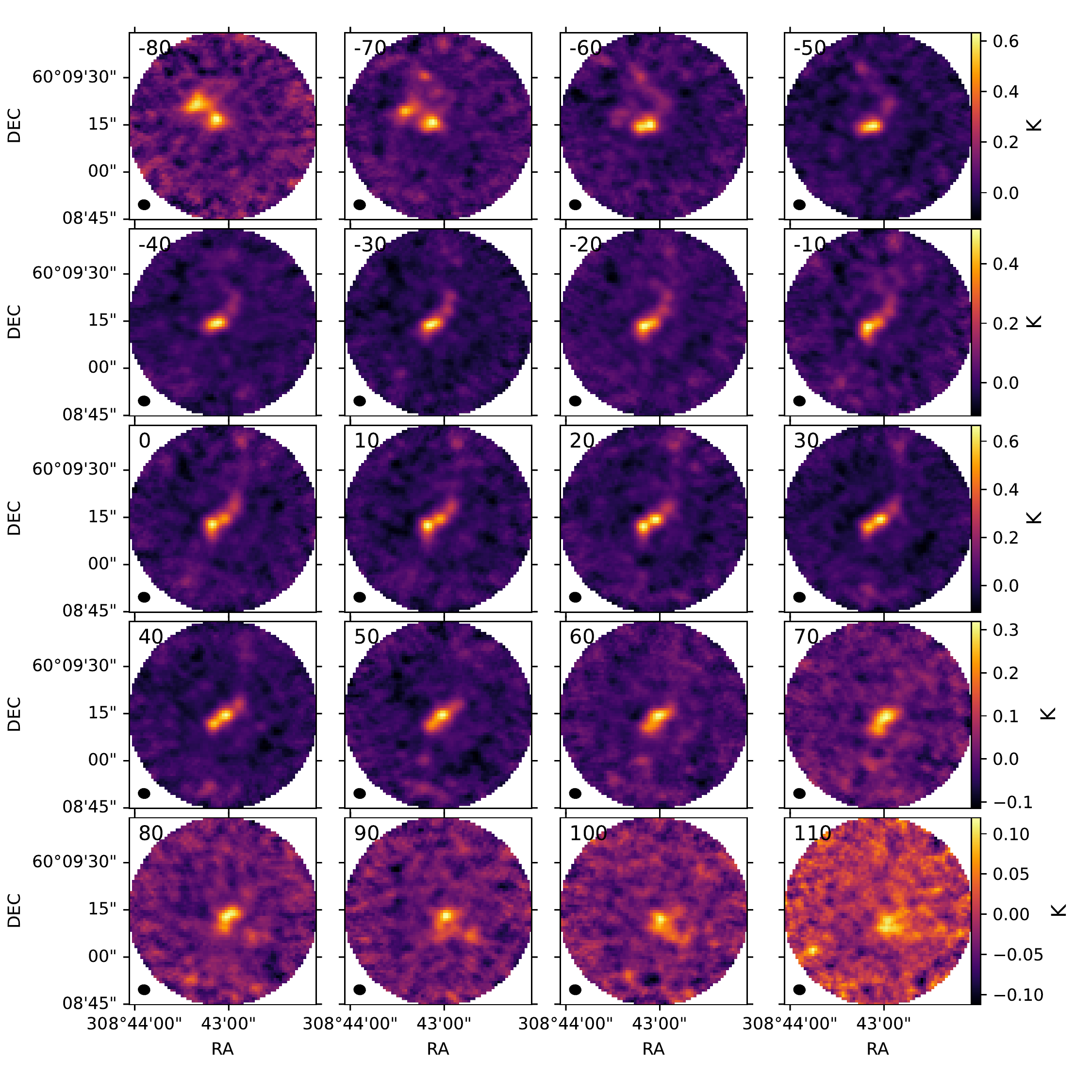}
    \caption{HNC channel maps, from -80 to 110 km s$^{-1}$.}
\end{figure*}
\begin{figure*}
    \centering
    \includegraphics[width=1.0\textwidth]{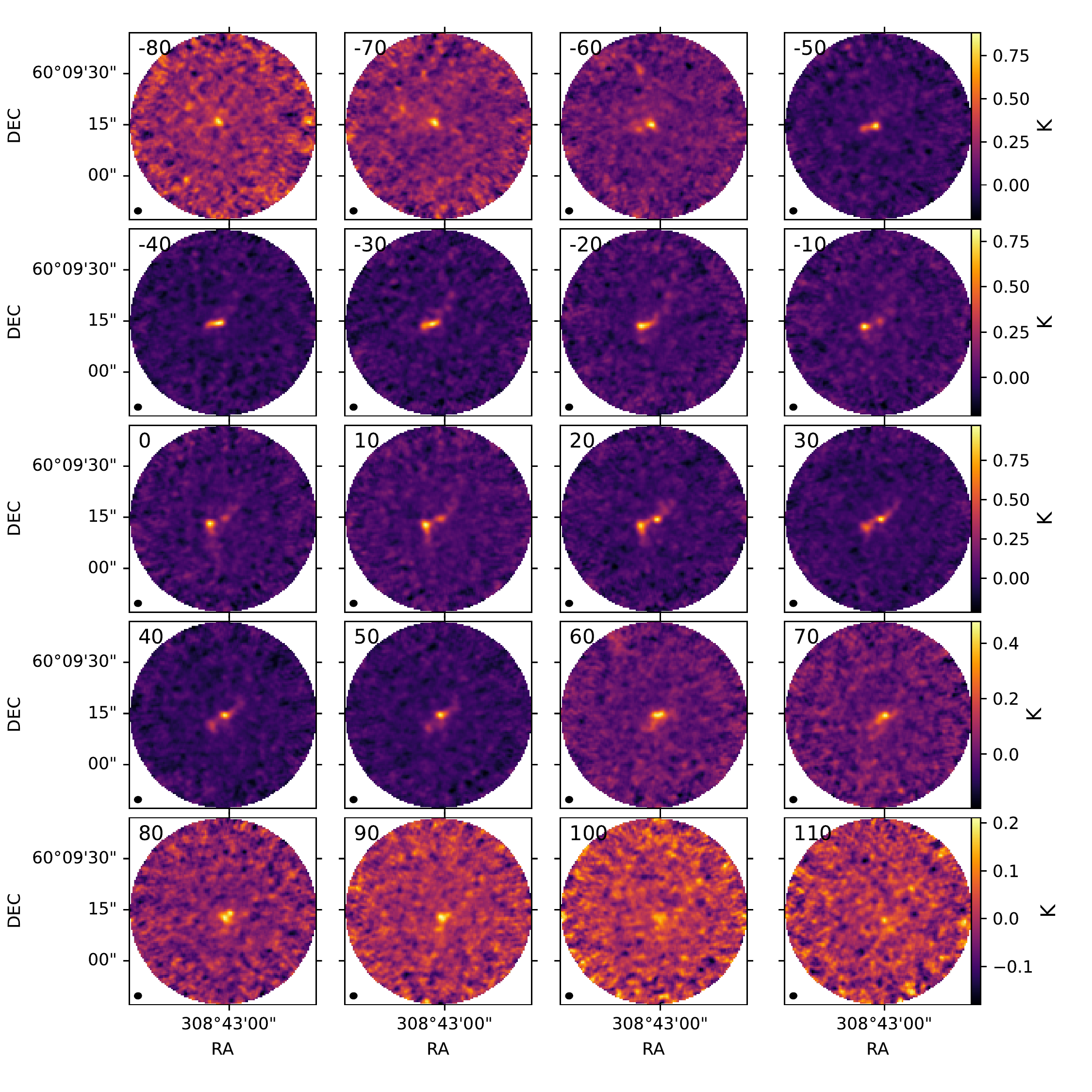}
    \caption{CS(2-1) channel maps, from -80 to 110 km s$^{-1}$.}
\end{figure*}
\begin{figure*}
    \centering
    \includegraphics[width=1.0\textwidth]{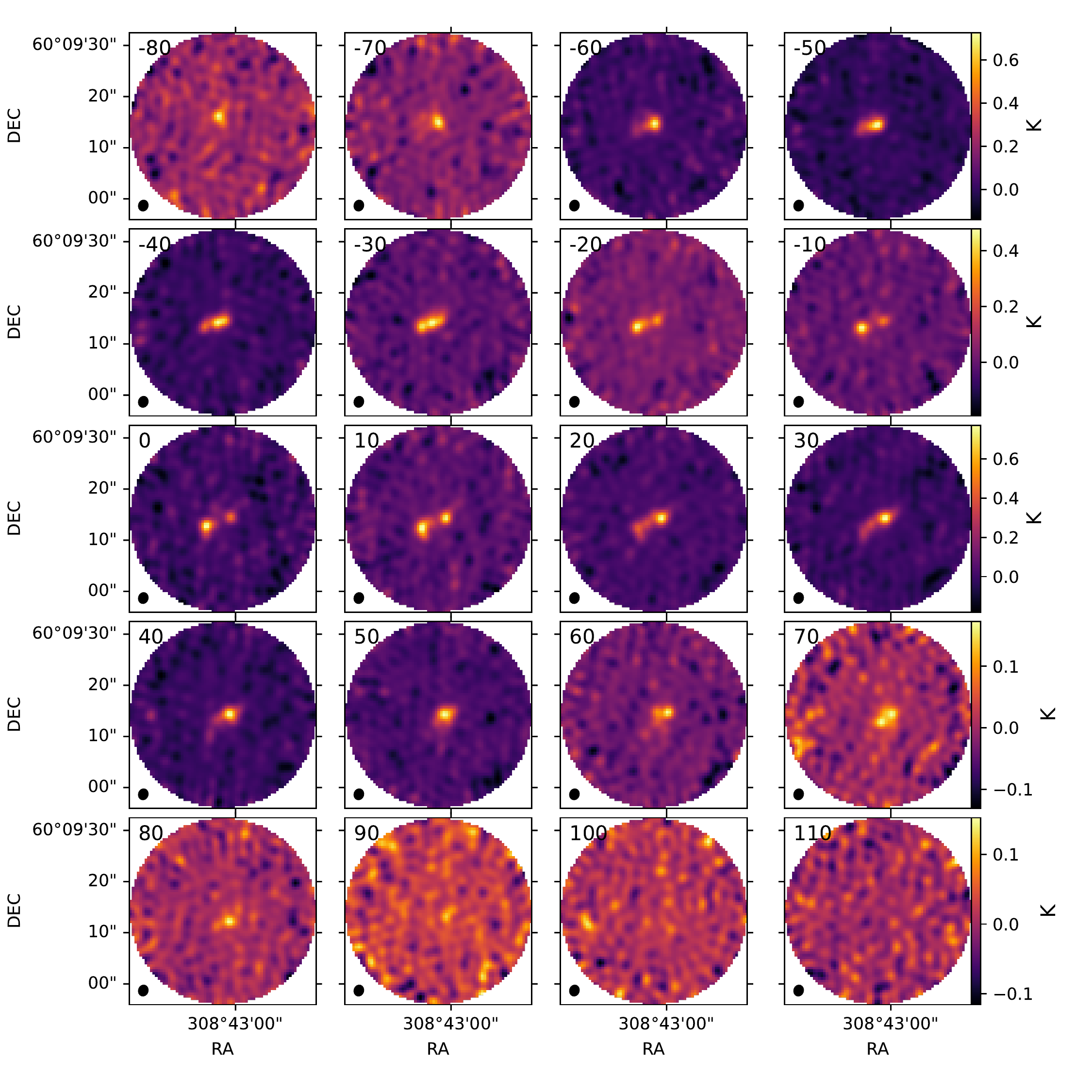}
    \caption{CS(3-2) channel maps, from -80 to 110 km s$^{-1}$.}
\end{figure*}
\begin{figure*}
    \centering
    \includegraphics[width=1.0\textwidth]{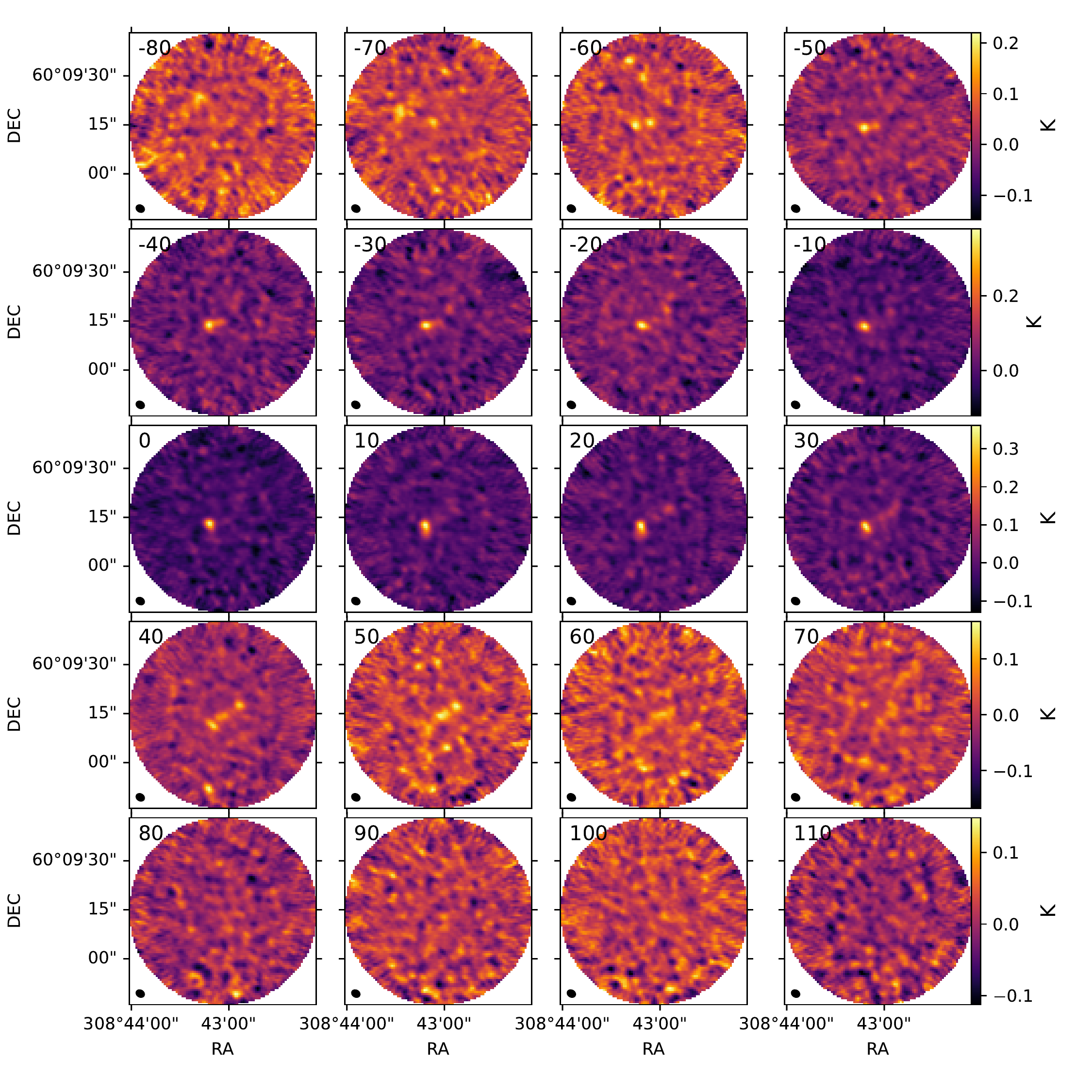}
    \caption{N$_2$H$^+$ channel maps, from -80 to 110 km s$^{-1}$.}
\end{figure*}
\begin{figure*}
    \centering
    \includegraphics[width=1.0\textwidth]{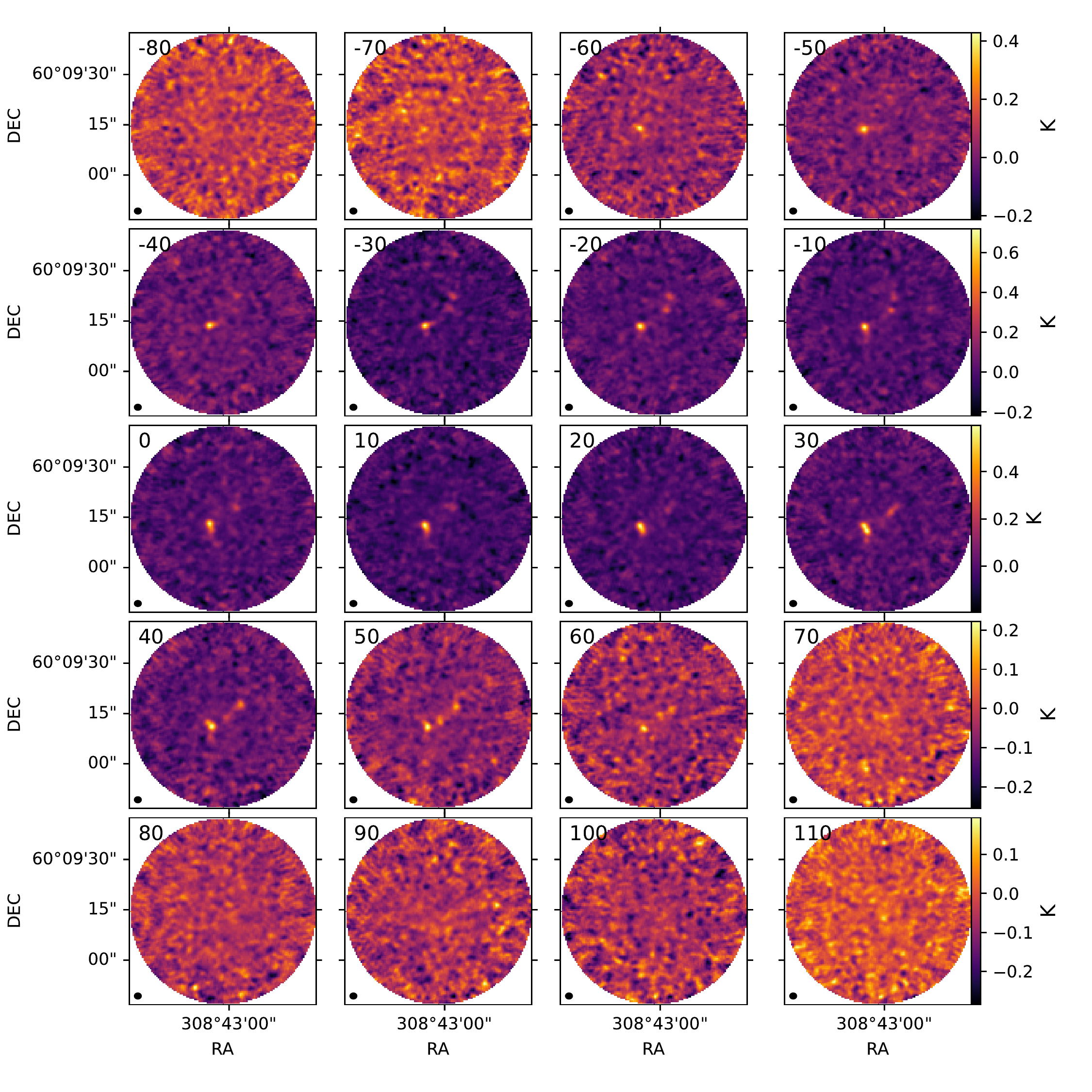}
    \caption{CH$_3$OH(2k-1k) channel maps, from -80 to 110 km s$^{-1}$.}
\end{figure*}
\begin{figure*}
    \centering
    \includegraphics[width=1.0\textwidth]{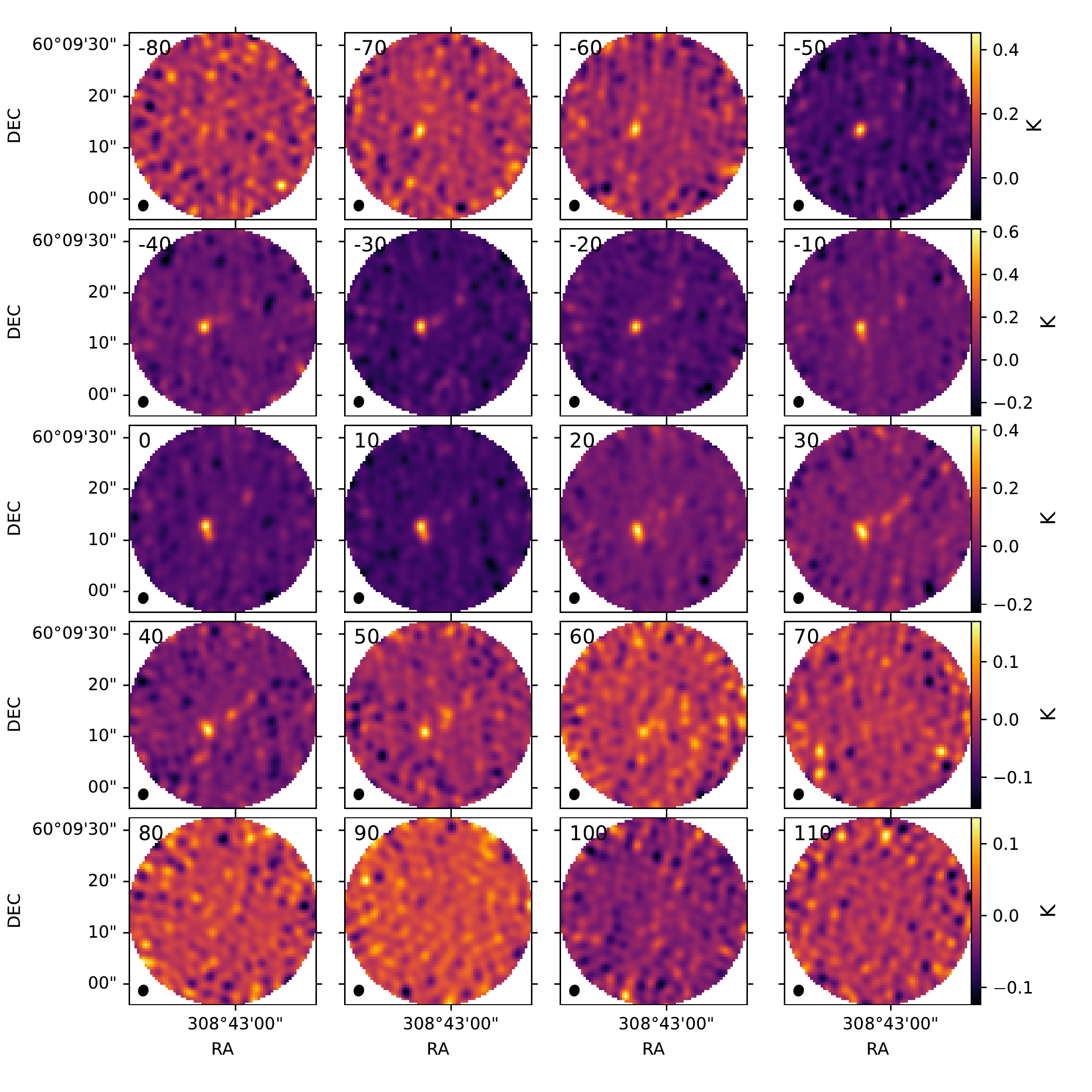}
    \caption{CH$_3$OH(3k-2k) channel maps, from -80 to 110 km s$^{-1}$.}
\end{figure*}
\begin{figure*}
    \centering
    \includegraphics[width=1.0\textwidth]{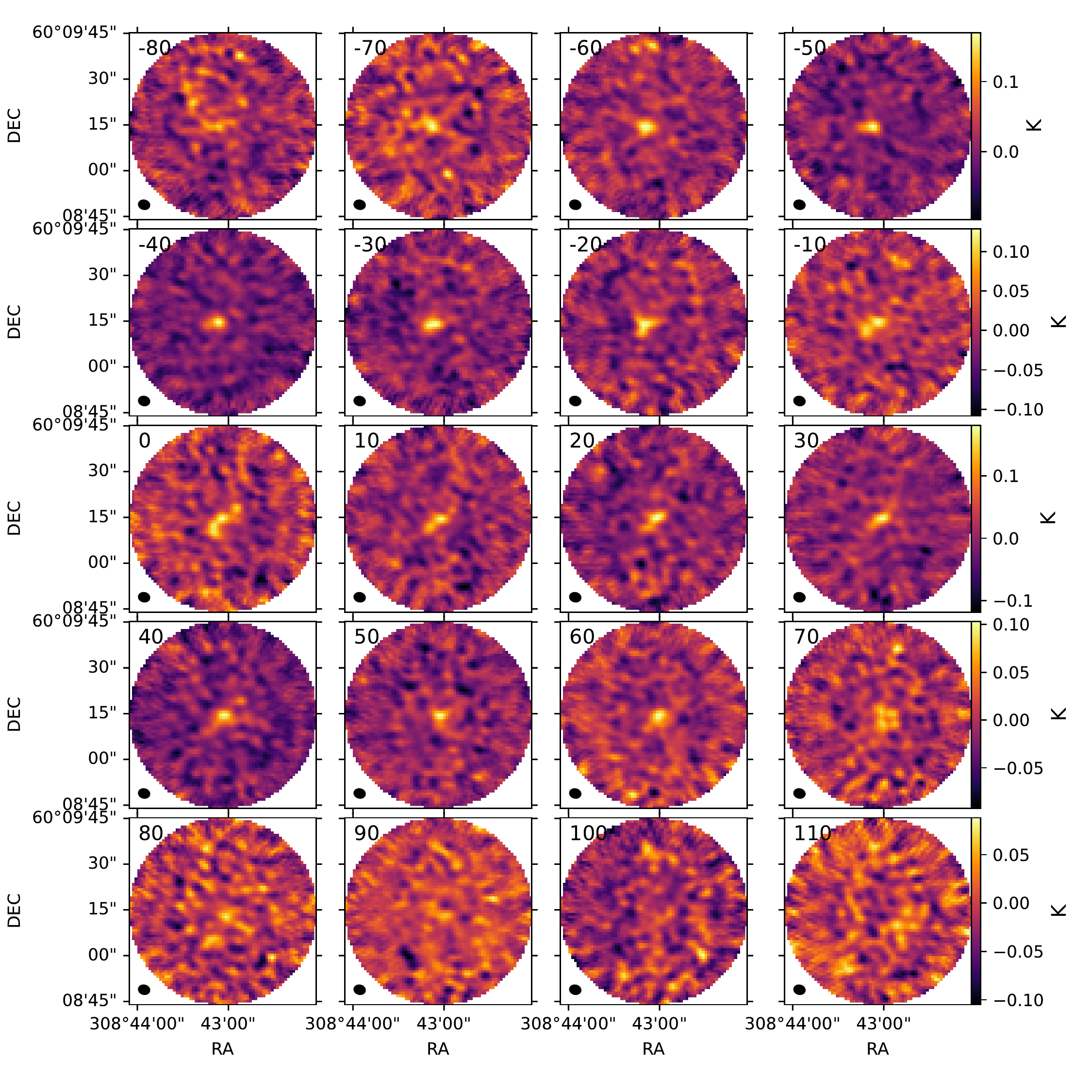}
    \caption{C$_2$H(1-0) channel maps, from -80 to 110 km s$^{-1}$.}
\end{figure*}
\begin{figure*}
    \centering
    \includegraphics[width=1.0\textwidth]{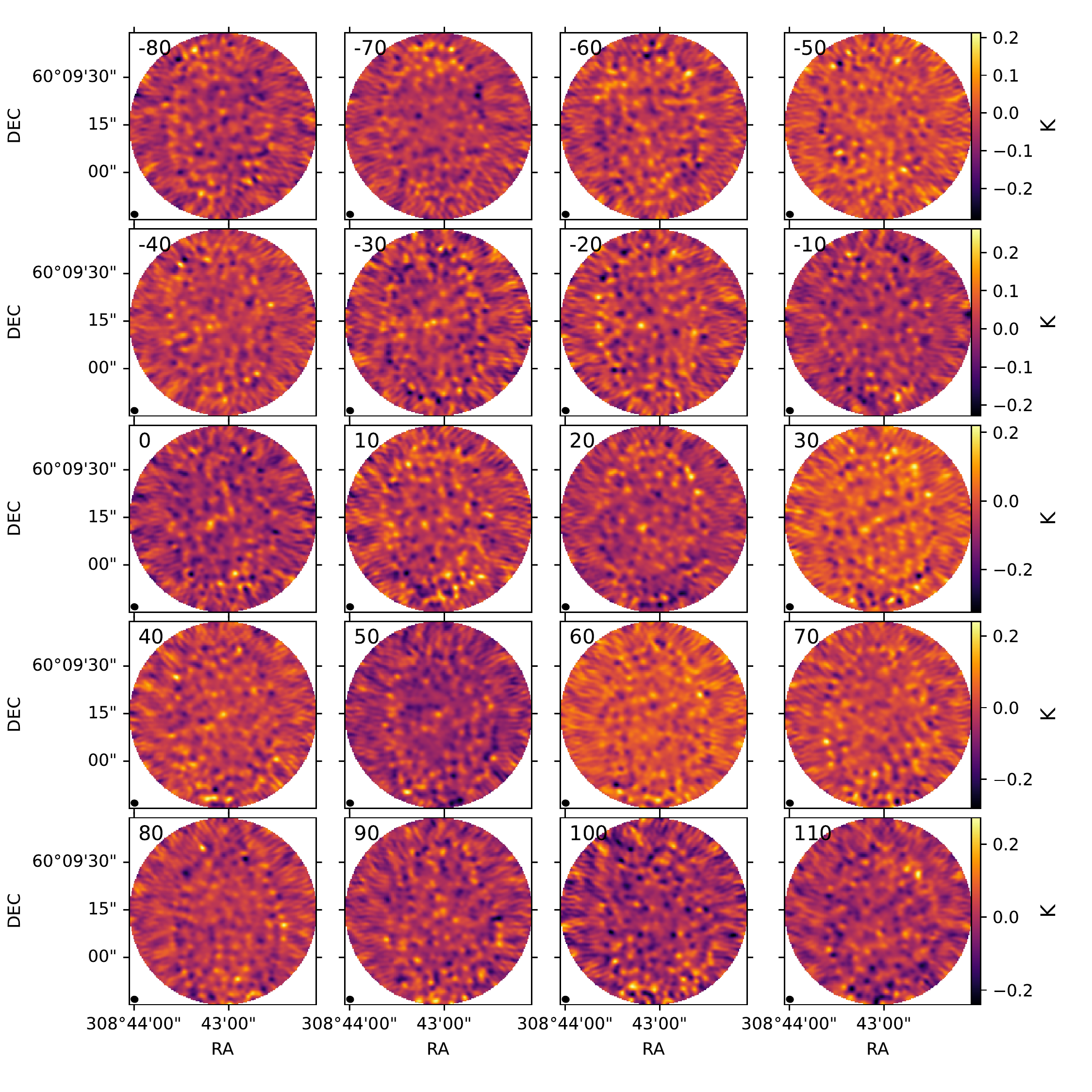}
    \caption{HC$_3$N(10-9) channel maps, from -80 to 110 km s$^{-1}$.}
\end{figure*}
\begin{figure*}
    \centering
    \includegraphics[width=1.0\textwidth]{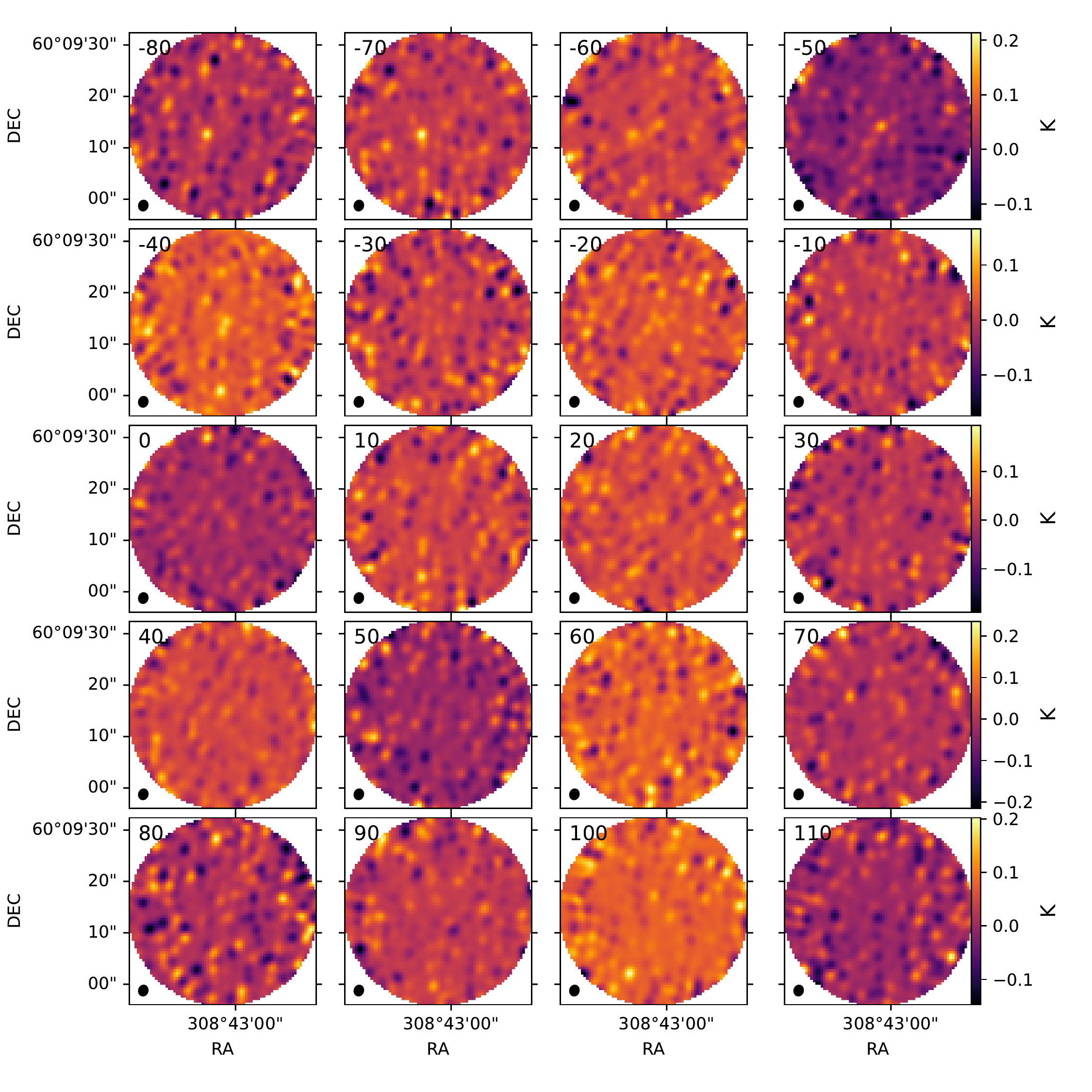}
    \caption{HC$_3$N(16-15) channel maps, from -80 to 110 km s$^{-1}$.}
\end{figure*}
\begin{figure*}
    \centering
    \includegraphics[width=1.0\textwidth]{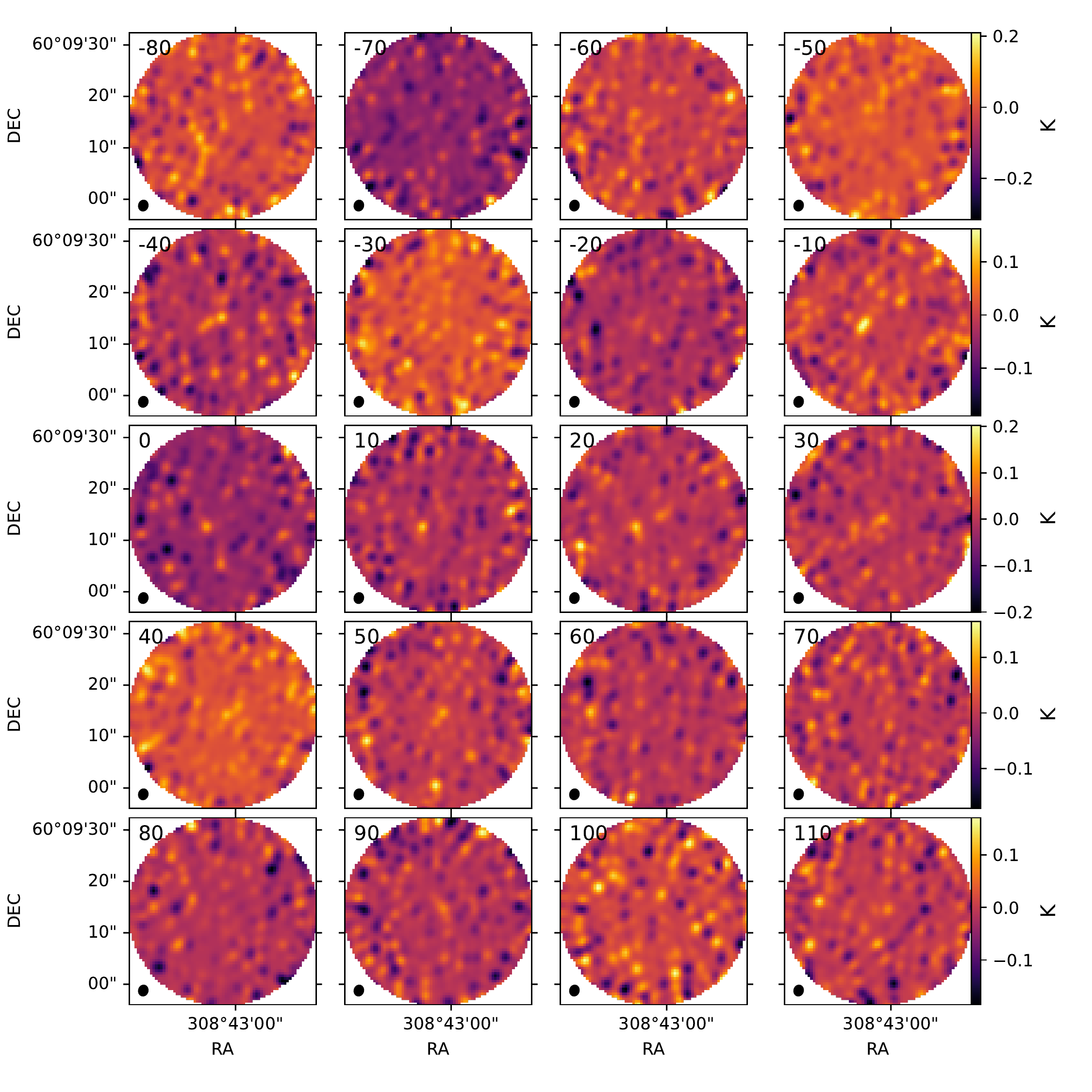}
    \caption{H$_2$CO(2-1) channel maps, from -80 to 110 km s$^{-1}$.}
    \label{app:channelmap_last}
\end{figure*}

\end{document}